\documentclass[%
reprint,
amsmath,amssymb,
aps,
onecolumn,
prfluids,
longbibliography,
]{revtex4-2}

\usepackage{graphicx}
\usepackage{dcolumn}
\usepackage{bm}
\usepackage{dsfont}

\newcommand{\Ov}[1]{{\overline{#1}}}
\newcommand{\EVh}[1]{\mathbb{E}_{h}\mkern-5mu\left[#1\right]}
\newcommand{\EV}[1]{\mathbb{E}\mkern-4mu\left[#1\right]}
\newcommand{\pd}[1]{{\partial_{#1}}}
\newcommand{\PD}[2]{\frac{\partial #1}{\partial #2}}

\newcommand{\dd}{{\mathrm{d}}}
\newcommand{\I}{{\mathrm{i}}}
\newcommand{\Rey}{{\mathrm{Re}}}

\usepackage{xcolor}
\usepackage[normalem]{ulem}

\begin{document}

\preprint{APS/123-QED}

\title{Fluctuation–dissipation relations in isotropic
turbulence\\ from Kraichnan’s fully resolved DIA
closure}

\author{Elias Kohler}
\email{elias.kohler@uni-konstanz.de}
\affiliation{Fachbereich Physik, Universit\"at Konstanz, 78457 Konstanz, Germany}
\author{Matthias Fuchs}
\affiliation{Fachbereich Physik, Universit\"at Konstanz, 78457 Konstanz, Germany}

\date{\today}

\begin{abstract}
Kraichnan’s Direct Interaction Approximation (DIA) provides a classical closure framework for homogeneous turbulence, yet its quantitative predictions across all scales remain incompletely understood. We present a fully resolved numerical solution of the forced DIA equations without additional modelling assumptions, enabling a consistent analysis of stationary spectral properties and temporal dynamics across all wavenumbers. The main focus is on the fluctuation–dissipation relation between the normalised velocity correlation function and the Green function of linear response. This relation holds only at large scales and is systematically violated in the inertial and dissipation ranges. The non-Gaussian nature of this violation is encoded in the DIA through its non-Markovian equations, even though the DIA can be derived using a Gaussian mean-field approximation in the stochastic field functional. In the inertial range, characteristic decay times agree quantitatively with those obtained from direct numerical simulations, while deviations arise in the dissipation range. These results highlight the ability of the DIA to capture temporal dynamics despite its limitations in predicting spectral scaling.
\end{abstract}

\maketitle


\section{Introduction}
\label{sec:introduction}

Owing to the effective unpredictability of individual realisations of the velocity field in turbulent flows, Reynolds introduced a statistical description of turbulence in 1895 \cite{Reynolds1895}. Since then, turbulence has been widely studied using stochastic field theory \citep{Pope2000,Davidson2004,Zhou2021,mccomb2014}. Within this framework, the Martin--Siggia--Rose--De Dominicis--Janssen (MSRDJ) path integral formalism \citep{Martin1973,Janssen1976,Dominicis1976} provides a unified  description of dynamical systems in terms of statistical and quantum field theories \citep{Kamenev2011}. In this formulation, the central observables are the propagators, viz.~the two-point correlation function and the two-point Green function, which characterise the statistical properties of the fields and their response to small external perturbations. Closely related mathematical structures also arise in large-deviation descriptions of dynamical systems,  e.g.~in response theories for climate and geophysical dynamics \cite{Leith1975,Leith1978,Bell1980,North1993,Lacorata2007}. In particular, wave kinetic theories of weak turbulence can be formulated in a manner that parallels the MSRDJ path integral \citep{Guioth2022,Onuki2024}.

In turbulence studies, the MSRDJ approach underlies two important theoretical developments. The first is the field-theoretic description of the stochastically forced Navier--Stokes equations, which provides a statistical framework for fully developed turbulence \citep{Tomassini1997,Canet2022}. Although not originally formulated in this language, the direct interaction approximation (DIA) introduced by Kraichnan in 1959 \cite{Kraichnan1959} can be interpreted as the corresponding mean-field theory.  For homogeneous and isotropic turbulence (HIT), the DIA yields three coupled integro-differential equations governing the two-point velocity correlation functions at equal and unequal times together with the two-point Green’s function in wavenumber space. The second development concerns the field-theoretic description of the directed percolation phase transition. Pomeau \cite{Pomeau1986} proposed that the transition from laminar to turbulent flow can be interpreted as a directed percolation process. This scenario has since been supported by numerous experimental and numerical studies, as reviewed by Hof \cite{Hof2023}. Within the MSRDJ framework, mean-field and renormalization-group analyses of directed percolation have provided important insights into this class of non-equilibrium phase transitions \citep{Hinrichsen2000}.

More generally, MSRDJ functionals can be formulated for abstract non-equilibrium dynamical systems, such as an elastic manifold evolving in a random potential \citep{Cugliandolo1996,Cugliandolo2002}. For certain classes of random potentials, exact dynamical mean-field type equations  can be derived in the limit of large dimensions. This technique can also be used to derive dynamical mean-field equations for the spherical $p$-spin model, for which a fully numerical solution was obtained in Ref.~\cite{Kim2001}. These equations exhibit a structure closely related to Kraichnan’s direct interaction approximation.  In such systems,  the degree of non-equilibrium is commonly characterised through violations of the fluctuation–dissipation theorem (FDT) \citep{Cugliandolo1997,Sollich2002,Buhot2002}.  In thermal equilibrium, the FDT provides an exact relation between the Green’s function and the time derivative of the correlation function \cite{Umberto2008}. For Gaussian statistics, a simpler fluctuation–dissipation relation (FDR) holds that directly relates the Green’s function to the correlation function \cite{Umberto2008}.

FDRs also have a long history in turbulence theory. Kraichnan derived an elementary equilibrium FDR under the assumption of a canonical probability distribution \cite{Kraichnan1959FDR,Kraichnan2000FDR}. In Gaussian turbulence models, a similar relation identifies the retarded Green function with the appropriately normalised two-time correlation function, so that the two quantities are proportional \cite{Kaneda1981,Kida1997,Mccomb1989,McComb2005,Kraichnan1964decay,Koniges1987,Mccomb2017}. Hereafter, an \emph{FDR violation} refers specifically to the failure of this Gaussian correlation--response proportionality, rather than to the absence of every possible generalised response relation. Such a violation provides an operational signature of nonequilibrium dynamics and directly tests the statistical assumptions underlying turbulence closures. 

Despite its importance, the temporal dynamics of the two-point velocity correlation function and the Green’s function in the stationary, forced Navier–Stokes equations have rarely been investigated systematically using direct numerical simulations (DNS). 
In studies of stationary turbulence, energy is typically injected near a forcing wavenumber $\kappa_c$ and transferred towards the viscous range by the nonlinear Navier--Stokes interaction. Modes with $k\ll\kappa_c$ correspond to scales larger than the directly forced structures and lie outside the forward cascade. Recent numerical results indicate that these modes display equilibrium-like statistics and quantitatively satisfy the turbulent FDR \cite{Alexakis2023}. By contrast, in the forward-cascade range,  impulse-response measurements show that the velocity correlation function and the Green function are not proportional in either shell models \cite{Biferale2001,Matsumoto2014} or forced Navier--Stokes turbulence \cite{Carini2010,Matsumoto2021}. In this so-called inertial range, however, universal properties are expected and the FDR is of particular interest. The DNS of Matsumoto \textit{et al.} demonstrates the FDR mismatch across the inertial range in both Eulerian and Lagrangian coordinates \cite{Matsumoto2021}. These observations prompt us to ask why an equilibrium-like FDR is recovered at scales larger than the forcing scale and which mechanism sustains its violation in the forward-cascade range.

From the perspective of statistical field theory, the stationary DIA provides a natural starting point for investigating this issue, as it represents the simplest mean-field approximation of the MSRDJ functional for the stationary, forced Navier–Stokes equations. The functional renormalization group (FRG) for turbulence also starts from the same MSRDJ functional and exploits exact symmetry Ward identities and a nonperturbative RG flow  to derive the large-wave-number structure of correlation and response functions. A leading-order approximation then proves useful \cite{Tomassini1997,Canet2017,Tarpin2018,Gorbunova2021,Canet2022}. As FRG does not yet provide a complete, numerically tractable closure across all scales, we use the DIA for the present study.

We solve the homogeneous, isotropic, and stationary DIA equations numerically for several prescribed forcing spectra, without introducing further closure approximations. Previous numerical studies of the DIA and related two-time closures focused primarily on freely decaying isotropic turbulence \cite{Kraichnan1964decay,Orszag1972,Mccomb1989}. For stationary turbulence, Kraichnan formulated the DIA in the wavenumber--frequency domain, but replaced the full two-time problem with characteristic-frequency approximations rather than solving it numerically \cite{Kraichnan1964steady}. In the forced case, McComb and Quinn \cite{Mccomb2003} instead integrated the nonstationary equations through the transient evolution until a statistically stationary state was reached, reporting results at moderate Taylor--Reynolds numbers. Here, we solve the stationary DIA equations directly for a prescribed forcing spectrum. This avoids resolving the complete transient history and permits the simultaneous determination of the spectrum, correlation, and Green function over wider scale separations and Reynolds-number ranges. To the best of our knowledge, no systematic direct solution of the unabridged stationary DIA equations for prescribed forcing, including both independent two-time functions and comparison with DNS response measurements, has previously been reported.

As a mean-field theory, the DIA is known to exhibit quantitative discrepancies with experiments and direct numerical simulations. In particular, Kraichnan \cite{Kraichnan1959} showed that the DIA predicts an inertial-range energy spectrum $E(k)\propto k^{-3/2}$, whereas turbulent flows follow the Kolmogorov scaling $E(k)\propto k^{-5/3}$ \citep{Kolmogorov1941,Grant1962,MydlarskiWarhaft1996,Ishihara2016,Ishihara2020,Kuchler2023}. The origin of this discrepancy has been discussed extensively in the literature \citep{Kraichnan1964,Edwards1964} and is primarily attributed to the Eulerian convective time $(k v_{\mathrm{rms}})^{-1}$ entering the DIA energy-transfer balance, even though it should not control the local energy cascade. This deficiency motivated Lagrangian-history and related closures,  as Lagrangian velocity correlation times follow the inertial-range turnover scaling $k^{-2/3}$ \cite{Kraichnan1965,Kraichnan1966,Kaneda1981,Kida1997}.
Sweeping nevertheless contributes physically to Eulerian temporal decorrelation \cite{Tennekes1975,Chen1989,He2017,Matsumoto2021}, so the incorrect DIA energy spectrum does not necessarily imply inaccurate two-time dynamics. Several Eulerian and Lagrangian theories impose an FDR-type relation and use it to eliminate the response, or predict closely related connections between correlations and propagators \cite{Kaneda1981,Kida1997,Mccomb2017}. DNS, however, shows that the measured functions are not simply proportional in either coordinate description \cite{Matsumoto2021}.  We show that the stationary DIA solution reproduces the qualitative correlation--response mismatch and predicts quantitatively accurate characteristic times over a substantial part of the wavenumber range resolved by DNS. The comparison also reveals where spectral deficiencies affect the two-time dynamics, most clearly in the dissipation range. 

The paper is organised as follows. Section~\ref{sec:Model} introduces the stochastically forced Navier--Stokes model. Section~\ref{sec:MSRDJ} formulates the MSRDJ path integral, defines the correlation and response functions, and derives the Gaussian mean-field closure corresponding to the Eulerian DIA. Technical details are deferred to the Appendices. Section~\ref{sec:stationary_DIA} specialises the closure to stationary HIT and discusses its temporal and spectral scales. Section~\ref{sec:NumericalSolution} presents the numerical spectrum and analyses the correlation and response functions, their scaling forms, characteristic times, and mismatch. Section~\ref{sec:comparison_dns} compares the DIA predictions with the DNS of Matsumoto \textit{et al.}. Section~\ref{sec:discussion} discusses the results, and Sec.~\ref{sec:conclusions} summarises the implications for fluctuation--response dynamics and Eulerian turbulence closures.

\section{Stochastically forced Navier--Stokes model}\label{sec:Model}

To investigate the velocity correlation and Green functions within a common analytical framework, we consider homogeneous and isotropic turbulence maintained by a Gaussian random force. The resulting stochastic Navier--Stokes equation defines the stationary ensemble and provides the starting point for the MSRDJ functional.

We consider an incompressible fluid with kinematic viscosity $\nu_0$ in the cubic domain $[0,L]^3$ with periodic boundary conditions, corresponding to the torus $\mathbb{T}_L^3$, with reciprocal lattice $Z_L^3:=(2\pi/L)\mathbb{Z}^3$. Our Fourier conventions and the corresponding infinite-volume limit are summarized in Appendix~\ref{app:fourier_half_lattice}. Greek indices denote Cartesian components $\alpha,\beta,\ldots\in I=\{1,2,3\}$, and repeated indices are summed.

In Fourier space, the incompressible Navier--Stokes equation reads \cite{mccomb2014}
\begin{equation}
\partial_t u_t^\alpha(\bm{k})=
Q_t^\alpha[u](\bm{k})
-\nu_0 k^2u_t^\alpha(\bm{k})
+P_{\alpha\beta}(\bm{k})f_t^\beta(\bm{k}),
\label{equ:NavierStokesEq_uk}
\end{equation}
where $k=\lVert\bm{k}\rVert$,
\begin{equation}
Q_t^\alpha[u](\bm{k})
=
M_{\alpha\beta\gamma}(\bm{k})
\sum_{\bm{p}\in Z_L^3}
u_t^\beta(\bm{k}-\bm{p})
u_t^\gamma(\bm{p}) \frac{1}{L^3},
\label{def:Q}
\end{equation}
and
\begin{align}
M_{\alpha\beta\gamma}(\bm{k})
&=
\frac{1}{2\I}
\left[
k_\beta P_{\alpha\gamma}(\bm{k})
+k_\gamma P_{\alpha\beta}(\bm{k})
\right],
\label{def:M_alpha_beta_gamma}\\
P_{\alpha\beta}(\bm{k})
&=
\delta_{\alpha\beta}
-\frac{k_\alpha k_\beta}{k^2}.
\label{def:P_alpha_beta}
\end{align}
The projector $P_{\alpha\beta}$ enforces the incompressibility condition $k_\alpha u_t^\alpha(\bm{k})=0$.

We set the spatially uniform mode to zero, $u_t^\alpha(\bm{0})=0$, and denote the remaining Fourier lattice by $\mathbb{G}=Z_L^3\setminus\{\bm{0}\}$. Since the reality condition couples the Fourier modes at $\bm{k}$ and $-\bm{k}$, a formulation in terms of independent variables requires restricting the stochastic dynamics to a half-lattice $\mathbb{G}_+\subset\mathbb{G}$, whose precise definition is given in Appendix~\ref{app:fourier_half_lattice}. The fields on the complementary half-lattice are then fixed by $u_t^\alpha(-\bm{k})=\Ov{u}_t^\alpha(\bm{k})$, where an overline denotes complex conjugation.

We drive the fluid with homogeneous, isotropic Gaussian white noise. For each $\bm{k}\in\mathbb{G}_+$, let $W_t^\alpha(\bm{k})$ be an independent complex Wiener process satisfying
\begin{equation}
\mathbb{E}\!\left[
\dd W_t^\alpha(\bm{k})
\dd\Ov{W}_t^\beta(\bm{p})
\right]=\delta_{\alpha\beta}\delta_{\bm{k},\bm{p}}\dd t,\qquad
\mathbb{E}[\dd W_t^\alpha(\bm{k})]=0.
\label{equ:Wiener_covariance}
\end{equation}

The stochastic Navier--Stokes equation then reads
\begin{equation}
\dd U_t^\alpha(\bm{k})=\left(Q_t^\alpha[U](\bm{k})-\nu_0k^2U_t^\alpha(\bm{k})\right)\dd t+\sqrt{L^3F(\Vert\bm{k}\Vert)}P_{\alpha\beta}(\bm{k})\dd W_t^\beta(\bm{k}),
\label{equ:NavierStokesEq_stochastic}
\end{equation}
where $F(\Vert\bm{k}\Vert)\geq 0$ is the prescribed forcing spectrum and constitutes the only statistical input to the model, while the bare viscosity $\nu_0$ serves as the control parameter. Equation \eqref{equ:NavierStokesEq_stochastic} is interpreted in the It\^{o} sense. Capital letters distinguish stochastic fields from their realisations. Starting from $U_{t_0}=0$, the system reaches a statistically stationary state after the initial transient, or formally in the limit $t_0\to-\infty$.

\section{MSRDJ functional}
\label{sec:MSRDJ}

The stochastic Navier--Stokes equation \eqref{equ:NavierStokesEq_stochastic} generates, for each component $\alpha$ and wavevector $\bm{k}\in\mathbb{G}_+$, a random path $U^\alpha(\bm{k})$ on the time interval $[t_0,t]$. We denote a realisation of this path by $u^\alpha(\bm{k})$. Expectation values of functionals of these paths can be represented using the Martin--Siggia--Rose--De Dominicis--Janssen (MSRDJ) formalism \cite{Martin1973,Janssen1976,Dominicis1976}. General introductions can be found in Refs.~\cite{Kamenev2011,Helias2020}, while a detailed derivation for the Navier--Stokes equation in real space is given by Canet \cite{Canet2022}. Since the construction is standard, we state only the resulting functional.

For every velocity path realisation $u^\alpha(\bm{k})$ generated by the stochastic Navier--Stokes equation, we introduce a complex response path $\psi^\alpha(\bm{k})$ satisfying $\psi_\tau^\alpha(-\bm{k})=\Ov{\psi}_\tau^\alpha(\bm{k})$. The MSRDJ path space $\Omega_{(t_0,t]}$ comprises all velocity and response paths $u^\alpha(\bm{k})$ and $\psi^\alpha(\bm{k})$ on the evolution interval $(t_0,t]$. Its precise discretized definition and the associated integration measure $\dd\omega$ are given in Appendix~\ref{app:MSRDJ_discretization}.

The expectation value of an integrable path functional $f:\Omega_{(t_0,t]}\to\mathbb{C}$ is represented as
\begin{equation}
\EV{f}=\int_{\Omega_{(t_0,t]}}f(\omega)e^{-S[\omega]}\dd\omega,
\qquad
\EV{1}=1.
\label{def:ExpectedValueMSRDJ}
\end{equation}
The action is decomposed as
\begin{equation}
S[\omega]=S_{\mathrm{lam}}[\omega]+S_{\mathrm{tur}}[\omega].
\end{equation}
Using the reality conditions for the velocity and response fields, the laminar part can be written in the compact continuum notation
\begin{equation}
\begin{split}
S_{\mathrm{lam}}[\omega] = \int_{t_0}^{t} \sum_{\bm{k}\in\mathbb{G}}
\bigg( \frac{1}{2}\Ov{\psi}_\tau^\alpha(\bm{k})F(\Vert\bm{k}\Vert)P_{\alpha\beta}(\bm{k})\psi_\tau^\beta(\bm{k})-\I\Ov{\psi}_\tau^\alpha(\bm{k})\left[\pd{\tau}u_\tau^\alpha(\bm{k})+\nu_0k^2u_\tau^\alpha(\bm{k})\right]\bigg)\frac{\dd\tau}{L^3},
\end{split}
\label{equ:Slam}
\end{equation}
while the turbulent part is
\begin{equation}
S_{\mathrm{tur}}[\omega]=\int_{t_0}^{t}\sum_{\bm{k}\in\mathbb{G}}\I\Ov{\psi}_\tau^\alpha(\bm{k})Q_\tau^\alpha[u](\bm{k})\frac{\dd\tau}{L^3}.
\label{equ:Stur}
\end{equation}

The separation of the action into laminar and turbulent contributions has a physical basis, since setting $Q_\tau^\alpha[u]=0$ recovers the laminar case. The time integrals in \eqref{equ:Slam} and \eqref{equ:Stur} are to be understood as the discrete Riemann sums defined in Appendix~\ref{app:MSRDJ_discretization}. In particular, the continuum expressions are shorthand for the discrete actions \eqref{equ:Slam_discrete} and \eqref{equ:Stur_discrete}. Time derivatives, equal-time limits, and response functions are therefore always interpreted according to the underlying It\^{o} prescription.

\subsection{Two-point correlation and response functions}
\label{sec:two_point_functions}

Using the general expectation value defined in Eq.~\eqref{def:ExpectedValueMSRDJ}, we introduce the two central statistical observables, describing how velocity modes lose memory and respond to perturbations. For $t\geq s$, the two-time velocity correlation function is defined by
\begin{equation}
\EV{U_t^\alpha(\bm{k})\Ov{U}_s^\beta(\bm{p})}
:=\int_{\Omega_{(t_0,t]}}u_t^\alpha(\bm{k})\Ov{u}_s^\beta(\bm{p})e^{-S[\omega]}\dd\omega,
\label{def:UU}
\end{equation}
and the two-time MSRDJ response function by
\begin{equation}
\EV{U_t^\alpha(\bm{k})\Ov{\Psi}_s^\beta(\bm{p})}
:=\int_{\Omega_{(t_0,t]}}u_t^\alpha(\bm{k})\Ov{\psi}_s^\beta(\bm{p})e^{-S[\omega]}\dd\omega.
\label{def:UPsi}
\end{equation}
Here and throughout, capital letters denote stochastic fields, whereas lowercase letters denote their realisations on the MSRDJ path space.

The response function is causal,
\begin{equation}
\EV{U_t^\alpha(\bm{k})\Ov{\Psi}_s^\beta(\bm{p})}=0,
\qquad
t<s,
\label{equ:response_causality}
\end{equation}
while its equal-time value depends on the underlying It\^{o} prescription and will be determined below. Correlations between two response fields vanish for all times,
\begin{equation}
\EV{\Psi_t^\alpha(\bm{k})\Ov{\Psi}_s^\beta(\bm{p})}=0.
\label{equ:EV_psi_psi}
\end{equation}
We prove the relations \eqref{equ:response_causality} and \eqref{equ:EV_psi_psi} in Appendix \ref{app:discrete_Green_function}.

The physical Green function is obtained by perturbing Eq.~\eqref{equ:NavierStokesEq_stochastic} with an infinitesimal external source $P_{\alpha\beta}(\bm{k})h_\tau^\beta(\bm{k})$. We denote expectation values in the perturbed ensemble by $\EVh{\cdot}$. In the MSRDJ formalism, the retarded linear response of the mean velocity is represented as
\begin{equation}
G_{t,s}^{\alpha\beta}(\bm{k},\bm{p})=-\I\EV{U_t^\alpha(\bm{k})\Ov{\Psi}_s^\gamma(\bm{p})}P_{\gamma\beta}(\bm{p}).
\label{def:G}
\end{equation}
The precise discrete definition underlying Eq.~\eqref{def:G} is given in Appendix~\ref{app:discrete_Green_function}. In the continuum limit, it is the usual retarded functional derivative with respect to the source at time $s$. Up to the Fourier-space normalization adopted here, it corresponds to the mean linear response measured for stochastically forced turbulence by Matsumoto \textit{et al.}~\cite{Matsumoto2021}.

For later use, we introduce the compact notation
\begin{equation}
C_{t,s}^{\alpha\beta}(\bm{k},\bm{p}):=\EV{U_t^\alpha(\bm{k})\Ov{U}_s^\beta(\bm{p})}.
\label{def:C}
\end{equation}
The two-time velocity correlation function \eqref{def:C} measures the statistical persistence of a Fourier mode, whereas the retarded Green function \eqref{def:G} is its mean linear response to an infinitesimal external force.

\subsection{Exact relations}
\label{sec:exact_relations}

The MSRDJ functional yields exact relations between two-point and three-point correlation functions. These relations reflect the turbulence closure problem, in which the evolution equation for a correlation function involves unknown correlations of higher order. They can also be derived using several other formalisms \cite{Karman1938,mccomb1990,mccomb2014}. Since equal-time response functions and contact terms depend on the chosen temporal discretization, their derivation using the discrete MSRDJ functional is recalled in Appendix~\ref{app:exact_relations_discrete}. Here, we state the resulting continuum relations.

For strictly separated times $t>s$, causality eliminates the advanced response contribution and gives
\begin{equation}
\begin{split}
\left(\pd{t}+\nu_0k^2 \right) \EV{U_t^\alpha(\bm{k}) \Ov{U}_s^\beta(\bm{p})} =
\sum_{\bm{q}\in\mathbb{G}} M_{\alpha\gamma\rho}(\bm{k}) \EV{U_t^\gamma(\bm{k}-\bm{q}) U_t^\rho(\bm{q}) \Ov{U}_s^\beta(\bm{p})}\frac{1}{L^3}.
\end{split}
\label{equ:C2_C3_ts}
\end{equation}
Similarly, the response function satisfies
\begin{equation}
\begin{split}
\left(
\pd{t}+\nu_0k^2
\right)
\EV{
U_t^\alpha(\bm{k})
\Ov{\Psi}_s^\beta(\bm{p})
}
=
\sum_{\bm{q}\in\mathbb{G}}
M_{\alpha\gamma\rho}(\bm{k})
\EV{
U_t^\gamma(\bm{k}-\bm{q})
U_t^\rho(\bm{q})
\Ov{\Psi}_s^\beta(\bm{p})
}\frac{1}{L^3}.
\end{split}
\label{equ:G2_G3_ts}
\end{equation}
The underlying It\^{o} prescription fixes the equal-time response to
\begin{equation}
\EV{U_t^\alpha(\bm{k})\Ov{\Psi}_t^\beta(\bm{p})}=\I L^3\delta_{\alpha\beta}\delta_{\bm{k},\bm{p}}.
\label{equ:equal_time_response}
\end{equation}
For the equal-time velocity correlation, the corresponding exact relation is
\begin{equation}
\begin{split}
\left(
\pd{t}
+\nu_0k^2
+\nu_0p^2
\right)
\EV{
U_t^\alpha(\bm{k})
\Ov{U}_t^\beta(\bm{p})
}
&=
F(\Vert\bm{k}\Vert)
L^3
P_{\alpha\beta}(\bm{k})
\delta_{\bm{k},\bm{p}}
\\
&\qquad
+
\sum_{\bm{q}\in\mathbb{G}}
M_{\alpha\gamma\rho}(\bm{k})
\EV{
U_t^\gamma(\bm{k}-\bm{q})
U_t^\rho(\bm{q})
\Ov{U}_t^\beta(\bm{p})
}\frac{1}{L^3}
\\
&\qquad
+
\sum_{\bm{q}\in\mathbb{G}}
\Ov{M}_{\beta\gamma\rho}(\bm{p})
\EV{
U_t^\alpha(\bm{k})
\Ov{U}_t^\gamma(\bm{p}-\bm{q})
\Ov{U}_t^\rho(\bm{q})
}\frac{1}{L^3}.
\end{split}
\label{equ:C2_C3_tt}
\end{equation}
The forcing contribution in Eq.~\eqref{equ:C2_C3_tt} appears only once, as required by the It\^{o} prescription.

Equations \eqref{equ:C2_C3_ts}--\eqref{equ:C2_C3_tt} display the closure problem directly. The evolution of a two-point correlation depends on three-point correlations that are not determined by the two-point function itself. Their evolution equations would in turn involve fourth-order correlations, generating an infinite hierarchy. Thus, the relations are exact but not closed without an additional approximation.

In statistically stationary turbulence, the equal-time correlation is independent of time, and the time derivative in Eq.~\eqref{equ:C2_C3_tt} therefore vanishes. Upon setting $\bm{k}=\bm{p}$ and contracting over $\alpha=\beta$, the nonlinear contribution on the right-hand side of Eq.~\eqref{equ:C2_C3_tt} is twice the corresponding contribution in Eq.~\eqref{equ:C2_C3_ts} in the limit $t\to s^+$. Combining these results and using $P_{\alpha\alpha}(\bm{k})=2$ yields the exact initial slope of the stationary two-time velocity correlation function,
\begin{equation}
\left.\pd{t}\EV{U_t^\alpha(\bm{k})\Ov{U}_s^\alpha(\bm{k})}\right|_{t=s^+}
=-F(\Vert\bm{k}\Vert)L^3.
\label{equ:stationary_equal_time_slope}
\end{equation}

\subsection{Self-consistent Gaussian mean-field approximation}
\label{sec:mean_field}

The direct evaluation of the velocity correlation and response functions is complicated by the cubic turbulent action $S_{\mathrm{tur}}$, which leads to the appearance of higher-order correlations in Eqs.~\eqref{equ:C2_C3_ts}, \eqref{equ:G2_G3_ts} and \eqref{equ:C2_C3_tt}. We therefore introduce an effective Gaussian action that collects the dressed Gaussian fluctuations of the full MSRDJ path space into a self-consistently determined quadratic vertex kernel $\bm{\Gamma}_{\mathrm{eff}}$. At this stage, $\bm{\Gamma}_{\mathrm{eff}}$ is not identified with the full two-time functions. Rather, its inverse generates effective velocity-correlation and response kernels.

The correlation and response sectors are treated as independent. In particular, no fluctuation--dissipation relation is imposed between the velocity correlation and Green functions. Any relation, or violation thereof, between them is therefore an output of the approximation rather than an assumption entering the closure.

The two- and three-point functions in Eqs.~\eqref{equ:C2_C3_ts}, \eqref{equ:G2_G3_ts}, and \eqref{equ:C2_C3_tt} are expanded around the effective Gaussian action. Each expansion is truncated after its first non-vanishing contribution. The two-point functions are retained at Gaussian order, whereas the first nonzero contribution to the three-point functions arises through a single insertion of the cubic turbulent action. This moment-wise truncation defines the self-consistent Gaussian mean-field approximation. Details of the derivation are given in Appendix~\ref{app:mean_field_derivation}.

Applying Wick's theorem to the leading contributions expresses the three-point functions in terms of the velocity correlation $C$ and Green function $G$. For $t>s$, the resulting two-time correlation equation is
\begin{equation}
\left(
\pd{t}+\nu_0k^2
\right)
C_{t,s}^{\alpha\beta}(\bm{k},\bm{p})
=
\int_{t_0}^{t}
H_{t,s,\tau}^{\alpha\beta}[C,G](\bm{k},\bm{p})
\dd\tau,
\label{equ:C2_ts}
\end{equation}
while the Green function satisfies
\begin{equation}
\left(
\pd{t}+\nu_0k^2
\right)
G_{t,s}^{\alpha\beta}(\bm{k},\bm{p})
=
\int_{s}^{t}
L_{t,s,\tau}^{\alpha\beta}[C,G](\bm{k},\bm{p})
\dd\tau.
\label{equ:G2_ts}
\end{equation}
The nonlinear integral operators on the right-hand sides depend only on $C$ and $G$ and thereby close the two-time equations. The explicit forms of the nonlinear kernels $H[C,G]$ and $L[C,G]$ are given in Appendix~\ref{app:mean_field_derivation}.

The corresponding equal-time equation is
\begin{equation}
\begin{split}
\left(\pd{t}+\nu_0k^2+\nu_0p^2\right)C_{t,t}^{\alpha\beta}(\bm{k},\bm{p})
&=
F(\Vert\bm{k}\Vert)L^3 P_{\alpha\beta}(\bm{k}) \delta_{\bm{k},\bm{p}}+
\int_{t_0}^{t} \left[H_{t,t,\tau}^{\alpha\beta}[C,G](\bm{k},\bm{p})+\Ov{H_{t,t,\tau}^{\beta\alpha}[C,G](\bm{p},\bm{k})}\,\right]\dd\tau,
\end{split}
\label{equ:C2_tt}
\end{equation}
supplemented by
\begin{equation}
G_{t,t}^{\alpha\beta}(\bm{k},\bm{p})=L^3 \delta_{\bm{k},\bm{p}} P_{\alpha\beta}(\bm{k}).
\label{equ:G2_tt}
\end{equation}
Equation~\eqref{equ:G2_tt} follows directly from Eq.~\eqref{equ:equal_time_response}. 

The resulting Gaussian mean-field equations  are the Eulerian direct-interaction approximation (DIA) introduced by Kraichnan \cite{Kraichnan1959}. In this sense, the
Eulerian DIA is the simplest self-consistent Gaussian mean-field
approximation of the MSRDJ functional. Diagrammatically, the closed equations have the structure of a self-consistent one-loop approximation \cite{Leslie1973,Woodruff1992,Berera2013,mccomb2014}.

\section{Stationary homogeneous and isotropic DIA}
\label{sec:stationary_DIA}

The DIA serves as a simple model for studying FDR violations because it allows us to determine the velocity correlation and retarded Green functions independently. 
We now specialise the derived  correlation and response equations to stationary, homogeneous, and isotropic turbulence (HIT) and derive the spectral and temporal scales used below, including in the comparison with DNS. 

Homogeneity, isotropy, and incompressibility imply
\begin{align}
C_{t,s}^{\alpha\beta}(\bm{k},\bm{p})
&=
P_{\alpha\beta}(\bm{k})
C(\Vert\bm{k}\Vert,t,s)
L^3\delta_{\bm{k},\bm{p}},
\label{equ:isotropic_C}
\\
G_{t,s}^{\alpha\beta}(\bm{k},\bm{p})
&=
P_{\alpha\beta}(\bm{k})
G(\Vert\bm{k}\Vert,t,s)
L^3\delta_{\bm{k},\bm{p}}.
\label{equ:isotropic_G}
\end{align}
In the stationary state, the scalar functions depend only on the time
difference. We write
\begin{align}
C(k,t,s)
&=
C(k)c(k,\vert t-s\vert),
\label{equ:PropertyStat1}
\\
G(k,t,s)
&=
g(k,t-s),
\qquad t\geq s,
\label{equ:PropertyStat2}
\end{align}
where $k=\Vert\bm{k}\Vert$. The normalisations are $c(k,0)=g(k,0)=1$, so that $C(k)=C(k,t,t)$. The Green function vanishes for negative time
differences by causality.

The energy and work spectra are defined by
\begin{equation}
E(k)
=
\frac{k^2}{2\pi^2}C(k),
\qquad
W(k)
=
\frac{k^2}{2\pi^2}F(k).
\label{def:energy_work_spectra}
\end{equation}
The total kinetic energy per unit mass $\mathcal E$, the mean dissipation rate $\varepsilon_d$, and the work rate   $\varepsilon_w$
are
\begin{align}
\mathcal{E}&=\int_0^\infty E(k)\dd k, \label{def:E_tot} \\
\varepsilon_w &= \int_0^\infty W(k)\dd k, \label{def:epsilon_w} \\
\varepsilon_d &=\int_0^\infty  2\nu_0  k^2E(k)\dd k. \label{def:epsilon_d}
\end{align}

We characterise the turbulent state by the dissipation wavenumber and
the Taylor--Reynolds number,
\begin{equation}
k_d= \left(\frac{\varepsilon_d}{\nu_0^3}\right)^{1/4},\qquad
\Rey_\lambda = \frac{v_{\mathrm{rms}} \lambda}{\nu_0}= \sqrt{\frac{20}{3}}\frac{\mathcal{E}}{\sqrt{\varepsilon_d\nu_0}}.
\label{equ:Def_Re_lambda}
\end{equation}
Here, $v_{\mathrm{rms}}^2=2\mathcal{E}/3$ is the mean-square velocity of a single component, and the Taylor microscale $\lambda$ is defined by $\lambda^2=15\nu_0v_{\mathrm{rms}}^2/\varepsilon_d$. The latter is defined through the curvature of the normalised second-order structure function $S_2(r)/v_{\mathrm{rms}}^2$ at the origin \cite{mccomb2014}.

We assume that the general DIA equations admit a stationary solution
whose two-time functions decay as their time separation tends to
infinity. Taking $t_0\to-\infty$ and subsequently the
infinite-volume limit, the tensor equations  \eqref{equ:C2_ts}, \eqref{equ:G2_ts} and \eqref{equ:C2_tt} reduce to three coupled
scalar equations for $c(k,\tau)$, $g(k,\tau)$, and $E(k)$. The angular
reduction of the interaction vertex is standard and is given in
Refs.~\cite{mccomb1990,mccomb2014}. The resulting equations are
\begin{align}
\left(\pd{\tau}+\nu_0k^2\right)c(k,\tau) &= -\int_0^\tau M(k,\tau-s)c(k,s)\dd s +\frac{\Phi(k,\tau)}{E(k)},
\label{equ:DIA_ckt}
\\
\left(\pd{\tau}+\nu_0k^2 \right)g(k,\tau) &= -\int_0^\tau M(k,\tau-s)g(k,s)\dd s,
\label{equ:DIA_gkt}
\\
2\nu_0k^2E(k) &= W(k)+2\Phi(k,0).
\label{equ:DIA_Ek}
\end{align}
The stationary-history term is
\begin{equation}
\begin{split}
\Phi(k,\tau)
=
k^2
\int_0^\infty
N(k,\tau+s)g(k,s)\dd s
-
E(k)
\int_0^\infty
M(k,\tau+s)c(k,s)\dd s.
\end{split}
\label{equ:DIA_Phikt}
\end{equation}
The two memory kernels are
\begin{align}
M(k,\tau)&=\frac{1}{2}\int_0^\infty E(p)c(p,\tau) \int_{-1}^{1} V(k,p,\mu)g(q,\tau) \dd\mu \dd p, \label{equ:DIA_Mg}
\\
N(k,\tau)&=\frac{1}{2}\int_0^\infty E(p)c(p,\tau) \int_{-1}^{1} V(k,p,\mu)c(q,\tau) \frac{E(q)}{q^2} \dd\mu\dd p,
\label{equ:DIA_Mc}
\end{align}
where
\begin{equation}
q=q(k,p,\mu) = \sqrt{k^2-2kp\mu+p^2} = \Vert\bm{k}-\bm{p}\Vert
\end{equation}
and $\mu$ is the cosine of the angle between $\bm{k}$ and $\bm{p}$.
The isotropic vertex is \cite[p. 65]{mccomb2014} 
\begin{equation}
V(k,p,\mu)= \frac{\left(k^4-2k^3p\mu+kp^3\mu \right)\left(1-\mu^2\right)}{k^2-2kp\mu+p^2}=\frac{\left( \Vert\bm{k}\Vert^4-2(\bm{k}\cdot\bm{p})\Vert\bm{k}\Vert^2+(\bm{k}\cdot\bm{p})\Vert\bm{p}\Vert^2\right)\left(\Vert\bm{k}\Vert^2\Vert\bm{p}\Vert^2 -(\bm{k}\cdot\bm{p})^2\right)}{\Vert\bm{k}\Vert^2 \Vert\bm{p}\Vert^2 \Vert\bm{k}-\bm{p}\Vert^2}.
\label{equ:DIA_V_kpmu}
\end{equation}

For prescribed viscosity $\nu_0$ and work spectrum $W(k)$,
Eqs.~\eqref{equ:DIA_ckt}--\eqref{equ:DIA_Mc} contain no adjustable
closure parameters. They determine the stationary spectrum and the
two normalised time functions self-consistently.

The memory-kernel formulation is also convenient numerically. For
fixed kernels and energy spectrum,
Eqs.~\eqref{equ:DIA_ckt} and \eqref{equ:DIA_gkt} are linear
integro-differential equations, while Eq.~\eqref{equ:DIA_Ek} is
algebraic in $E(k)$. This structure suggests a fixed-point iteration
in which the two-time functions, memory kernels, and energy spectrum
are updated successively. Details of the discretisation and
iteration procedure are given in
Appendix~\ref{app:NumericalDiscretisation}.

\subsection{Characteristic time scales}
\label{subsec:discussion_timescales}

Equations \eqref{equ:DIA_ckt} and \eqref{equ:DIA_gkt} contain two
distinct characteristic time scales. The viscous term defines the
dissipation time
\begin{equation}
\tau_{\mathrm{dis}}(k)
:=
\frac{1}{\nu_0k^2},
\label{def:tau_dis}
\end{equation}
while the nonlinear memory kernels contain the Eulerian convective or
sweeping time \cite{Kraichnan1959,Kraichnan1964}
\begin{equation}
\tau_{\mathrm{con}}(k)
:=
\frac{1}{v_{\mathrm{rms}}k}.
\label{def:tau_con}
\end{equation}
The appearance of the latter is seen directly from the equal-time
memory kernel. Expanding the interaction vertex for wavenumbers below
and above the energy-containing range gives
\begin{equation}
M(k,0)
\simeq
\begin{cases}
\displaystyle
\frac{2}{5}k^2v_{\mathrm{rms}}^2,
& k\ll k_E,
\\[2mm]
\displaystyle
k^2v_{\mathrm{rms}}^2,
& k\gg k_E,
\end{cases}
\label{equ:Mg0_asymptotics}
\end{equation}
where $k_E$ denotes a characteristic wavenumber of the
energy-containing range. Thus,
$M(k,0)\sim\tau_{\mathrm{con}}(k)^{-2}$ in both limits. The asymptotic
expressions Eq.~\eqref{equ:Mg0_asymptotics} are derived in Appendix \ref{app:approximation_Mg0}. 

In terms of the dissipation wavenumber and Taylor--Reynolds number,
the two time scales are
\begin{equation}
\begin{split}
\tau_{\mathrm{dis}}(k)
&=
\varepsilon^{-1/3}k_d^{-2/3}
(k_d/k)^2,
\\
\tau_{\mathrm{con}}(k)
&=
\varepsilon^{-1/3}k_d^{-2/3}
15^{1/4}\Rey_\lambda^{-1/2}
(k_d/k).
\end{split}
\label{equ:timescales_con_dis}
\end{equation}
Their ratio is therefore
\begin{equation}
\eta_k
:=
\frac{\tau_{\mathrm{con}}(k)}
{\tau_{\mathrm{dis}}(k)}
=
15^{1/4}\Rey_\lambda^{-1/2}
\frac{k}{k_d}.
\label{def:ratio_con_dis_eta}
\end{equation}

Although the viscous contribution has the same form in
Eqs.~\eqref{equ:DIA_ckt} and \eqref{equ:DIA_gkt}, the short-time
behaviour of $c$ and $g$ is different. Evaluating the equations at
$t=0$ and using the stationary balance
\eqref{equ:DIA_Ek} gives
\begin{equation}
\left.
\pd{t}c(k,t)
\right|_{t=0^+}
=
-\frac{W(k)}{2E(k)},
\qquad
\left.
\pd{t}g(k,t)
\right|_{t=0^+}
=
-\nu_0k^2.
\label{equ:starttime_ckt_gkt}
\end{equation}
The first identity agrees with the exact stationary relation
\eqref{equ:stationary_equal_time_slope}. Hence, the DIA reproduces the
exact right-sided initial slope of the velocity correlation function.
In an unforced wavenumber range, $W(k)=0$ and the initial slope of
$c(k,t)$ vanishes, while the Green function initially decays at the
viscous rate.

\subsection{Stationary energy balance}
\label{subsec:stationary_energy_balance}

The third DIA equation, \eqref{equ:DIA_Ek}, is the spectral energy
balance of stationary turbulence. Introducing the transfer spectrum $T(k):=2\Phi(k,0)$, it can be written as
\begin{equation}
2\nu_0k^2E(k) = W(k)+T(k).
\label{equ:spectral_energy_balance}
\end{equation}

The nonlinear Navier--Stokes interaction conserves the total kinetic
energy and can only redistribute it among wavenumbers. The DIA
preserves this property, so that \cite{Kraichnan1959,mccomb2014}
\begin{equation}
\int_0^\infty T(k)\dd k
=
2\int_0^\infty\Phi(k,0)\dd k
=
0.
\label{equ:int_Tk_vanishes}
\end{equation}
Integration of Eq.~\eqref{equ:spectral_energy_balance} consequently
gives $\varepsilon_d=\varepsilon_w=:\varepsilon.$ An explicit proof of Eq.~\eqref{equ:int_Tk_vanishes} in the present
formulation is given in Appendix~\ref{app:EnergyConservation}.

\subsection{DIA inertial-range spectrum}
\label{subsec:DIA_energy_spectrum}

Kraichnan showed that the Eulerian DIA predicts the inertial-range spectrum $E(k)\propto \sqrt{\varepsilon v_{\mathrm{rms}}}\,k^{-3/2}$ \cite{Kraichnan1959,Kraichnan1964}, a result that we confirm numerically in Sec.~\ref{subsubsec:discussion_ir}. Using Eq.~\eqref{equ:Def_Re_lambda}, this result
can be written as
\begin{equation}
E_{\mathrm{ir}}(k)
=
\alpha_{\mathrm{Kr}}
\Rey_\lambda^{1/4}
\varepsilon^{2/3}
k_d^{-1/6}
k^{-3/2},
\label{equ:Kraichnanscaling}
\end{equation}
where $\alpha_{\mathrm{Kr}}$ is a dimensionless constant.

This prediction differs from the experimentally and numerically
established Kolmogorov inertial-range scaling
\cite{Kolmogorov1941},
\begin{equation}
E_{\mathrm{Ko}}(k)
=
\alpha_{\mathrm{Ko}}
\varepsilon^{2/3}k^{-5/3}.
\label{equ:Kolmogorov_scaling}
\end{equation}

The two spectra also imply different characteristic wavenumbers for
the onset of viscous dissipation. Following Kraichnan
\cite{Kraichnan1959}, we estimate these scales by extending each
inertial-range law to a sharp cutoff and determining the cutoff from
the dissipation balance. For the DIA spectrum,
\begin{equation}
\begin{split}
\varepsilon
\simeq
2\nu_0
\int_0^{k_{\mathrm{Kr}}}
k^2E_{\mathrm{ir}}(k)\dd k
=
\frac{4}{3}
\nu_0\alpha_{\mathrm{Kr}}
\Rey_\lambda^{1/4}
\varepsilon^{2/3}
k_d^{-1/6}
k_{\mathrm{Kr}}^{3/2},
\end{split}
\end{equation}
which gives
\begin{equation}
k_{\mathrm{Kr}}
=
\left(
\frac{3}{4\alpha_{\mathrm{Kr}}}
\right)^{2/3}
\Rey_\lambda^{-1/6}k_d.
\label{equ:k_Kraichnan}
\end{equation}
Applying the same estimate to the Kolmogorov spectrum gives
\begin{equation}
k_{\mathrm{Ko}}
=
\left(
\frac{2}{3\alpha_{\mathrm{Ko}}}
\right)^{3/4}
k_d.
\label{equ:k_Kolmogorov}
\end{equation}
Thus, apart from the dimensionless spectral prefactor, the appropriate
DIA dissipation scale is
\begin{equation}
\kappa_d
:=
\Rey_\lambda^{-1/6}k_d,
\label{def:kappa_d}
\end{equation}
whereas the Kolmogorov dissipation scale is proportional to $k_d$.
The rescaling by $\kappa_d$ will be used below when comparing DIA
spectra at different Reynolds numbers.

\subsection{Eulerian dynamics and alternative closures}
\label{subsec:alternative_closures}

The incorrect $k^{-3/2}$ inertial-range spectrum of the Eulerian DIA is commonly attributed to its treatment of large-scale sweeping. Eulerian velocity modes
decorrelate on the convective time $(kv_{\mathrm{rms}})^{-1}$, which is the physically relevant time scale for Eulerian two-time quantities \cite{Tennekes1975,Chen1989,He2017,Matsumoto2021}. In the DIA, however, this sweeping time also enters the stationary energy-transfer balance. The resulting energy flux depends on the large-scale velocity $v_{\mathrm{rms}}$ and produces the
Kraichnan rather than the Kolmogorov spectrum \cite{Kraichnan1964}.

Kraichnan related this problem to random Galilean invariance and
developed the Lagrangian-history DIA
\cite{Kraichnan1965,Kraichnan1966}. By removing the common advection
by large-scale motions from the dynamics relevant to energy transfer,
Lagrangian closures recover the local turnover time $\varepsilon^{-1/3}k^{-2/3}$ and the Kolmogorov $k^{-5/3}$ spectrum. Related developments include
the Lagrangian renormalised approximations
\cite{Kaneda1981,Kida1997,kaneda2007,Inagaki2021}, the Lagrangian
closure of Okamura \cite{Okamura2018}, Kraichnan's test-field models
\cite{Kraichnan1971a,Kraichnan1971b}, and the EDQNM closure
\cite{Orszag1970}. These approaches successfully describe important
one-time and inertial-range properties. Several of them either impose
a fluctuation--dissipation-type relation, do not retain an independent
response function, or replace the full two-time dynamics by a modelled
decorrelation rate. They are therefore not designed to predict the
difference between the Eulerian velocity correlation and response
functions considered here.

An alternative Eulerian viewpoint was developed by McComb, who
questioned whether random Galilean invariance should be imposed as an
independent statistical principle \cite{Mccomb1989}. The resulting
local-energy-transfer theory is compatible with Kolmogorov scaling,
but its two-time formulation closes the covariance and propagator
through a time-ordered fluctuation--dissipation relation. More
recently, the FRG has provided a
nonperturbative Eulerian formulation based on the MSRDJ functional. It reproduces Kolmogorov equal-time scaling together
with sweeping-dominated Eulerian time correlations, although existing
applications have focused primarily on velocity correlations rather than on a
direct quantitative comparison with the retarded response \cite{Tomassini1997,Canet2017,Tarpin2018,Gorbunova2021}. At large wavenumbers $k\gg \kappa_c$, the FRG predicts Gaussian sweeping and the asymptotic restoration of Kraichnan’s FDR. See \cite{Canet2022} for a comprehensive review. We revisit this point in Sec. \ref{subsec:dns_time_scales}.

The incorrect DIA spectrum does not, by itself, imply an incorrect description of the Eulerian two-time dynamics. Its spectral failure reflects the improper cancellation of sweeping contributions from the energy-transfer equation, whereas sweeping remains a genuine physical mechanism of Eulerian temporal decorrelation. Indeed, the DNS study by Matsumoto \textit{et al.} \cite{Matsumoto2021} shows that the characteristic times of both the Eulerian velocity correlation and response functions scale as $(kv_{\mathrm{rms}})^{-1}$, while their Lagrangian counterparts follow the Kolmogorov timescale. Motivated by this distinction, we examine the temporal dynamics of the Eulerian DIA in detail.

\section{Numerical solution of the stationary DIA}
\label{sec:NumericalSolution}

We solve the stationary DIA equations
\eqref{equ:DIA_ckt}--\eqref{equ:DIA_Ek} for three representative work
spectra,
\begin{equation}
W_i(k) = \frac{\phi_i(k/\kappa_c)}{\displaystyle\int_0^\infty\phi_i(x)\dd x}
\frac{\varepsilon}{\kappa_c},\qquad i\in\{1,2,3\}.
\label{equ:Forcing}
\end{equation}
Here, $\kappa_c$ is the reference wavenumber of the forcing and
$\varepsilon$ is the prescribed mean energy-input rate. The
normalisation ensures
\begin{equation}
\int_0^\infty W_i(k)\dd k = \varepsilon
\end{equation}
for all three spectra. Their shapes and the locations $k_i$ of their
maxima are listed in Table~\ref{tab:Forcing}.

\begin{table}
  \centering
  \caption{Test functions $\phi_i(x)$ defining the shapes of the work spectra $W_i(k)$ for $i\in\{1,2,3\}$. The wavenumber $k_i$ is the point at which the work spectrum $W_i(k)$ has its maximum. The value of $k_3$ is only a numerical approximation.}
  \label{tab:Forcing}
  \begin{tabular}{ccc}
    \hline\hline
$i$ & $\phi_i(x)$ & $k_i/\kappa_c$ \\
\hline
$1$ & $x^4e^{-x^2}$ & $\sqrt{2}$ \\
$2$ & $x^4e^{-x}$ & $4$ \\
$3$ & $x^6e^{-x^4}$ & $1.1067$ \\
    \hline\hline
  \end{tabular}
\end{table}

We use $\kappa_c$ and $\varepsilon$ to define the characteristic
length and time scales. The corresponding natural units are
\begin{equation}
[t] = \varepsilon^{-1/3}\kappa_c^{-2/3}, \qquad [E(k)]=\varepsilon^{2/3}\kappa_c^{-5/3}, \qquad [\mathcal{E}] = \varepsilon^{2/3}\kappa_c^{-2/3}.
\label{equ:numerical_units}
\end{equation}
The viscosity is parametrised by the dissipation wavenumber,
\begin{equation}
\nu_0 = \left(\frac{\varepsilon}{k_d^4}\right)^{1/3}.
\end{equation}
After nondimensionalisation, the ratio $k_d/\kappa_c$ is the remaining
control parameter for a fixed forcing shape. The Taylor--Reynolds
number is not prescribed independently. It is determined from the
resulting total energy through
Eq.~\eqref{equ:Def_Re_lambda}.

The numerical solution uses the memory-kernel representation of the
stationary DIA. At each wavenumber, time is rescaled by $\xi(k)=M(k,0)^{-1/2}$,
which is updated together with the memory kernel during the iteration.
For fixed values of the functions entering the kernels,
Eqs.~\eqref{equ:DIA_ckt} and \eqref{equ:DIA_gkt} are linear
integro-differential equations for $c(k,t)$ and $g(k,t)$, while
Eq.~\eqref{equ:DIA_Ek} yields a linear system for $E(k)$. This
pseudo-linear structure permits a fixed-point iteration in which the
two-time functions, memory kernels, and energy spectrum are updated
successively until convergence.

Numerical solutions of DIA and related two-time closures have
previously been considered for freely decaying and forced isotropic
turbulence \cite{Mccomb2003}, while stationary Lagrangian closures
have been studied in Refs.~\cite{kaneda2007,Okamura2018}. The present
calculation focuses on the simultaneous stationary solution for
$E(k)$, $c(k,t)$, and $g(k,t)$ for several forcing spectra and a range
of $k_d/\kappa_c$. Details of the wavenumber and time discretisations,
the fixed-point iteration, and the convergence criteria are given in
Appendix~\ref{app:NumericalDiscretisation}.

The numerical results are presented in two stages. We first examine the stationary energy spectra and their dependence on the Reynolds number and forcing shape. We then analyse the correlation and Green functions, their characteristic timescales, and the resulting violation of the fluctuation--dissipation relation.

\subsection{Energy spectra}
\label{sec:EnergySpectra}

The stationary energy spectrum is analysed here for two specific purposes. First, it provides an internal validation of the numerical solution through the known spectral predictions and conservation properties of Eulerian DIA. Second, it identifies the wavenumber ranges and DIA dissipation scale governing the memory kernels and the subsequent two-time correlation–response dynamics.

\begin{figure}
\centerline{\includegraphics{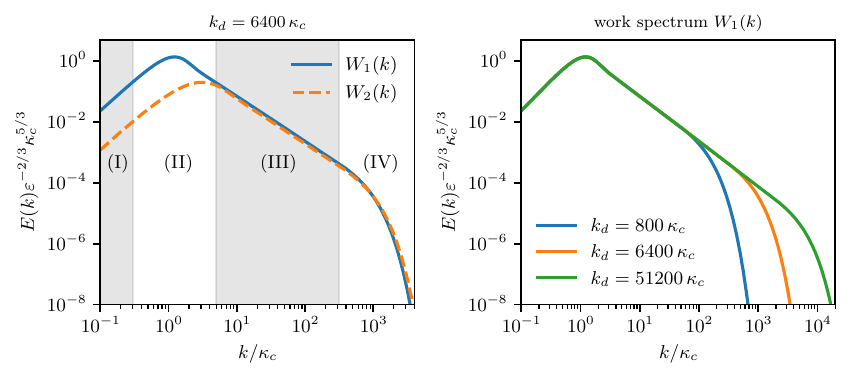}}
\caption{Stationary DIA energy spectra. The left panel compares the energy spectra obtained with the work spectra $W_1$ and $W_2$ at fixed
$k_d=6400\kappa_c$. The right panel shows energy spectra for $W_1$ at three
increasing dissipation wavenumbers.}
\label{fig:Overview_Ek}
\end{figure}

Figure~\ref{fig:Overview_Ek} provides an
overview of the numerical spectra. The left panel compares the energy spectra obtained with the work spectra $W_1$ and $W_2$ at the fixed dissipation wavenumber
$k_d=6400\kappa_c$. Their different forcing shapes mainly
affect the low-wavenumber part of the energy spectrum. The right panel
shows spectra for the fixed work spectrum $W_1$ and increasing values
of $k_d$. 

For the subsequent analysis, we distinguish four wavenumber ranges:
\begin{itemize}
\item[$\mathrm{(I)}$]
\textit{Large-scale range} $\mathrm{(lr)}$:
$k\lesssim 0.3\kappa_c$.
\item[$\mathrm{(II)}$]
\textit{Energy-containing range} $\mathrm{(er)}$:
$0.3\kappa_c\lesssim k\lesssim k_{\mathrm{bot}}$.
\item[$\mathrm{(III)}$]
\textit{Inertial range} $\mathrm{(ir)}$:
$k_{\mathrm{bot}}\lesssim k\lesssim k_{\mathrm{top}}$.
\item[$\mathrm{(IV)}$]
\textit{Dissipation range} $\mathrm{(dr)}$:
$k\gtrsim k_{\mathrm{top}}$.
\end{itemize}
For the work spectrum $W_1$, the lower inertial-range boundary is
approximately $k_{\mathrm{bot}}\simeq 5\kappa_c$. In the large-scale range, the spectrum behaves as $E(k)\propto k^2$, independent of $W_i$. The position and shape of the energy-containing range depend on the work spectrum. Increasing $k_d$ leaves these two
low-wavenumber ranges approximately unchanged but moves
$k_{\mathrm{top}}$ to larger wavenumbers, thereby extending the
inertial range.

All work spectra considered here decay exponentially or faster outside their
respective forcing ranges. Direct energy input is therefore negligible
at sufficiently large wavenumbers, although the spectrum can retain an
indirect forcing dependence through nonlinear transfer. In the
following subsections, we quantify the inertial-range power law and
the dissipation-range decay and examine whether they exhibit asymptotic universality, viz.~whether their dependence on
forcing shape and Reynolds number can be removed by an appropriate
rescaling.

\subsubsection{Inertial range}
\label{subsubsec:discussion_ir}

At finite Reynolds number, we represent the inertial-range spectrum as
\begin{equation}
E_{\mathrm{ir}}(k)= \alpha(\Rey_\lambda) \Rey_\lambda^{2/9} \varepsilon^{2/3}
\kappa_d^{-5/3}\left(\frac{k}{\kappa_d}\right)^{-\beta(\Rey_\lambda)}.
\label{equ:Ek_powerlaw_ir}
\end{equation}
To determine the effective exponent, we introduce the local
logarithmic slope
\begin{equation}
\beta_{\mathrm{loc}}(k)
=
-\frac{\dd\log E(k)}{\dd\log k}.
\label{def:beta_local}
\end{equation}

\begin{figure}
\centerline{\includegraphics{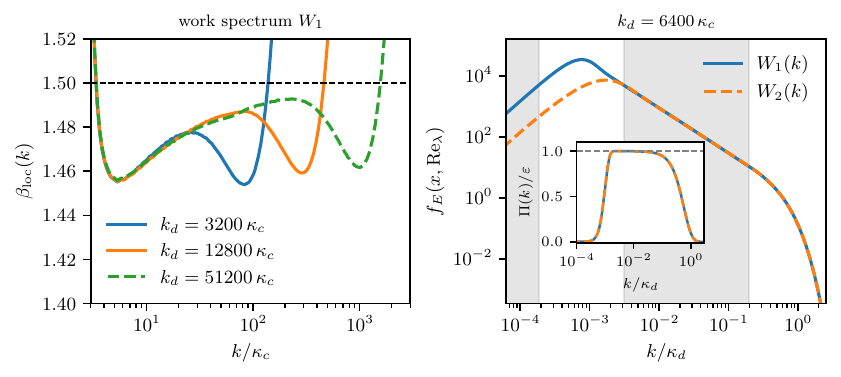}}
\caption{Identification and rescaling of the DIA inertial range. The
left panel shows the local exponent $\beta_{\mathrm{loc}}(k)$ for the
work spectrum $W_1$ and three increasing dissipation wavenumbers. The
 local maximum in the inertial range defines
$\beta(\Rey_\lambda)$. The right panel shows the rescaled energy spectra
for $W_1$ and $W_2$ at
$k_d=6400\kappa_c$. The inset displays the corresponding normalised
energy flux $\Pi(k)/\varepsilon$.}
\label{fig:betaLoc_rescale}
\end{figure}

The left panel of Fig.~\ref{fig:betaLoc_rescale} shows
$\beta_{\mathrm{loc}}(k)$ for the work spectrum $W_1$ and three
increasing values of $k_d$. At low wavenumbers, the local exponent is
influenced by the energy-containing range. At larger wavenumbers, it
is modified by the crossover to the dissipation range. Between these
regions, a local maximum develops and becomes increasingly pronounced
as $k_d$ increases. We define $\beta(\Rey_\lambda)$ as the local
maximum of $\beta_{\mathrm{loc}}(k)$ in the inertial
range. This provides an operational estimate of the inertial-range
exponent with minimal influence from the adjacent spectral ranges.

Let $k_*$ denote the position of this maximum. The corresponding
prefactor is obtained from Eq.~\eqref{equ:Ek_powerlaw_ir} as
\begin{equation}
\alpha(\Rey_\lambda)=E(k_*)
\Rey_\lambda^{-2/9}
\varepsilon^{-2/3}
\kappa_d^{5/3}
\left(
\frac{k_*}{\kappa_d}
\right)^{\beta(\Rey_\lambda)}.
\label{def:alpha_effective}
\end{equation}

To examine the collapse across Reynolds numbers, we define the
rescaled spectrum
\begin{equation}
f_E(x,\Rey_\lambda):=\frac{E(k)}{\alpha(\Rey_\lambda)\Rey_\lambda^{2/9}\varepsilon^{2/3}\kappa_d^{-5/3}}, \qquad x=\frac{k}{\kappa_d}.
\label{def:fE_finite_Re}
\end{equation}
The right panel of Fig.~\ref{fig:betaLoc_rescale} shows the rescaled spectra $f_E(x,\Rey_\lambda)$ obtained with $W_1$ and $W_2$ at fixed $k_d=6400 \kappa_c$. The two spectra collapse beyond their respective energy-containing
ranges. Numerically, the upper inertial-range boundary is located at
approximately $k_{\mathrm{top}} \simeq 0.2\kappa_d$.

As an independent diagnostic of the inertial range, we calculate the
spectral energy flux \cite{mccomb2014}
\begin{equation}
\Pi(k)
=
\int_k^\infty T(p)\dd p.
\label{def:energy_flux}
\end{equation}
The inset of the right panel shows that
$\Pi(k)/\varepsilon\simeq1$ throughout the range identified as
inertial. The nearly constant flux confirms that the power-law region
corresponds to an approximately conservative transfer of energy
through wavenumber space.

\begin{figure}
\centerline{\includegraphics{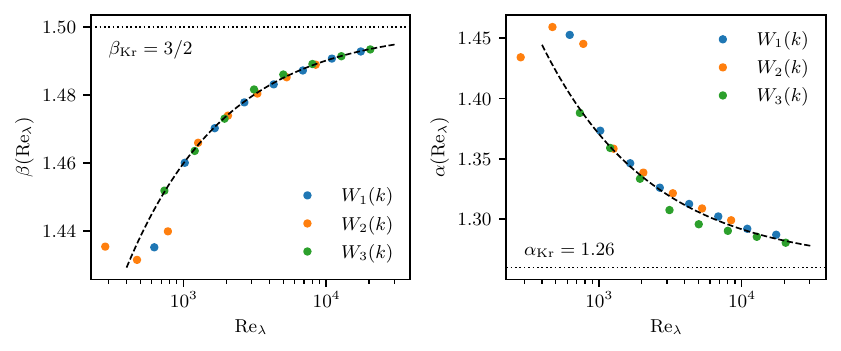}}
\caption{Reynolds-number dependence of the DIA inertial-range
parameters for the three work spectra $W_i$. The left panel shows
$\beta(\Rey_\lambda)$ and the right panel shows
$\alpha(\Rey_\lambda)$. Dashed lines represent the empirical fits
\eqref{equ:beta_Kr_Re} and \eqref{equ:alpha_Kr_Re}.}
\label{fig:betaKr_alphaKr}
\end{figure}

Figure~\ref{fig:betaKr_alphaKr} shows the resulting exponent and
prefactor for all three work spectra and the full range of
Taylor--Reynolds numbers considered. The data obtained from the
different forcing shapes approach the same limiting values. Their
finite-Reynolds-number dependence is described empirically by
\begin{align}
\beta(\Rey_\lambda)
&\simeq
1.50-2.95\,\Rey_\lambda^{-0.62},
\label{equ:beta_Kr_Re}
\\
\alpha(\Rey_\lambda)
&\simeq
1.26+5.42\,\Rey_\lambda^{-0.57}.
\label{equ:alpha_Kr_Re}
\end{align}
These fits are shown as dashed lines in
Fig.~\ref{fig:betaKr_alphaKr}. They describe the present numerical
solutions but are not implied by Kraichnan's asymptotic scaling
argument.

In the large-Reynolds-number limit, the numerical results approach $\beta_{\mathrm{Kr}}=1.50$ and $\alpha_{\mathrm{Kr}}=1.26$. The limiting exponent confirms Kraichnan's analytical prediction
$E(k)\propto k^{-3/2}$ \cite{Kraichnan1959}. The value
$\alpha_{\mathrm{Kr}}=1.26$ is the corresponding numerical estimate
of the asymptotic prefactor in the normalisation adopted in
Eq.~\eqref{equ:Ek_powerlaw_ir}. Within the tested set of work spectra,
both asymptotic values are insensitive to the forcing shape and can thus be considered universal.

\subsubsection{Dissipation range and spectral collapse}
\label{subsubsec:discussion_dr}

In the dissipation range, the energy spectrum is rapidly cut off by viscous damping and takes a (stretched) exponential form. A detailed analysis of the dissipation-range exponents is given in Appendix~\ref{app:dissipation_range_analysis}. In the near-dissipation range $0.2\lesssim x\lesssim4$, the local stretched-exponential exponent varies continuously, whereas for $x\gtrsim4$ it becomes numerically indistinguishable from unity. The far-dissipation spectrum is therefore described by a power law multiplied by a simple exponential, consistent with Kraichnan's asymptotic prediction \cite{Kraichnan1959}.

The rescaled spectra $f_E(x,\Rey_\lambda)$ defined in Eq.~\eqref{def:fE_finite_Re} collapse onto a common master curve outside the energy-containing range. In the large-Reynolds-number limit, $\beta(\Rey_\lambda)\to3/2$ and $\alpha(\Rey_\lambda)\to\alpha_{\mathrm{Kr}}$, giving
\begin{equation}
E(k)=\alpha_{\mathrm{Kr}}\Rey_\lambda^{2/9}\varepsilon^{2/3}\kappa_d^{-5/3}f_E(x).
\label{equ:universal_E_scaling}
\end{equation}
The master function connects the inertial-range scaling
\begin{equation}
f_E(x)\sim x^{-3/2},\qquad x\lesssim0.2,
\label{equ:fE_ir}
\end{equation}
to the fitted far-dissipation asymptote
\begin{equation}
f_E(x)\simeq 61 x^2e^{-6.36x},\qquad x\gtrsim4.
\label{equ:fE_dr}
\end{equation}

\begin{figure}
\centerline{\includegraphics{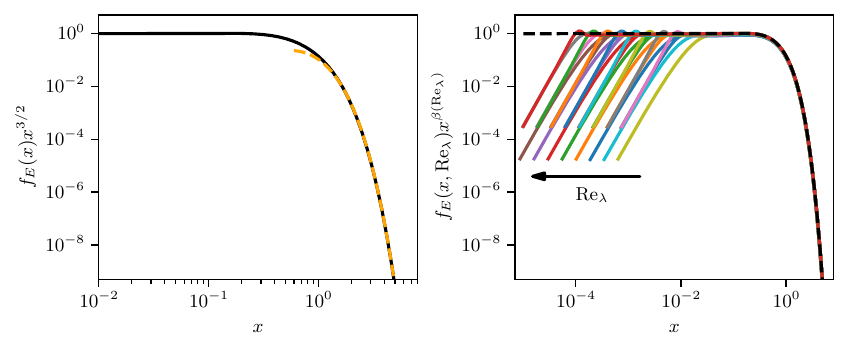}}
\caption{DIA spectral master function. The left panel shows the compensated master function $f_E(x)x^{3/2}$ together with the far-dissipation-range asymptote \eqref{equ:fE_dr}, indicated by the orange dashed line. The right panel shows the compensated spectra $f_E(x,\Rey_\lambda)x^{\beta(\Rey_\lambda)}$ computed in this work for Reynolds numbers $300\lesssim\Rey_\lambda\lesssim 20000$. The black arrow indicates the direction of increasing $\Rey_\lambda$, while the black dashed line represents the master function shown in the left panel.}
\label{fig:fE}
\end{figure}

The left panel of Fig.~\ref{fig:fE} shows the DIA master function $f_E(x)x^{3/2}$ together with the far-dissipation asymptote \eqref{equ:fE_dr}. The right panel demonstrates the spectral collapse across the Reynolds numbers and forcing spectra considered here.

\subsubsection{Relation to DNS and experiments}
\label{subsubsec:spectrum_comparison}

The preceding analysis characterises the stationary background governing the DIA two-time dynamics. In the inertial range, the DIA exponent approaches $3/2$, whereas experiments and DNS support the Kolmogorov value $5/3$, subject to finite-Reynolds-number and intermittency corrections \cite{MydlarskiWarhaft1996,Ishihara2016,Ishihara2020,Kuchler2023}. The finite-Reynolds-number corrections predicted by the DIA do not eliminate this asymptotic discrepancy. The crossover to the dissipation range also differs between the DIA and DNS. In the DNS, it occurs at $k\simeq0.2k_d$, whereas in the DIA it occurs at $k\simeq0.2\kappa_d$, where $\kappa_d=\Rey_\lambda^{-1/6}k_d$. This distinction is particularly important when rescaling the energy spectra and identifying the wavenumber range over which inertial-range scaling is expected.

Differences also arise in the dissipation range. DNS and experiments commonly exhibit a stretched-exponential near-dissipation spectrum with an effective exponent below or close to unity \cite{Khurshid2018,Gorbunova2020}. By contrast, the local decay exponent of the DIA spectrum remains above unity throughout most of the near-dissipation range and therefore produces a faster spectral decay. Only in the far-dissipation range do both descriptions approach a simple exponential, although their characteristic wavenumbers and exponential coefficients remain quantitatively different. The detailed analysis of the local DIA exponent is given in Appendix~\ref{app:dissipation_range_analysis}.

\subsection{Temporal correlation functions}
\label{subsec:time_dependent_functions}

The stationary energy spectrum separates the dynamics into the four
wavenumber ranges introduced in Sec.~\ref{sec:EnergySpectra}. We now
examine the time dependence of the velocity correlation
$c(k,t)$ and the Green function $g(k,t)$ in each range, using both
numerical results and asymptotic arguments.

\begin{figure}
\centerline{\includegraphics{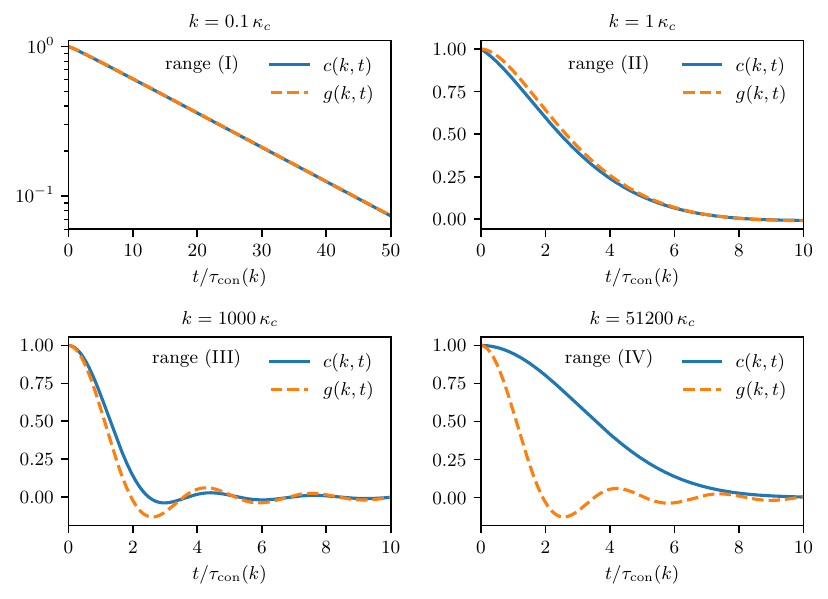}}
\caption{Representative two-time DIA functions for
$k_d=51200\kappa_c$ and the work spectrum $W_1$. The panels show
$c(k,t)$ and $g(k,t)$ in the large-scale range $\mathrm{(I)}$, the
energy-containing range $\mathrm{(II)}$, the inertial range
$\mathrm{(III)}$, and the dissipation range $\mathrm{(IV)}$,
respectively. Time is normalised by
$\tau_{\mathrm{con}}(k)$.}
\label{fig:Overview_decay_ckt_gk}
\end{figure}

Figure~\ref{fig:Overview_decay_ckt_gk} provides a representative
overview for $k_d=51200\kappa_c$ and the work spectrum $W_1$. Time is
normalised by the convective scale
$\tau_{\mathrm{con}}(k)$ defined in
Eq.~\eqref{def:tau_con}. The same regimes are
observed for the other dissipation wavenumbers $k_d$ and work spectra $W_i$
considered.

In the large-scale range $\mathrm{(I)}$, shown in the upper left
panel of Fig.~\ref{fig:Overview_decay_ckt_gk}, both functions decay exponentially and satisfy
$c(k,t)=g(k,t)$ to numerical accuracy. This behaviour will be derived
analytically in Sec.~\ref{subsub:decay_ckt_gkt_llr}. The upper right
panel shows a wavenumber in the forced energy-containing range
$\mathrm{(II)}$. Here, $c(k,t)$ and $g(k,t)$ already differ at short
times, as shown by Eq. \eqref{equ:starttime_ckt_gkt}. Both functions also
develop weakly damped oscillations and multiple zero crossings.

The lower left panel shows the inertial range $\mathrm{(III)}$.
The two functions have similar but distinct decay times, while the
oscillations of $c(k,t)$ are more strongly damped than those of
$g(k,t)$. In the dissipation range $\mathrm{(IV)}$, shown in the lower
right panel, the decay of $g(k,t)$ remains qualitatively similar to
its inertial-range form. By contrast, $c(k,t)$ becomes approximately
Gaussian and decays on a substantially longer characteristic time.

The overview demonstrates that $c(k,t)$ and $g(k,t)$ generally possess
different characteristic decay times. We examine whether the velocity
correlation and the Green function admit the
scaling forms
\begin{align}
c(k,t) &= f_c(t/\xi_c(k)),\label{def:fc}
\\
g(k,t)&= f_g(t/\xi_g(k)).
\label{def:fg}
\end{align}
The characteristic halving times $\xi_c $ and $ \xi_g$  are defined by
\begin{equation}
c(k,\xi_c(k))=g(k,\xi_g(k))=\frac{1}{2}.
\label{def:halving_times}
\end{equation}

Numerically, we resolve wavenumbers up to $k=4k_d$. For Taylor--Reynolds numbers $\Rey_\lambda\gtrsim6500$, the ratio of the convective to the dissipative timescale satisfies
\begin{equation}
\eta_k=\frac{\tau_{\mathrm{con}}(k)}{\tau_{\mathrm{dis}}(k)}\lesssim8\Rey_\lambda^{-1/2}\simeq0.1,\qquad k\leq4k_d.
\label{equ:StrictEstimation_eta_k_4kd}
\end{equation}
This estimate follows from condition \eqref{def:ratio_con_dis_eta}. We investigate the dynamics of $c(k,t)$ and $g(k,t)$ in this high-Reynolds-number regime, where $\tau_{\mathrm{dis}}(k) \gg \tau_{\mathrm{con}}(k)$. 

\begin{figure}
\centerline{\includegraphics{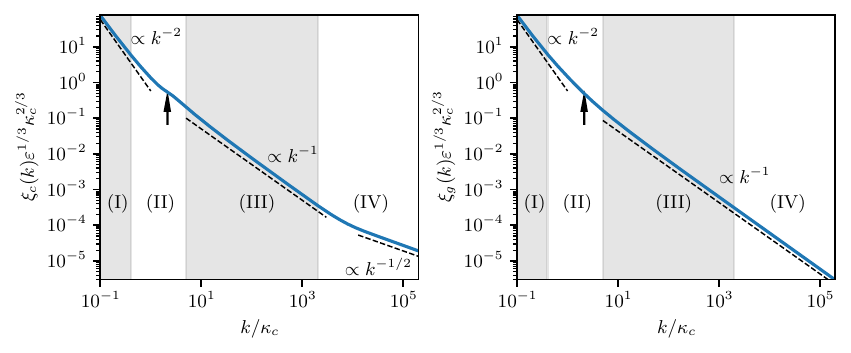}}
\caption{Characteristic times $\xi_c(k)$ of the velocity correlation
and $\xi_g(k)$ of the Green function for
$k_d=51200\kappa_c$ and $W_1$. The arrow marks the forcing-dominated
region near $k\simeq2\kappa_c$. The four spectral ranges are
indicated by the same numbers as in Fig.~\ref{fig:Overview_Ek}.}
\label{fig:xic_xig}
\end{figure}

Figure~\ref{fig:xic_xig} shows $\xi_c(k)$ and $\xi_g(k)$ for
$k_d=51200\kappa_c$ and $W_1$. In the large-scale range
$k\ll\kappa_c$, both characteristic times scale as $k^{-2}$. This
behaviour will be derived in
Sec.~\ref{subsub:decay_ckt_gkt_llr}.

In the energy-containing range, $\xi_c(k)$ is strongly affected by the
work spectrum, as indicated by the arrow in the left panel. This
dependence follows from the exact short-time relation \eqref{equ:starttime_ckt_gkt}. The forcing dependence of
$\xi_g(k)$ is weaker because the work spectrum affects the Green
function only indirectly through the stationary energy spectrum and memory
kernel.

In the inertial range, both characteristic times follow the sweeping
scaling $\xi_c(k)\propto k^{-1}$ and $\xi_g(k)\propto k^{-1}$. Their prefactors are nevertheless different. In the dissipation
range, the two time scales separate more strongly. The correlation
time crosses over to $\xi_c(k)\propto k^{-1/2}$, whereas the Green-function time retains $\xi_g(k)\propto k^{-1}$. Within the parameter range investigated here, these exponents are independent of the work spectrum and dissipation wavenumber. The forcing and Taylor--Reynolds number instead determine the prefactors and the boundaries between the spectral ranges.

The following four subsections examine the scaling functions
$f_c$ and $f_g$ in ranges $\mathrm{(I)}$--$\mathrm{(IV)}$. We also
derive the observed wavenumber dependence of $\xi_c$ and $\xi_g$
where analytical asymptotic arguments are available.

\subsubsection{Decay in the large-scale range}
\label{subsub:decay_ckt_gkt_llr}

In the large-scale range $\mathrm{(I)}$, the numerical results indicate
a common exponential decay of the velocity correlation and Green
functions. We now derive this behaviour analytically in the limit
$k\ll\kappa_c$.

The energy entering the memory kernels is concentrated at
wavenumbers $p$ comparable to the forcing wavenumber $\kappa_c$.
Consequently, the dominant interactions at small $k$ satisfy
$k\ll p$ and $q=\sqrt{k^2-2kp\mu+p^2}\simeq p$. Expanding the interaction vertex \eqref{equ:DIA_V_kpmu} in $k/p$ and performing the angular
integration gives
\begin{equation}
M(k,t)=\frac{2k^2}{15}\int_0^\infty E(p)c(p,t)\left(2g(p,t)-p\partial_p g(p,t)\right)\mathrm dp+o(k^2).
\label{equ:Mg_small_k_time}
\end{equation}
An analogous expansion shows that $N(k,t)$ decays on the same
short time scale.

The modes contributing to the integrals in
Eq.~\eqref{equ:Mg_small_k_time} decorrelate much faster than the
large-scale mode at $k$. Hence, there exists a memory time
$\xi_M(k)$ satisfying $\xi_M(k) \ll \xi_c(k),\xi_g(k)$, such that both memory kernels are negligible for $t\gtrsim\xi_M(k)$. On the slow large-scale time scale, the convolution in the DIA equations \eqref{equ:DIA_ckt} and \eqref{equ:DIA_gkt} can therefore be approximated in the Markovian limit by
\begin{equation}
\begin{split}
\int_0^t M(k,t-s)c(k,s)\dd s = \int_0^t M(k,s)c(k,t-s)\dd s \simeq c(k,t) \int_0^\infty M(k,s)\dd s.
\end{split}
\label{equ:Markov_large_scale}
\end{equation}
The stationary-history term also becomes negligible for
$t\gtrsim\xi_M(k)$ because
$\Phi(k,t)$ only contains memory kernels evaluated at times greater
than or equal to $t$.

We therefore define the DIA-renormalised viscosity 
\begin{equation}
\mu_0 :=
\lim_{k\to0^+}
\frac{1}{k^2}
\int_0^\infty M(k,s)\dd s
= \frac{2}{15}
\int_0^\infty \int_0^\infty E(p)c(p,s)\left(2g(p,s)-p\partial_p g(p,s)\right)\mathrm dp\dd s. \label{def:mu0}
\end{equation}
In the long-wavelength limit, the nonlinear DIA memory term thus has
the same form as viscous damping, and the molecular viscosity
$\nu_0$ is replaced by the effective viscosity $\nu_0+\mu_0$.
In this restricted sense, $\mu_0$ may also be interpreted as an
eddy-viscosity contribution \cite{Carnevale1983}. The large-scale DIA equations then reduce to
\begin{align}
\left(
\pd{t}+\nu_0k^2
\right)c(k,t)
&\simeq
-\mu_0k^2c(k,t),
\label{equ:c_small_k}
\\
\left(
\pd{t}+\nu_0k^2
\right)g(k,t)
&\simeq
-\mu_0k^2g(k,t).
\label{equ:g_small_k}
\end{align}
Apart from a short initial-memory transient, their solutions are
therefore
\begin{equation}
c(k,t) \simeq g(k,t) \simeq e^{-(\nu_0+\mu_0)k^2t}.
\label{equ:cg_large_scale}
\end{equation}
Thus, the simple relation $c(k,t)=g(k,t)$ is recovered in the
large-scale limit. 

Using the definitions \eqref{def:fc}--\eqref{def:fg}, the corresponding
large-scale master functions are
\begin{equation}
f_c^{\mathrm{lr}}(x)=f_g^{\mathrm{lr}}(x)=e^{-\log(2)x}
\label{equ:fc_fg_llr}
\end{equation}
with the characteristic decay times
\begin{equation}
\xi_c^{\mathrm{lr}}(k)=\xi_g^{\mathrm{lr}}(k)=\frac{\log(2)}{(\nu_0+\mu_0)k^2}.
\end{equation}

The coefficient $\mu_0$ depends on the forcing through the stationary
spectrum and the two-time functions in
Eq.~\eqref{def:mu0}. For $W_1$, the numerical solutions give $\mu_0 \simeq 0.9\varepsilon^{1/3}\kappa_c^{-4/3}.$ By contrast, $\nu_0=\varepsilon^{1/3}k_d^{-4/3}$. Hence, $\mu_0\gg\nu_0$ when $k_d\gg\kappa_c$. Although $\mu_0$ depends on the forcing, the exponent and the normalised exponential decay are independent of it within the
forcing spectra considered here.

\begin{figure}
\centerline{\includegraphics{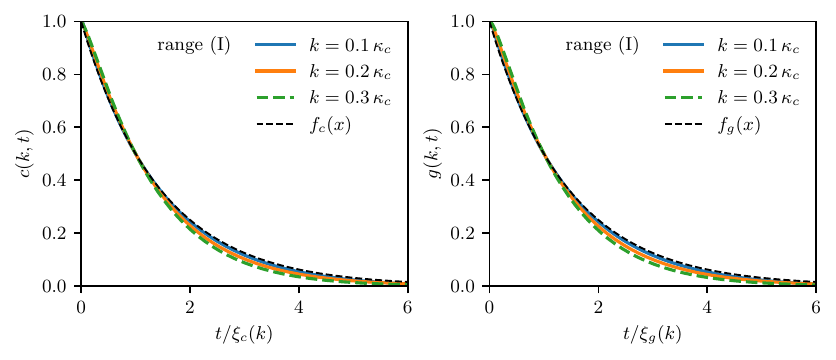}}
\caption{Large-scale decay functions for
$k_d=51200\kappa_c$ and $W_1$. The left panel shows
$c(k,t)$ as a function of $t/\xi_c(k)$ for three wavenumbers in range
$\mathrm{(I)}$. The right panel shows the Green function $g(k,t)$ plotted against $t/\xi_g(k)$. Black dashed lines represent the exponential master
function \eqref{equ:fc_fg_llr}.}
\label{fig:fc_fg_llr}
\end{figure}

Figure~\ref{fig:fc_fg_llr} tests the predicted master function. The
left panel shows $c(k,t)$ for three wavenumbers in range
$\mathrm{(I)}$ as a function of $t/\xi_c(k)$. The right panel shows
the Green function $g(k,t)$ as a function of
$t/\xi_g(k)$. At the smallest wavenumbers, both sets of curves collapse
onto the exponential prediction
\eqref{equ:fc_fg_llr}. Deviations increase as $k$ approaches the
energy-containing range, where the scale separation underlying
Eq.~\eqref{equ:Markov_large_scale} becomes less accurate.

The Markovian approximation therefore explains both the diffusive
large-scale scaling
$\xi_c^{\mathrm{lr}}(k),\xi_g^{\mathrm{lr}}(k)\propto k^{-2}$ observed in
Fig.~\ref{fig:xic_xig} and the recovery of
$c(k,t)=g(k,t)$ as $k/\kappa_c\to0^+$.

\subsubsection{Decay in the energy-containing range}
\label{subsub:decay_ckt_gkt_ecr}

\begin{figure}
  \centerline{\includegraphics{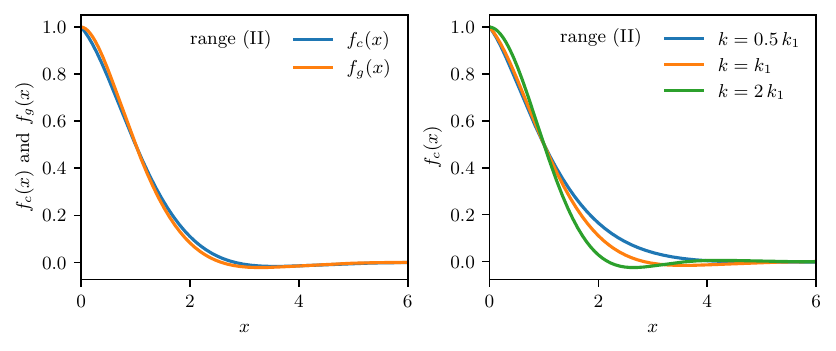}}
  \caption{Scaling functions in the energy-containing range for
  $W_1$ and $k_d=51200\kappa_c$. The left panel compares
  $f_c(x)$ and $f_g(x)$ at the forcing maximum $k_1$. The right panel
  shows $f_c(x)$ at three neighbouring wavenumbers around $k_1$.}
  \label{fig:fc_fg_ecr}
\end{figure}

In the energy-containing range $\mathrm{(II)}$, the temporal decay
retains a direct dependence on the work spectrum and is therefore not
expected to be universal. Figure~\ref{fig:fc_fg_ecr} illustrates this
behaviour for $W_1(k)$ and $k_d=51200\kappa_c$. The results are
independent of $k_d$ when $k_d\gg\kappa_c$.

At the forcing maximum $k_1$, the scaling functions $f_c(x)$ and
$f_g(x)$ differ only slightly, as shown in the left panel. For all
work spectra considered here, their maximum deviation near the
corresponding forcing maximum $k_i$ is approximately $4\%$. Despite
this close agreement, the right panel shows that the shape of
$f_c$ varies appreciably between neighbouring wavenumbers. The
same behaviour is found for $f_g$. Thus, $f_c$ and $f_g$ remain
similar to each other in the energy-containing range, but neither
defines a wavenumber-independent universal scaling function.

\subsubsection{Large-wavenumber asymptotics}
\label{subsub:decay_ckt_gkt_large_k}

The expected universality of inertial-range temporal correlations motivates the following asymptotic analysis. For $k\gg\kappa_c$, the DIA memory kernels are accurately approximated by
\begin{equation}
N(k,t)\simeq N(k,0)c(k,t),\qquad M(k,t)\simeq M(k,0)g(k,t).
\label{equ:approx_Mct_Mgt_k_large}
\end{equation}
A derivation is given in Appendix~\ref{app:approximation_Mg_Mc}. Together with $M(k,0)\simeq k^2v_{\mathrm{rms}}^2$, these approximations reduce the stationary DIA equations to
\begin{align}
\left(\pd{t}+\nu_0k^2\right)c(k,t)&=-k^2v_{\mathrm{rms}}^2\int_0^t g(k,t-s)c(k,s)\dd s+\frac{\Phi(k,t)}{E(k)},
\label{equ:DIA_ckt_approx_k_large}\\
\left(\pd{t}+\nu_0k^2\right)g(k,t)&=-k^2v_{\mathrm{rms}}^2\int_0^t g(k,t-s)g(k,s)\dd s,
\label{equ:DIA_gkt_approx_k_large}\\
\nu_0k^2E(k)&=\Phi(k,0),
\label{equ:DIA_Ek_approx_k_large}
\end{align}
where
\begin{equation}
\begin{split}
\Phi(k,t)={}&k^2 N(k,0)\int_0^\infty c(k,t+s)g(k,s)\dd s
\\
&-E(k)k^2v_{\mathrm{rms}}^2\int_0^\infty g(k,t+s)c(k,s)\dd s.
\end{split}
\label{equ:DIA_Phikt_approx_k_large}
\end{equation}
The work spectrum does not appear because the forcing is confined to small wavenumbers.

The analogous approximation $N(k,0)\simeq E(k)v_{\mathrm{rms}}^2$ is not admissible. It would imply $\Phi(k,0)=0$ and hence, through Eq.~\eqref{equ:DIA_Ek_approx_k_large}, $E(k)=0$ for $\nu_0>0$. The detailed wavenumber dependence of $N(k,0)$ must therefore be retained.

Equation~\eqref{equ:DIA_gkt_approx_k_large} is closed and can be solved by a Laplace transformation \cite{Kraichnan1959}. The result is
\begin{equation}
g(k,t)=\frac{J_1(2kv_{\mathrm{rms}}t)}{kv_{\mathrm{rms}}t}e^{-\nu_0k^2t}.
\label{equ:DIA_gkt_solution_k_large}
\end{equation}
In the regime $\tau_{\mathrm{dis}}(k)\gg\tau_{\mathrm{con}}(k)$, the corresponding Green-function master curve is therefore
\begin{equation}
f_g(x)\simeq\frac{J_1(2\alpha_gx)}{\alpha_gx},
\label{equ:fg_Kraichnan}
\end{equation}
with
\begin{equation}
\xi_g(k)=\alpha_g\tau_{\mathrm{con}}(k),\qquad \alpha_g\simeq1.108.
\label{equ:xig_Kraichnan}
\end{equation}
Here, $\alpha_g$ is determined by $J_1(2\alpha_g)/\alpha_g=1/2$.

The resulting large-wavenumber reduction does not uniquely determine the velocity correlation function. In the inviscid limit, $\nu_0\to 0$, the choice $c(k,t)=g(k,t)$ is one possible solution, as originally proposed by Kraichnan \cite{Kraichnan1959}. He later recognized that the solution of the reduced equation is not unique and stated that $\xi_c(k)/\xi_g(k)$ approaches a constant different from unity in the inertial range \cite{Kraichnan1964decay}. In the following, we show that the large-wavenumber reduction also admits asymptotic branches for which $\xi_c(k)\propto\tau_{\mathrm{con}}(k)$ does not hold.

A particularly simple alternative solution can be identified directly. The constant function $c(k,t)=1$ also satisfies Eq.~\eqref{equ:DIA_ckt_approx_k_large}, because  
\begin{equation}
\begin{split}
\left(\pd{t}+\nu_0k^2\right)1=\nu_0k^2&=-k^2v_{\mathrm{rms}}^2\int_0^t g(k,s)\dd s+\frac{k^2 N(k,0)}{E(k)}\int_0^\infty g(k,s)\dd s-k^2v_{\mathrm{rms}}^2\int_t^\infty g(k,s)\dd s
\\
&=\frac{\Phi(k,0)}{E(k)}=\nu_0k^2.
\end{split}
\label{equ:constant_c_large_k}
\end{equation}
The last equality follows from Eq.~\eqref{equ:DIA_Ek_approx_k_large}. This formal solution is admissible because $g(k,t)$ decays sufficiently rapidly for all integrals in Eqs.~\eqref{equ:DIA_ckt_approx_k_large}--\eqref{equ:DIA_Ek_approx_k_large} to remain finite. In the limit $k/\kappa_c\to\infty$, a slowly decaying function such as $c(k,t)\simeq f_g(t/\tau_{\mathrm{con}}(\kappa_c))$ likewise provides an asymptotic solution satisfying the physically relevant condition $c(k,t)\to0$ as $t\to\infty$.

This formal non-uniqueness results from the complete decoupling of small and large wavenumbers in the approximation given in  Eq.~\eqref{equ:approx_Mct_Mgt_k_large}, which is derived for $k\gg\kappa_c$. Numerically, we find no evidence that the full DIA equation \eqref{equ:DIA_ckt} has a non-unique solution. Instead, the inertial-range correlation decays on the convective timescale, $\xi_c(k)\propto\tau_{\mathrm{con}}(k),$ whereas in the dissipation range, $\xi_c(k)\propto\left[\tau_{\mathrm{con}}(k)\tau_{\mathrm{con}}(\kappa_d)\right]^{1/2}$. The latter geometric-mean scaling reflects the influence of both the rapidly and slowly decaying asymptotic solutions. The numerical results are discussed in Sec.~\ref{subsub:decay_ckt_gkt_ir_dr}.

\subsubsection{Decay in the inertial and dissipation ranges}
\label{subsub:decay_ckt_gkt_ir_dr}

We now compare the large-wavenumber predictions with the numerical
solutions of the full stationary DIA equations. Unless stated
otherwise, we use the work spectrum $W_1(k)$ and
$k_d=51200\kappa_c$. The same scaling behaviour is found for the
other work spectra and dissipation wavenumbers considered here,
provided that $k_d\gg\kappa_c$.

\begin{figure}
  \centerline{\includegraphics{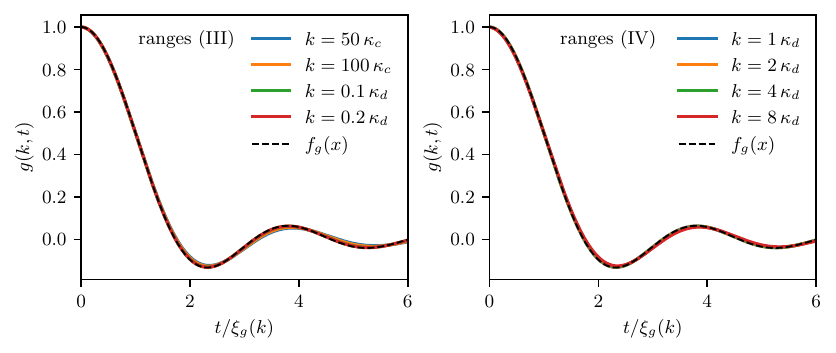}}
  \caption{Green function
  $g(k,t)$ as a function of
  $t/\xi_g(k)$ for four wavenumbers in the inertial range $\mathrm{(III)}$ in the left panel and in the dissipation range $\mathrm{(IV)}$ in the right panel. The black dashed curves show the analytical scaling function $f_g$ in Eq. \eqref{equ:fg_Kraichnan}.}
  \label{fig:fg_ir_dr}
\end{figure}

The left and right panels of Fig.~\ref{fig:fg_ir_dr} show the Green function $g(k,t)$ as a function of
$t/\xi_g(k)$ for wavenumbers in the inertial and dissipation ranges, respectively.
The curves collapse onto the common scaling function $f_g$, given in Eq.~\eqref{equ:fg_Kraichnan}. Thus, the same master function describes the Green function throughout both ranges $\mathrm{(III)}$ and $\mathrm{(IV)}$.

\begin{figure}
  \centerline{\includegraphics{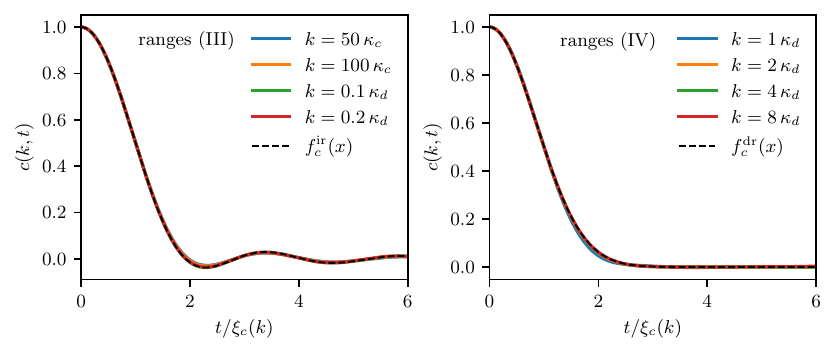}}
  \caption{The velocity correlation function $c(k,t)$ is shown as a function of $t/\xi_c(k)$ for four wavenumbers in the inertial range $\mathrm{(III)}$ in the left panel and in the dissipation range $\mathrm{(IV)}$ in the right panel. The black dashed lines represent the scaling functions $f_c^{\mathrm{ir}}$ and $f_c^{\mathrm{dr}}$.}
  \label{fig:fc_ir_dr}
\end{figure}

The velocity correlation function exhibits a different behaviour.
The left panel of Fig.~\ref{fig:fc_ir_dr} shows $c(k,t)$ as a function of
$t/\xi_c(k)$ for four wavenumbers in the inertial range. Within the
numerical accuracy, the curves collapse onto a common master function
$f_c^{\mathrm{ir}}(x)$. The master curve shown in the figure was
obtained at $k=1000\kappa_c$, in the centre of the inertial range. The numerically determined master function $f_c^{\mathrm{ir}}$ is valid for all work spectra considered here. Hence, the coupling to the energy-containing range selects a correlation function that is independent of the detailed forcing shape, although this coupling cannot be derived in the analytical
large-wavenumber limit.

The inertial-range functions $f_c^{\mathrm{ir}}$ and $f_g$ have
similar initial decays but are not identical. In particular, the
oscillations of $f_c^{\mathrm{ir}}(x)$ are more strongly damped than
those of $f_g$. Their quantitative comparison is deferred to
Sec.~\ref{sub:FDT}.

The right panel of Fig. \ref{fig:fc_ir_dr} presents the corresponding results in the dissipation range. The rescaled curves again collapse closely, but
onto a different master function. In this range, the numerical result
is accurately represented by the Gaussian form
\begin{equation}
f_c^{\mathrm{dr}}(x) = e^{-\log(2)x^2}.
\label{equ:fc_gauss}
\end{equation}
The normalisation follows from the definition of $\xi_c(k)$ as the
halving time. Thus, the velocity correlation possesses separate
master functions in the inertial and dissipation ranges,
whereas the Green function is described by the same scaling
function in both ranges.

The characteristic time of the Green function follows directly from
the large-wavenumber solution $\xi_g(k)= \alpha_g\tau_{\mathrm{con}}(k)$ with $\alpha_g\simeq 1.108$. The numerical results confirm this expression throughout the inertial
and dissipation ranges.

\begin{figure}
  \centerline{\includegraphics{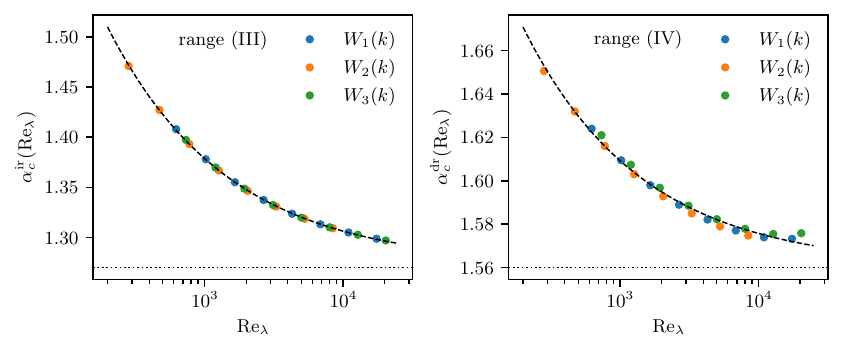}}
  \caption{Reynolds-number dependence of the correlation-time
  coefficients for all three work spectra. The left panel shows
  $\alpha_c^{\mathrm{ir}}$ and the right panel shows
  $\alpha_c^{\mathrm{dr}}$. The black dashed lines represent the
  empirical fits \eqref{equ:alpha_c_ir} and
  \eqref{equ:alpha_c_dr}.}
  \label{fig:a_ir_and_a_dr}
\end{figure}

In the inertial range, the characteristic time of the velocity
correlation has the same $k^{-1}$ scaling,
\begin{equation}
\xi_c^{\mathrm{ir}}(k)
=
\alpha_c^{\mathrm{ir}}(\Rey_\lambda)
\tau_{\mathrm{con}}(k).
\label{equ:xic_ir}
\end{equation}
As shown in the left panel of Fig.~\ref{fig:a_ir_and_a_dr}, the results
for all three work spectra are described by the empirical fit
\begin{equation}
\alpha_c^{\mathrm{ir}}(\Rey_\lambda)
\simeq
1.27+3.35\,\Rey_\lambda^{-0.50}.
\label{equ:alpha_c_ir}
\end{equation}
Consequently,
$\alpha_c^{\mathrm{ir}}\to1.27$ as
$\Rey_\lambda\to\infty$. This limiting value remains distinct from
$\alpha_g\simeq1.108$. The numerical stationary DIA solution
therefore does not approach the formal large-wavenumber solution
$c=g$, even at asymptotically large Reynolds numbers.

In the dissipation range, the correlation time is well
represented by
\begin{equation}
\xi_c^{\mathrm{dr}}(k)
=
\alpha_c^{\mathrm{dr}}(\Rey_\lambda)
\tau_{\mathrm{con}}(k)^{\eta}\tau_{\mathrm{con}}(\kappa_d)^{1-\eta},\qquad \eta\simeq 0.505
\label{equ:xic_dr}
\end{equation}
Since $\tau_{\mathrm{con}}(k)\propto k^{-1}$, this corresponds to
$\xi_c^{\mathrm{dr}}(k)\propto k^{-0.505}$, or approximately
$k^{-1/2}$. The right panel of Fig.~\ref{fig:a_ir_and_a_dr} shows that
the coefficient is described empirically by
\begin{equation}
\alpha_c^{\mathrm{dr}}(\Rey_\lambda)
\simeq
1.56+1.61\,\Rey_\lambda^{-0.51}.
\label{equ:alpha_c_dr}
\end{equation}
Both coefficients become independent of the forcing shape within the
numerical accuracy as $\Rey_\lambda$ increases.

\subsection{Fluctuation--dissipation relations}
\label{sub:FDT}

In Sec.~\ref{subsub:decay_ckt_gkt_llr}, we established that the normalised velocity correlation and Green functions coincide in the long-wavelength regime, $c(k,t)=g(k,t)$ for $k\ll\kappa_c$. This raises the central question of whether an analogous relation persists at larger wavenumbers.

More generally, any relation connecting spontaneous fluctuations to the response to an external perturbation is referred to as a fluctuation--dissipation relation (FDR). We reserve the term fluctuation--dissipation theorem (FDT) for the corresponding equilibrium identity. To illustrate it, consider the equilibrium Langevin dynamics $\dot{y}=-\gamma\partial_y H(y)+\eta+h$, where $H$ is a smooth potential and $\eta$ is Gaussian white noise with covariance $\langle\eta(t)\eta(s)\rangle=F\delta(t-s)$. The noise--dissipation balance $F=2\gamma k_{\mathrm B}T$ then implies
\begin{equation}
\Theta(\tau)\partial_\tau C(\tau)
=-\gamma k_{\mathrm B}T\,G(\tau),
\label{equ:equilibrium_FDT}
\end{equation}
where $C(\tau)=\langle y(\tau+s)y(s)\rangle$ and $G(\tau)=\delta\langle y(\tau+s)\rangle/\delta h(s)$ are the stationary correlation and response functions, respectively. Corresponding formulations also hold for vector-valued observables and Hamiltonian systems in thermal equilibrium \cite{Umberto2008}.

Stationary turbulence, by contrast, is a driven and dissipative nonequilibrium system, so the equilibrium FDT is not expected to hold in general. Nevertheless, generalized FDRs can remain informative far from equilibrium, as demonstrated, for example, by modified fluctuation--dissipation relations in glassy systems \cite{Cugliandolo1997,Sollich2002,Buhot2002}. We therefore use the DIA to determine the form, scale dependence, and dynamical origin of FDR violations in turbulence.

\subsubsection{Turbulent fluctuation--dissipation relation}

For the Navier--Stokes equations supplemented by thermal fluctuations, the equilibrium noise--dissipation balance fixes the covariance amplitude of the stochastic forcing to $F(k)=2\nu_0k^2k_{\mathrm B}T$ in the present normalization \cite[p.~44]{Zarate2006}. This condition yields an equilibrium distribution determined by the kinetic energy, together with detailed balance and the FDT, and is not restricted to the linearized or laminar dynamics \cite{Bandak2022}.  It relates the thermal forcing to viscous dissipation in the fluctuating Navier--Stokes equations. Accordingly, the relaxation coefficient in Eq.~\eqref{equ:equilibrium_FDT} is $\gamma(k)=\nu_0k^2$. In the turbulent stationary state considered here, however, the external forcing is not constrained by this thermal noise--dissipation balance, and the resulting flux of energy across scales drives the system out of equilibrium \cite{Wu2020}. Motivated by the equilibrium result, we therefore test the effective FDT expression
\begin{equation}
\Theta(t)C(k)\partial_t c(k,t)
=-\nu_{\mathrm{eff}}k^2k_{\mathrm B}T_{\mathrm{eff}}(k)g(k,t), \qquad k > 0,
\label{equ:FDT_general}
\end{equation}
where $\nu_{\mathrm{eff}}$ and $T_{\mathrm{eff}}(k)$ denote an effective viscosity and temperature, respectively. The laminar equilibrium limit is recovered when $\nu_{\mathrm{eff}}\to\nu_0$ and $T_{\mathrm{eff}}\to T$ as $Q_t^\alpha[U](\bm{k})\to0$. Equivalently, Eq.~\eqref{equ:FDT_general} requires the ratio $\partial_t c(k,t)/g(k,t)$ to be independent of time over its range of validity.

According to Sec.~\ref{subsub:decay_ckt_gkt_llr}, the DIA predicts in the large-scale range $\mathrm{(I)}$ the choice
\begin{equation}
\nu_{\mathrm{eff}}=\nu_0+\mu_0, \qquad  k_{\mathrm B}T_{\mathrm{eff}}(k)=C(k),  \qquad k \ll \kappa_c.
\end{equation}
for the effective FDT \eqref{equ:FDT_general}, where $\mu_0$ is the renormalised viscosity.

The infrared limit $C(k)\to C_0>0$, corresponding to the equipartition spectrum $E(k)\propto k^2$, is well established by analytical arguments and DNS studies \cite{Kraichnan1973,Dallas2015,Alexakis2019}. The present DIA solution recovers this known equal-time behaviour and additionally yields $c(k,t)=g(k,t)$ in the same limit. Taken together, these results show that the infrared modes are consistent with a forcing-dependent effective equilibrium with $k_{\mathrm B}T_{\mathrm{eff}}=C_0[F]$. The forcing dependence of $C_0[F]$, however, means that this effective temperature cannot, on this basis alone, be identified with a thermodynamic temperature.

For $k\gtrsim\kappa_c$, the DIA results in Secs.~\ref{subsub:decay_ckt_gkt_ecr}--\ref{subsub:decay_ckt_gkt_ir_dr} generally show that the ratio $\partial_t c(k,t)/g(k,t)$ is strongly time dependent. Consequently, the correlation and response functions in this range cannot be related through Eq.~\eqref{equ:FDT_general} using a time-independent effective temperature. We therefore do not pursue this equilibrium-inspired parametrization beyond the long-wavelength regime.

Kraichnan's fluctuation--relaxation relation \cite{Kraichnan1959,Kraichnan1959FDR,Kraichnan2000FDR} makes a different and stronger assumption. It postulates that the normalised two-time velocity correlation and Green functions coincide,
\begin{equation}
c(k,t)=g(k,t), \qquad k> 0.
\label{equ:turbulent_FDR}
\end{equation}
We refer to this relation as the fluctuation--dissipation relation (FDR) in the following. Our results show that the FDR does not hold in the inertial range, where $\xi_c(k)/\xi_g(k)\simeq1.15$ for $\Rey_\lambda\gg1$. Nevertheless, the two characteristic times have the same wavenumber dependence, as discussed in Sec.~\ref{subsub:decay_ckt_gkt_ir_dr}. This result is consistent with Kraichnan's observation that the correlation and response times need not coincide \cite{Kraichnan1964decay}. The discrepancy becomes more pronounced in the dissipation range, where the two decay times exhibit different wavenumber scalings, $\xi_c(k)\propto k^{-1/2}$ and $\xi_g(k)\propto k^{-1}$.

\begin{figure}
    \centering
    \includegraphics{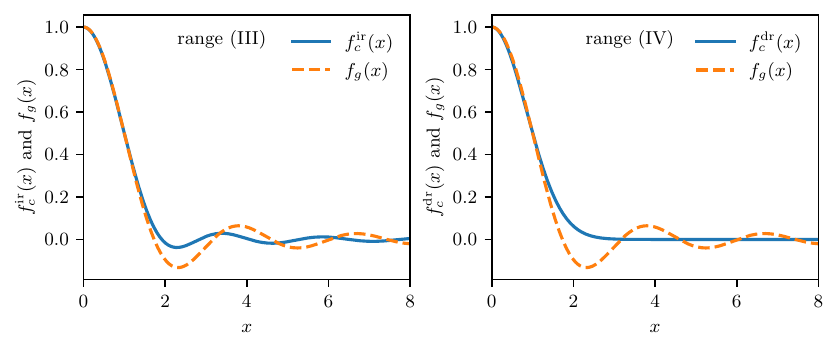}
    \caption{Scaling functions $f_c(x)$ and $f_g(x)$ in the inertial range $\mathrm{(III)}$ (left) and the dissipation range $\mathrm{(IV)}$ (right). Each function is plotted against time normalised by its respective characteristic decay time. The functions agree closely over the dominant initial decay but differ in their oscillatory long-time tails.}
    \label{fig:FDR_fc_fg}
\end{figure}

It is nevertheless instructive to compare the shapes of the corresponding scaling functions. Figure~\ref{fig:FDR_fc_fg} shows $f_c(x)$ and $f_g(x)$ in the inertial range $\mathrm{(III)}$ and the dissipation range $\mathrm{(IV)}$, with the functions expressed in terms of their respective dimensionless time variables, $x=t/\xi_c(k)$ and $x=t/\xi_g(k)$. After this rescaling, the functions agree closely for $x\lesssim 1.5$, by which point both have decayed to approximately $0.2$. At later times, $x\gtrsim 1.5$, the Green function develops substantially stronger oscillations, and the two scaling functions separate.

The violation of Kraichnan's FDR therefore has two distinct aspects. In physical time, the dominant discrepancy arises from the unequal characteristic decay times $\xi_c(k)$ and $\xi_g(k)$, whereas the normalised shapes remain remarkably similar over the dominant initial decay. The oscillatory long-time tails provide an additional, shape-dependent departure from Eq.~\eqref{equ:turbulent_FDR} and are particularly important when integral relaxation times are considered.

\subsubsection{Analytical characterization of the FDR violation}
\label{subsub:FDR}

The implicitly defined halving times $\xi_c(k)$ and $\xi_g(k)$ are not well suited to an analytical investigation of the FDR violation. We therefore introduce the integral relaxation times
\begin{equation}
\xi_c^{\mathrm{int}}(k):=\int_0^\infty c(k,t)\dd t,
\qquad
\xi_g^{\mathrm{int}}(k):=\int_0^\infty g(k,t)\dd t.
\label{def:integral_times}
\end{equation}
Unlike the halving times, these quantities are directly accessible through a Laplace transformation and capture the complete temporal decay, including any oscillatory tails. Note that $\xi_c^{\mathrm{int}}$ and $\xi_g^{\mathrm{int}}$ remain positive for all $k>0$ and exhibit the same qualitative wavenumber dependence as the corresponding halving times $\xi_c$ and $\xi_g$. 

Applying the Laplace transform $\mathcal L$ to Eqs.~\eqref{equ:DIA_ckt} and \eqref{equ:DIA_gkt} gives 
\begin{align}
-1+\left(\sigma+\nu_0k^2\right)\mathcal{L}[c(k,\cdot)](\sigma)
&=-\mathcal{L}[M(k,\cdot)](\sigma)\mathcal{L}[c(k,\cdot)](\sigma)
+\frac{1}{E(k)}\mathcal{L}[\Phi(k,\cdot)](\sigma),
\\
-1+\left(\sigma+\nu_0k^2\right)\mathcal{L}[g(k,\cdot)](\sigma)
&=-\mathcal{L}[M(k,\cdot)](\sigma)\mathcal{L}[g(k,\cdot)](\sigma),
\end{align}
where $\sigma$ denotes the Laplace frequency. Eliminating the common viscous and memory contributions yields the exact identity 
\begin{equation}
\frac{\mathcal{L}[c(k,\cdot)](\sigma)}{\mathcal{L}[g(k,\cdot)](\sigma)}-1
=\frac{1}{E(k)}\mathcal{L}[\Phi(k,\cdot)](\sigma).
\label{equ:FDR_Laplace_space}
\end{equation}
Provided that the relevant time integrals converge, setting $\sigma=0$ gives
\begin{equation}
\varphi(k):=\frac{\xi_c^{\mathrm{int}}(k)}{\xi_g^{\mathrm{int}}(k)}-1
=\frac{1}{E(k)}\int_0^\infty\Phi(k,t)\dd t.
\label{eq:integrated_fdr_violation}
\end{equation}
Thus, $\varphi(k)$ is an exact integrated measure of the violation of $c=g$, determined by the time integral of the noncausal DIA contribution $\Phi(k,t)$ normalised by the energy spectrum. A nonzero value of $\varphi(k)$ proves that $c$ and $g$ differ, whereas $\varphi(k)=0$ alone is insufficient to establish their pointwise equality.

\begin{figure}
\centerline{\includegraphics{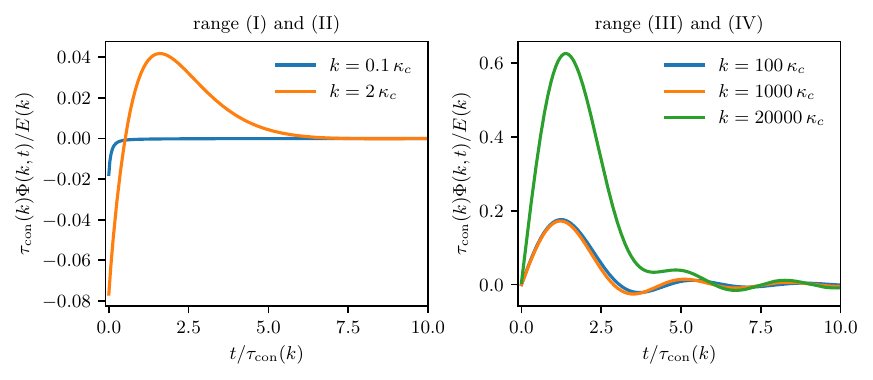}}
\caption{Temporal dependence of the dimensionless DIA contribution $\tau_{\mathrm{con}}(k)\Phi(k,t)/E(k)$ for the work spectrum $W_1(k)$ and $k_d=51200\kappa_c$. The left panel shows representative wavenumbers in the large-scale and energy-containing ranges. The right panel shows two inertial-range wavenumbers and one dissipation-range wavenumber. Time is normalised by $\tau_{\mathrm{con}}(k)$.}
\label{fig:Phi_kt}
\end{figure}

The temporal structure of $\Phi(k,t)$ is shown in Fig.~\ref{fig:Phi_kt} for the work spectrum $W_1(k)$ and $k_d=51200\kappa_c$. Its initial value follows from the stationary spectral balance,
\begin{equation}
\Phi(k,0)=\nu_0k^2E(k)-\frac{W(k)}{2}.
\label{equ:Phi_initial}
\end{equation}
In the forced range, where viscous dissipation is negligible, this expression reduces to $\Phi(k,0)\simeq-W(k)/2$.

At $k=0.1\kappa_c$, $\Phi(k,t)$ relaxes rapidly and monotonically on a timescale much shorter than $\tau_{\mathrm{con}}(k)$. At $k=2\kappa_c$, the decay becomes non-monotonic and occurs on a timescale comparable to the convective timescale. In the inertial range, the rescaled curves exhibit approximately the same temporal form. Because $W(k)\simeq 0$ and viscous dissipation is weak in this range, $\Phi(k,0)=\nu_0k^2E(k)$ is small. The function initially increases, reaching approximately
\begin{equation}
\Phi\left(k,1.2\tau_{\mathrm{con}}(k)\right)
\simeq 0.18\,\frac{E(k)}{\tau_{\mathrm{con}}(k)},
\end{equation}
and subsequently decays towards zero. The dissipation-range curve has the same qualitative form, but both its dimensionless amplitude and its decay time increase with $k$.

Substituting Eq.~\eqref{equ:DIA_Phikt} into Eq.~\eqref{eq:integrated_fdr_violation} gives
\begin{equation}
\begin{split}
\varphi(k)
=&\frac{1}{2}\int_0^\infty\int_{-1}^{1}\int_0^\infty\int_0^\infty V(k,p,\mu)E(p)c(p,t+s)
\\
&\qquad\quad\times\left[
\frac{k^2E(q)}{q^2E(k)}c(q,t+s)g(k,s)-g(q,t+s)c(k,s)
\right]\dd s\dd t\dd\mu\dd p.
\end{split}
\label{equ:explicit_integrated_fdr_violation}
\end{equation}

In the following, we use simple scaling arguments to show that, in the inertial range, the relevant contributions to $\varphi(k)$ arise from nonlocal mode couplings with $p\ll k$ and $q\simeq k$. These couplings connect the energy-containing range $\mathrm{(II)}$ with the inertial range $\mathrm{(III)}$. We further show that local interactions vanish asymptotically. Henceforth, let $k$ denote a wavenumber in the inertial range.

For local inertial-range triads, $p\sim q\sim k$, all spectra and two-time functions obey their inertial-range scaling forms. Rescaling $p$ and $q$ by $k$, the time arguments of $c$ and $g$ by $\tau_{\mathrm{con}}(k)$, and using $V\sim k^2$, we obtain
\begin{equation}
\varphi_{\mathrm{loc}}(k)
\sim kE(k)\,k^2\tau_{\mathrm{con}}^2(k)
=\frac{kE(k)}{v_{\mathrm{rms}}^2}
\propto k^{-1/2}.
\label{equ:phi_local_scaling}
\end{equation}
Here, the final proportionality follows from the inertial-range scaling of $E(k)$. This estimate assumes that no additional leading-order cancellation occurs between the two terms in Eq.~\eqref{equ:FDR_violation_Rkpq}. The local contribution therefore decreases as $k$ moves deeper into the inertial range.

For nonlocal triads with $p\ll k$ and $q\simeq k$, the mode $p$ lies predominantly in the energy-containing range and evolves slowly compared with the modes $k$ and $q$. Over time separations of order $\tau_{\mathrm{con}}(k)$, we may therefore set $c(p,t+s)\simeq1$. Rescaling the remaining two-time functions by $\tau_{\mathrm{con}}(k)$, using $V\sim k^2$, and noting that
\begin{equation}
\int_0^\infty E(p)\dd p\sim v_{\mathrm{rms}}^2,
\end{equation}
we find
\begin{equation}
\varphi_{\mathrm{nl}}(k)
\sim
v_{\mathrm{rms}}^2k^2 \tau_{\mathrm{con}}^2(k)
\sim O(1).
\label{equ:phi_nonlocal_scaling}
\end{equation}

The local contribution thus decreases as $k^{-1/2}$, whereas the nonlocal contribution from $p\ll k\simeq q$ can remain finite throughout the inertial range. Although the scaling argument does not determine the numerical coefficient, it identifies coupling to the energy-containing modes as the asymptotically dominant source of the inertial-range FDR violation. The violation is therefore generated primarily not by local inertial-range interactions, but by the nonlocal influence of the large-scale turbulent background on the correlation and response dynamics.

In Appendix \ref{app:InteractionsFDRviolation}, we further assess the asymptotic scaling predictions in Eqs.~\eqref{equ:phi_local_scaling} and \eqref{equ:phi_nonlocal_scaling} by numerically evaluating and plotting the local and nonlocal contributions to $\varphi(k)$ in the $(p,q)$--plane.

\section{Comparison with direct numerical simulation}
\label{sec:comparison_dns}

We next compare the stationary DIA spectrum and its two independent two-time functions with the DNS results of Matsumoto \textit{et al.} \cite{Matsumoto2021}. In their study, the velocity correlation function is obtained from temporal correlations of the velocity Fourier modes. The Green function is determined both by directly perturbing the velocity field and by correlating the velocity with an additional weak random forcing, with the two methods yielding consistent results.

\subsection{Forcing protocol}
\label{subsec:dns_forcing}

Matsumoto \textit{et al.} generate statistically stationary turbulence using a constant-injection, negative-damping force acting in the wavenumber band
$B=[\kappa_c,2.5\kappa_c]$ and proportional to the instantaneous velocity,
\begin{equation}
f^\alpha_t(\bm{k})=\gamma(t)\chi_B(k)u^\alpha_t(\bm{k}),\qquad \gamma(t)=\frac{\varepsilon}{2\mathcal E_B(t)},
\label{equ:dns_forcing}
\end{equation}
where $\chi_B(k)$ is the indicator function of the forced band, i.e.,
$\chi_B(k)=1$ for $k\in B$  and $\chi_B(k)=0$ otherwise. In our convention,
\begin{equation}
\mathcal E_B(t)= \sum_{\bm{k}\in\mathbb G_c} \frac{1}{2} \chi_B(k) u_t^\alpha(\bm{k}) \Ov{u}_t^\alpha(\bm{k}) \frac{1}{L_c^6}, \qquad L_c=2\pi/\kappa_c, \qquad \mathbb{G}_c= \kappa_c\mathbb{Z}^3 \setminus \{\bm{0}\}\label{def:Etot_in_band_b}
\end{equation}
is the instantaneous kinetic energy contained in the forced modes. By
construction, the corresponding instantaneous energy-injection rate is
$2\gamma(t)\mathcal E_B(t)=\varepsilon$.
In the DNS, the initial velocity field is drawn from an ensemble, after
which each realisation evolves deterministically according to the forced
Navier--Stokes equations. The coefficient $\gamma(t)$ therefore provides
a realisation-dependent feedback that maintains an exactly constant
energy-injection rate despite fluctuations of the energy contained in
the forced modes.

Within the DIA, the work spectrum encodes the forcing. For comparison with DNS, we represent the negative-damping force by a negative contribution to the viscous damping rate, with the field-dependent coefficient $\gamma[U](t)$. This coefficient requires a self-consistent closure. We therefore expand its reciprocal-energy
dependence about the effective Gaussian mean
$\EV{\mathcal E_B}_{\mathrm{eff}}$. As shown in
Appendix~\ref{app:forcing_DNS}, the contributions generated by
fluctuations about this mean are subleading at the Gaussian mean-field
level. In the stationary case, the leading contribution is consequently
the constant coefficient
\begin{equation}
\gamma_{\mathrm{DIA}}
=\frac{\varepsilon}{2\EV{\mathcal E_B}_{\mathrm{eff}}}, \qquad
\EV{\mathcal E_B}_{\mathrm{eff}}=\int_{\kappa_c}^{2.5\kappa_c}E_{\mathrm{DIA}}(k)\,\dd k.
\label{equ:gamma_dia}
\end{equation}
Thus, $\gamma_{\mathrm{DIA}}$ is not prescribed independently but is
determined self-consistently together with the DIA energy spectrum.

At this mean-field level, the negative-damping force enters both the DIA
correlation and response equations through the replacement 
\begin{equation}
\nu_0 k^2
\longrightarrow
\nu_0 k^2-\gamma_{\mathrm{DIA}}\chi_B(k),
\label{eq:dia_negative_damping}
\end{equation}
and gives the work spectrum
\begin{equation}
W_{\mathrm{DIA}}(k)= 2\gamma_{\mathrm{DIA}} \chi_B(k)E_{\mathrm{DIA}}(k). \label{equ:dia_negative_damping_Wk}
\end{equation}
By construction, the work rate defined in Eq.~\eqref{def:epsilon_w} therefore satisfies $\varepsilon_w^{\mathrm{DIA}}=\varepsilon$. 

In the DNS, the box size $L_c$ and the prescribed energy-injection rate $\varepsilon$ remain fixed throughout the time evolution. We therefore use $\kappa_c=2\pi/L_c$ and $\varepsilon$ as the reference units. With the forcing prescriptions in Eq.~\eqref{equ:dns_forcing} for the DNS and Eq.~\eqref{equ:dia_negative_damping_Wk} for the DIA, $\nu_0$ is the only remaining control parameter. Matsumoto \textit{et al.} set
$\nu_0=0.00114\,\varepsilon^{1/3}\kappa_c^{-4/3}$. In the statistically stationary state, the mean dissipation rate equals $\varepsilon$, yielding
$k_d=(\varepsilon/\nu_0^3)^{1/4}=161\kappa_c$. Accordingly, we compare the DNS and DIA at $k_d=161\kappa_c$. The DIA spectrum $E_{\mathrm{DIA}}(k)$ and the two-time functions $c_{\mathrm{DIA}}(k,t)$ and $g_{\mathrm{DIA}}(k,t)$ are thereby calculated without introducing any fitting parameters.

The DNS data comprise the stationary isotropic energy spectrum, the two-time velocity correlation function, and the Green function evaluated on discrete spherical shells with lower radial boundaries $k_n=n\kappa_c$, with $n\in\mathbb{N}$ and shell width $\Delta k=\kappa_c$. For direct comparison, the DIA predictions, which are defined for continuous wavenumbers, are averaged over the same shells and thereby mapped onto the discrete DNS wavenumbers. The shell-averaging procedure is described in detail in Appendix \ref{app:dns_shell-averaging}. Henceforth, we suppress the index $n$ and simply write $k$.

\subsection{Energy spectrum}
\label{subsec:dns_spectrum}

\begin{figure}
    \centerline{\includegraphics{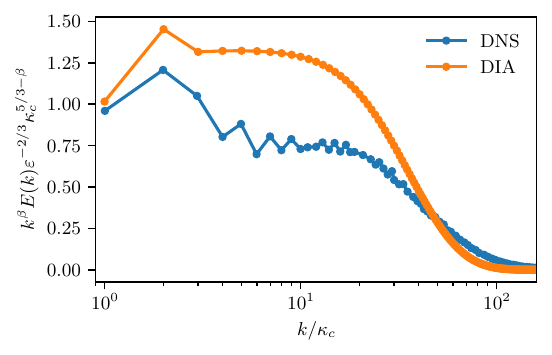}}
    \caption{Compensated energy spectra obtained from the DNS of Matsumoto \textit{et al.} \cite{Matsumoto2021} and from the stationary DIA, both at $k_d=161\kappa_c$. The compensation exponent $\beta=1.3$ describes the effective scaling over the finite wavenumber interval shown and is not interpreted as an asymptotic inertial-range exponent.}
    \label{fig:DNS_Ek}
\end{figure}

Figure~\ref{fig:DNS_Ek} compares the compensated shell spectra obtained using the effective exponent $\beta=1.3$ at $k_d=161\kappa_c$. This value gives the best joint  scaling in the inertial range. For $k=\kappa_c$, the DIA and DNS values differ by only approximately $6\%$. Here and below, relative deviations are measured with respect to the DNS values. For  $k=2\kappa_c$ and $k=3\kappa_c$, the deviations increase to approximately $23\%$. The close agreement at $k=\kappa_c$ is a genuine prediction of the DIA rather than a fitting artefact, since $\gamma_{\mathrm{DIA}}$ was determined self-consistently in Eq.~\eqref{equ:gamma_dia}. Wavenumbers $k\geq 3\kappa_c$ lie outside the directly forced wavenumber band.

Over a limited scaling interval, both spectra are
approximately described by
\begin{align}
E_{\mathrm{DNS}}(k) &\simeq 0.75 \left(\frac{k}{\kappa_c}\right)^{-\beta} \varepsilon^{2/3}\kappa_c^{-5/3}, \\
E_{\mathrm{DIA}}(k) &\simeq 1.32 \left(\frac{k}{\kappa_c}\right)^{-\beta} \varepsilon^{2/3}\kappa_c^{-5/3},  \label{eq:dns_dia_effective_spectra}
\end{align}
with $\beta=1.3$. The DIA prefactor therefore exceeds the DNS value by approximately $76\%$. The common value of $\beta$ should be interpreted as an effective finite-Reynolds-number exponent rather than as an asymptotic inertial-range scaling exponent.

The corresponding integral quantities are
\begin{align}
v_{\mathrm{rms}}^{\mathrm{DNS}} &\simeq 1.35\,\varepsilon^{1/3}\kappa_c^{-1/3}, & v_{\mathrm{rms}}^{\mathrm{DIA}} &\simeq 1.50\,\varepsilon^{1/3}\kappa_c^{-1/3}, \\
\langle \mathcal{E}_B\rangle_{\mathrm{DNS}} &\simeq 1.3\varepsilon^{2/3}\kappa_c^{-2/3}, & \EV{\mathcal{E}_B}_{\mathrm{eff}} &\simeq 1.39\,\varepsilon^{2/3}\kappa_c^{-2/3}.\label{eq:dns_dia_integral_quantities}
\end{align}
Thus, the DIA overestimates $v_{\mathrm{rms}}$ by approximately $11\%$ and underestimates the energy in the forcing band by approximately $7\%$.
With the DNS viscosity kept fixed, these values correspond to
$Re_\lambda^{\mathrm{DIA}}\simeq 260$, compared with
$Re_\lambda^{\mathrm{DNS}}\simeq 210$. The resulting difference of
approximately $24\%$ is therefore a prediction error of the DIA rather than a difference in the reference units.

At this moderate Reynolds number, the asymptotic spectral exponent of the
DIA is not yet clearly apparent over the available scaling interval.
Consequently, the present spectral comparison does not test the asymptotic
DIA spectrum. Its purpose is instead to establish the stationary energetic
background against which the DIA predictions for the correlation and
response functions are assessed.

\subsection{Characteristic time scales}
\label{subsec:dns_time_scales}

\begin{figure}
\centerline{\includegraphics{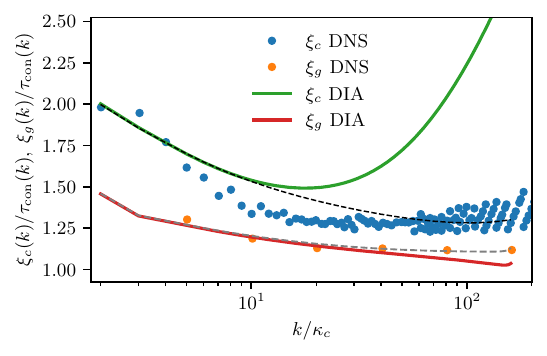}}
\caption{Halving times of the correlation and response functions obtained from DNS and DIA at $k_d=161\kappa_c$, normalised by $\tau_{\mathrm{con}}(k)$. The black dashed DIA curve for $k_d=1610\kappa_c$ illustrates the correlation time when the influence of the DIA dissipation range is shifted to larger wavenumbers. The corrected response time $\hat{\xi}_g(k)$ is defined in Eq.~\eqref{eq:corrected_response_time} and shown by the gray dashed curve.}
\label{fig:DNS_xi}
\end{figure}

Following Matsumoto \textit{et al.}, we characterise the decay of the shell-averaged correlation and response functions by their halving times $\xi_c(k)$ and $\xi_g(k)$ defined in Eq. \eqref{def:halving_times}. Figure~\ref{fig:DNS_xi} compares the DNS and DIA results normalised by the convective time scale $\tau_{\mathrm{con}}(k)$.

For the correlation time $\xi_c(k)$, the relative deviation between the DIA and DNS results is below approximately $6\%$ for $k \leq 5\kappa_c$. The deviation increases with $k$ but remains below approximately $15\%$ throughout the range $k\leq 20\kappa_c$. Thus, the DIA reproduces the DNS correlation times quantitatively in the energy-containing and inertial ranges.

At larger wavenumbers, the DIA correlation time is increasingly affected by its dissipation-range scaling. The crossover begins near $k\simeq0.2\kappa_d\simeq13\kappa_c$, while the asymptotic DIA behaviour $\xi_c(k)\propto k^{-1/2}$ becomes apparent for $k\gtrsim\kappa_d\simeq 64\kappa_c$. Consequently,
\begin{equation}
\frac{\xi_c(k)}{\tau_{\mathrm{con}}(k)}\propto k^{1/2}
\end{equation}
in the DIA dissipation range, whereas the DNS approaches the approximately constant value
\begin{equation}
\frac{\xi_c(k)}{\tau_{\mathrm{con}}(k)}\simeq1.3 .
\end{equation}
The increasing discrepancy at large $k$ therefore reflects the incorrect dissipation-range scaling of the Eulerian DIA. 

As a diagnostic test, we shift the DIA dissipation range to larger wavenumbers by choosing $k_d=1610\kappa_c$. The resulting correlation time is shown as the black dashed curve in Fig.~\ref{fig:DNS_xi}. In the resulting extended inertial range, the DIA approaches $\xi_c(k)/\tau_{\mathrm{con}}(k)\simeq 1.27$, close to the DNS value.

The agreement is even closer for the response time. For $k\leq40\kappa_c$, the relative deviation between the DIA and DNS values of $\xi_g(k)$ is approximately $2\%$. At larger wavenumbers, the DIA response is increasingly reduced by viscous damping, whereas the DNS continues to follow approximately
\begin{equation}
\xi_g(k)\simeq1.1\,\tau_{\mathrm{con}}(k).
\end{equation}
To isolate the leading viscous contribution to the DIA response, we define the corrected time $\hat{\xi}_g(k)$ by
\begin{equation}
g(k,\hat{\xi}_g(k))e^{\hat{\xi}_g(k)/\tau_{\mathrm{dis}}(k)}=\frac{1}{2}.
\label{eq:corrected_response_time}
\end{equation}
After this correction, the deviation from the DNS response time is approximately $1\%$ for all $k$. The corrected time $\hat{\xi}_g$ is shown in Fig. \ref{fig:DNS_xi} as a gray dashed curve. This suggests that the increasing discrepancy of the uncorrected response time is associated predominantly with the viscous damping factor predicted by the DIA. 

The difference between the correlation and response times can also be examined using the FRG. In the large-wavenumber limit $k/\kappa_c\to\infty$, FRG predicts the identical leading Gaussian sweeping form for the normalised correlation and response functions,
$c(k,t)\simeq g(k,t)\simeq\exp[-a_0(t/\tau_{\mathrm{con}}(k))^2]$, where $a_0$ is a nonuniversal constant determined by the forcing \cite{Canet2017,Tarpin2018,Gorbunova2021,Canet2022}. Thus, at the level of the leading time dependence, Kraichnan's FDR is asymptotically recovered and the difference between the halving times does not persist, with $\xi_c(k)/\xi_g(k)\to1$. The leading large-wavenumber FRG asymptotes therefore do not predict a persistent FDR violation.

\subsection{Normalised temporal shapes}
\label{subsec:dns_temporal_shapes}

Matsumoto \textit{et al.} found that the DNS correlation and response functions approximately collapse onto a Gaussian master curve when time is normalised by the corresponding halving time,
\begin{equation}
c_{\mathrm{DNS}}(k,t)\simeq f_{\mathrm{gauss}}\left(\frac{t}{\xi_c(k)}\right), \qquad g_{\mathrm{DNS}}(k,t)\simeq f_{\mathrm{gauss}}\left(\frac{t}{\xi_g(k)}\right),
\end{equation}
where
\begin{equation}
f_{\mathrm{gauss}}(x)=e^{-\log(2)x^2}.
\label{eq:gaussian_master_curve}
\end{equation}

\begin{figure}
\centerline{\includegraphics{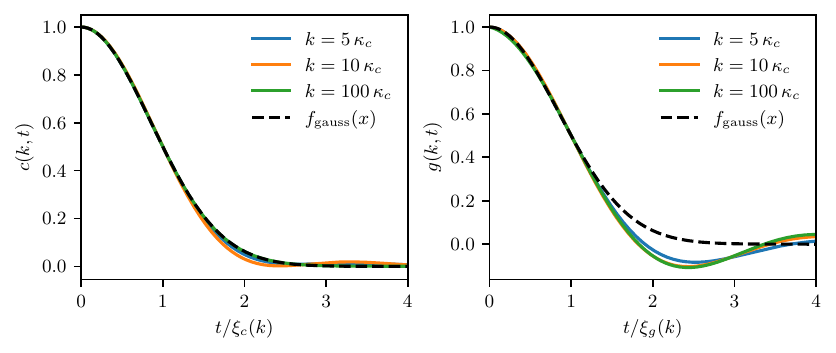}}
\caption{Shell-averaged DIA correlation functions (left) and response functions (right) plotted against time normalised by their respective halving times. Results are shown for $k=5\kappa_c$, $10\kappa_c$, and $100\kappa_c$. The black dashed reference curve is the Gaussian master function $f_{\mathrm{gauss}}$ observed in the DNS of Matsumoto \textit{et al.}}
\label{fig:Scaling_c_g}
\end{figure}

Figure~\ref{fig:Scaling_c_g} compares this DNS master curve with the DIA functions at $k=5\kappa_c$, $10\kappa_c$, and $100\kappa_c$, representing the inertial range, its upper end, and the dissipation range, respectively. The DIA correlation functions collapse closely onto $f_{\mathrm{gauss}}$ over the entire range shown. For $t/\xi_c(k)\lesssim 1.5$, the DIA curves agree closely with $f_{\mathrm{gauss}}$, while at later times their absolute deviation remains below approximately $5\%$ of the equal-time value.

A similarly good collapse is obtained for the DIA response functions for $t/\xi_g(k)\lesssim 1.5$. At later times, the DIA responses develop damped oscillations around the Gaussian master curve. Their amplitude remains below approximately $20\%$ of the equal-time response and decreases with increasing time. These oscillations are a characteristic feature of the DIA response and were discussed in Sec.~\ref{subsub:decay_ckt_gkt_ir_dr}.

Thus, over the time interval controlling the halving times, the DNS and DIA exhibit closely similar normalised decay profiles. Their correlation--response mismatch is therefore governed primarily by the difference between the characteristic times $\xi_c$ and $\xi_g$, rather than by qualitatively different temporal shapes. The late-time oscillations of the DIA response provide a secondary, closure-specific deviation from the approximately Gaussian DNS behaviour.

\section{Discussion}
\label{sec:discussion}

The central aim of this work was to determine whether a single self-consistent closure can account for the scale-dependent fluctuation--response behaviour observed in forced turbulence. Three distinct regimes emerge from the stationary DIA solution. At scales larger than the forcing scale, Kraichnan's  FDR $c(k,t)=g(k,t)$ is recovered and the dynamics are consistent with an effective thermal equilibrium. In the inertial range, correlation and response share the same convective scaling but have slightly different characteristic decay times, leading to a small systematic FDR violation. In the dissipation range, the mismatch grows, particularly between the characteristic times, and the DIA predicts a strong FDR violation.

\subsection{Effective thermal equilibrium at large scales}

Kraichnan showed that the turbulent FDR follows in thermal equilibrium when the stationary probability distribution is canonical and energy is equipartitioned among Fourier modes \cite{Kraichnan1959FDR,Kraichnan2000FDR,Umberto2008}. The DNS of Alexakis \textit{et al.} \cite{Alexakis2023} found both equilibrium-like large-scale statistics and a time-integrated FDR below the forcing range. The latter is a dynamical constraint and therefore provides stronger evidence for equilibrium-like behaviour than the equipartition spectrum alone.

The infrared DIA solution derived in Sec.~\ref{subsub:decay_ckt_gkt_llr} provides a physical interpretation of these observations. For $k\ll\kappa_c$, the infrared modes evolve much more slowly than the energy-containing modes. Their nonlinear memory consequently becomes effectively local on the infrared relaxation time and acts as a renormalised viscosity $\nu_0+\mu_0$. To leading order in $k/\kappa_c$, and apart from a short initial-memory transient, the resulting evolution is purely diffusive, with $c(k,t)=g(k,t)\simeq\exp[-(\nu_0+\mu_0)k^2t]$. The faster turbulent modes thus provide a bath-like background that determines both the diffusive damping and the stationary fluctuations of the slow modes.

This scale separation explains why static and dynamical equilibrium signatures emerge together. As shown in Secs.~\ref{subsub:decay_ckt_gkt_llr} and \ref{sub:FDT}, the constant infrared covariance gives $E(k)\propto k^2$, consistent with analytical, numerical and experimental results \cite{Kraichnan1973,Dallas2015,Alexakis2019,Gorce2022}, while the same reduced dynamics yields the pointwise relation $c(k,t)=g(k,t)$. This relation implies the time-integrated FDR observed in DNS but is stronger, since equality of the integrals does not require equality at every time separation. Together with the renormalised diffusive relaxation, it is also consistent with the derived FDT when the covariance is interpreted as an effective temperature.

The qualifier \emph{effective} remains essential. The complete turbulent system is driven, dissipative, and globally out of equilibrium \cite{Wu2020}. The infrared modes remain coupled to the forced scales. Both the renormalised viscosity $\mu_0$ and the covariance level $C_0$ are selected through this coupling, and the resulting effective temperature is therefore forcing-dependent rather than universal. The DIA result thus describes an emergent equilibrium-like balance in the large-scale two-point dynamics, not a restoration of thermodynamic equilibrium or detailed balance in the complete turbulent state.

\subsection{Sweeping and the forward-cascade FDR violation}

For $k\gg\kappa_c$, impulse--response measurements show that the correlation and Green functions are not proportional, although both exhibit the Eulerian $k^{-1}$ scaling associated with random sweeping by the energy-containing modes \cite{Carini2010,Matsumoto2021}. This distinction is central. A common characteristic scaling identifies the dominant decorrelation mechanism, but does not imply that spontaneous memory loss and mean relaxation are dynamically equivalent.

Kraichnan argued that sweeping could produce an inertial-range FDR if both functions were governed solely by statistically independent advection through the same large-scale velocity field \cite{Kraichnan1959FDR,Kraichnan2000FDR}. Under this assumption, correlation and response acquire the same ensemble-averaged sweeping factor. He nevertheless recognized that the DIA itself does not generally satisfy $c(k,t)=g(k,t)$ and found different correlation and response times in decaying turbulence \cite{Kraichnan1964decay}. The sweeping argument is therefore conditional and must be distinguished from the dynamics selected by the closure.

The stationary results of Secs.~\ref{subsub:decay_ckt_gkt_ir_dr} and \ref{sub:FDT} demonstrate the limitation of the purely kinematic argument. The DIA retains the same energy-containing modes as the leading source of the Eulerian $k^{-1}$ timescale, while its independent correlation and response equations select decay times with $\xi_c(k)/\xi_g(k)\simeq1.15$ at large Reynolds numbers. This finite separation constitutes an FDR violation despite the common sweeping scaling and agrees closely with the ordering and magnitude found in DNS \cite{Matsumoto2021}. Random sweeping therefore determines the leading dimensional timescale but not the relation between correlation and response.

The close collapse obtained after rescaling each function by its own halving time does not alter this conclusion. It represents a time-rescaled similarity of the dominant decay profiles, whereas an FDR compares both functions at the same physical time. Their unequal timescales already violate proportionality. The stronger oscillatory tail of the Green function provides an additional shape-dependent difference.

The asymptotic analysis in Sec.~\ref{subsub:decay_ckt_gkt_large_k} further shows why the common sweeping scale does not uniquely determine the correlation dynamics. Once the energy-containing range is removed from the reduced high-wavenumber equations, $c(k,t)=g(k,t)$ becomes only one of several admissible solutions. Formally subleading cross-scale terms are therefore essential for selecting the solution of the complete stationary DIA. In the dissipation range, the same coupling selects a second correlation regime whose timescale is influenced by the DIA spectral crossover. The resulting $k^{-1/2}$ correlation-time scaling should thus be interpreted not as a consequence of sweeping alone, but as feedback from the self-consistent DIA spectrum onto the correlation memory.

\subsection{Non-Markovian origin of the FDR violation}

As derived in Sec.~\ref{subsub:FDR}, we quantify the integrated FDR violation by $\varphi(k)=\xi_c^{\mathrm{int}}(k)/\xi_g^{\mathrm{int}}(k)-1$. Its magnitude is determined by the time integral of the non-Markovian function $\Phi(k,t)$, normalised by the energy spectrum $E(k)$, as given in Eq.~\eqref{eq:integrated_fdr_violation}. This function represents the stationary-history contribution that distinguishes the correlation equation from the response equation. A nonzero time integral of $\Phi(k,t)$ therefore measures the dynamical memory responsible for the correlation--response mismatch.

At equal times, $T(k)=2\Phi(k,0)$ connects this function to the energy-transfer spectrum. The instantaneous transfer alone, however, does not determine the two-time FDR violation. In the inertial range, $\Phi(k,0)$ is small, whereas at finite time $\Phi(k,t)$ reaches a maximum near
$t\simeq1.2\,\tau_{\mathrm{con}}(k)$ and then decays on a time scale
comparable to that of  $c(k,t)$. In the DIA framework, the FDR violation is consequently a genuinely non-Markovian property and cannot be inferred from the stationary spectral balance alone.

The scale decomposition in Sec.~\ref{subsub:FDR} identifies the interactions that sustain this memory. Contributions from local inertial-range triads, $p\sim q\sim k$, decrease as $\varphi_{\mathrm{loc}}(k)\propto(k/\kappa_c)^{-1/2}$, whereas nonlocal triads with $p\sim\kappa_c\ll q\sim k$ produce an $O(1)$ contribution. The forward-cascade FDR violation is therefore governed predominantly by coupling to the energy-containing range rather than by local cascade interactions. These large-scale modes set the leading sweeping time of both functions, but enter the correlation and response memories differently, producing unequal decay times and, at later times, unequal temporal shapes. Sweeping thus explains their common scaling without enforcing an FDR.

\subsection{Comparison with DNS and limitations of the DIA}

The comparison with the DNS of Matsumoto \textit{et al.} \cite{Matsumoto2021} tests the DIA interpretation at finite Reynolds number, $\Rey_\lambda\simeq210$. As detailed in Sec.~\ref{subsec:dns_forcing}, the DNS forcing is represented by replacing its fluctuating feedback coefficient with a mean negative-damping rate. No fit to the DNS spectrum or to either two-time function is introduced, making the subsequent temporal comparison a genuine prediction of the closure.

Over the limited intermediate-wavenumber interval, both spectra exhibit an effective exponent $\beta\simeq1.3$. This agreement should not be interpreted asymptotically, since the interval is influenced by both forcing and dissipation and the spectral amplitudes differ by approximately $76\%$. The finite-Reynolds-number spectral agreement is therefore qualitative rather than quantitative.

The strongest result is the agreement of the inertial-range characteristic times. The DNS gives $\xi_g^{\mathrm{DNS}}(k)\simeq1.1\,\tau_{\mathrm{con}}(k)$ and $\xi_c^{\mathrm{DNS}}(k)\simeq 1.3\,\tau_{\mathrm{con}}(k)$, close to the large-Reynolds-number DIA predictions $\xi_g^{\mathrm{DIA}}(k)\simeq1.108\,\tau_{\mathrm{con}}(k)$ and $\xi_c^{\mathrm{DIA}}(k)\simeq1.27\,\tau_{\mathrm{con}}(k)$. This close correspondence of the asymptotic prefactors should, however, not be confused with uniform agreement at the moderate Reynolds number of the DNS. Because the premature DIA dissipation-range crossover already affects part of the resolved inertial interval, the finite-Reynolds-number deviation between $\xi_c^{\mathrm{DNS}}(k)$ and $\xi_c^{\mathrm{DIA}}(k)$ reaches approximately $15\%$. The DIA nevertheless captures the common sweeping scaling, the ordering $\xi_c(k)>\xi_g(k)$, and the magnitude approached as the inertial range becomes better separated from dissipation.

The discrepancy becomes more pronounced in the dissipation range, where the DIA predicts $\xi_c^{\mathrm{DIA}}(k)\propto k^{-1/2}$, whereas the DNS retains the scaling $\xi_c^{\mathrm{DNS}}(k)\propto k^{-1}$. Because the spectrum enters both memory kernels, its incorrect crossover changes the self-consistent temporal background and provides a common explanation for the spectral and correlation-time errors. The DIA spectral deficiency therefore limits the finite-Reynolds-number and dissipation-range predictions without invalidating the asymptotic inertial-range timescales.

\section{Conclusion}
\label{sec:conclusions}

We have derived the Eulerian DIA as the simplest nontrivial self-consistent Gaussian mean-field closure of the MSRDJ turbulence functional in which covariance and response remain independent. The derivation does not require stationarity, homogeneity, or isotropy. The stationary HIT equations solved here are a subsequent specialisation. Most importantly, no FDR is imposed, so the relation between correlation and response is predicted by the closure.

At scales larger than the forcing scale, the DIA yields equipartition and  renormalised diffusive dynamics satisfying $c(k,t)=g(k,t)$, consistent with an effective infrared equilibrium. In the forward-cascade range, correlation and response share the Eulerian sweeping scaling, although their characteristic times differ slightly but noticeably. The resulting FDR violation is a non-Markovian effect sustained predominantly by nonlocal coupling to the energy-containing modes.

The DIA reproduces the inertial-range correlation and response times with notable accuracy. Its incorrect spectrum, however, produces a premature crossover to the dissipation range and an inaccurate $k^{-1/2}$ scaling of the correlation time. The two-time solution therefore identifies both a nontrivial success of the DIA and the point at which its spectral deficiency becomes dynamically relevant.

More generally, the spectrum is not merely an equal-time diagnostic but the self-consistent background entering both memory kernels. Neither two-time function can therefore be eliminated in favour of the other. A closure intended to describe fluctuation--response dynamics across the inertial and dissipation ranges must also retain, explicitly or through consistent renormalisation, their nonlocal coupling to the energy-containing scales. Although such cross-scale terms may appear subleading in a formal large-wavenumber expansion, they remain essential for selecting the physical correlation dynamics.

\begin{acknowledgments}
We thank Rudolf Haussmann for fruitful discussions and the Core Facility Scientific Compute Cluster (SCCKN) of the University of Konstanz for providing computational resources. The work was partially supported by the Deutsche Forschungsgemeinschaft (DFG) via SFB 1432.
\end{acknowledgments}
\vspace{0.2cm}

\paragraph*{Author contributions} M.F. conceived the research question and supervised the project. E.K. performed the numerical work, including method development, implementation, and analysis, carried out the analytical calculations, and wrote the original draft of the manuscript. Both authors contributed equally to the interpretation of the results and to the final version of the paper.
\vspace{0.2cm}

\appendix

\section{Fourier conventions and half-lattice decomposition}\label{app:fourier_half_lattice}

The Fourier transform of the Eulerian velocity field on the torus $\mathbb{T}_L^3$ is defined by
\begin{equation}
u_t^\alpha(\bm{k})=\int_{\mathbb{T}_L^3}u_t^\alpha(\bm{x})e^{-\I\bm{k}\cdot\bm{x}}\dd\bm{x},\qquad \bm{k}\in Z_L^3:=\frac{2\pi}{L}\mathbb{Z}^3.
\label{def:DFT}
\end{equation}
The inverse transform is
\begin{equation}
u_t^\alpha(\bm{x})=\sum_{\bm{k}\in Z_L^3}u_t^\alpha(\bm{k})e^{\I\bm{k}\cdot\bm{x}}\frac{1}{L^3}.
\label{def:iDFT}
\end{equation}
With these conventions, the infinite-volume limit is represented formally by
\begin{equation}
\sum_{\bm{k}\in Z_L^3}\frac{1}{L^3}
\longrightarrow
\int_{\mathbb{R}^3}\frac{\dd\bm{k}}{(2\pi)^3}.
\label{equ:continuum_limit}
\end{equation}

Since the velocity field is real, its Fourier coefficients satisfy $u_t^\alpha(-\bm{k})=\Ov{u}_t^\alpha(\bm{k})$. After removing the spatially uniform mode, the remaining Fourier lattice $\mathbb{G}=Z_L^3\setminus\{\bm{0}\}$ is decomposed into two disjoint half-lattices $\mathbb{G}_+$ and $\mathbb{G}_-=-\mathbb{G}_+$. Each half-lattice contains one wavevector from every pair $\{\bm{k},-\bm{k}\}$, so that
\begin{equation}
\mathbb{G}=\mathbb{G}_+\cup\mathbb{G}_-,
\qquad
\mathbb{G}_+\cap\mathbb{G}_-=\emptyset.
\end{equation}
A concrete choice is 
\begin{equation}
\begin{split}
\mathbb{G}_+={}&\{\bm{k}\in\mathbb{G}:k_3>0\}
\cup\{\bm{k}\in\mathbb{G}:k_3=0,\ k_2>0\}\cup\{\bm{k}\in\mathbb{G}:k_3=k_2=0,\ k_1>0\}.
\end{split}
\label{def:Gplus}
\end{equation}
Consequently, it is sufficient to formulate the stochastic dynamics on $\mathbb{G}_+$. The fields on $\mathbb{G}_-$ are then fixed by the reality condition.

\section{MSRDJ functional}

\subsection{Discrete construction of the MSRDJ functional}\label{app:MSRDJ_discretization}

We define the MSRDJ functional on the equidistant time grid
\begin{equation}
\Pi_{[t_0,t]}=\left\{t_0+n\Delta t:n=0,\ldots,N_t\right\},
\qquad
\Delta t=\frac{t-t_0}{N_t},
\label{def:time_partition}
\end{equation}
and introduce the abbreviation
\begin{equation}
\Pi_{(t_0,t]}:=\Pi_{[t_0,t]}\setminus\{t_0\}.
\label{def:open_time_partition}
\end{equation}
The Fourier lattice is initially restricted to a finite and inversion-symmetric Galerkin set. The corresponding spectral cutoff is suppressed in the notation. At finite $\Delta t$, no temporal regularity is imposed on the path variables.

In addition to the velocity variables $u_\tau^\alpha(\bm{k})$, we introduce complex response variables $\psi_\tau^\alpha(\bm{k})$. For a fixed initial condition at $t_0$, the discretized path space is
\begin{equation}
\Omega_{(t_0,t]}=\prod_{\tau\in\Pi_{(t_0,t]}} \prod_{\bm{k}\in\mathbb{G}_+} \prod_{\alpha\in I} \left(\mathbb{C}\times\mathbb{C}\right),
\label{def:path_space}
\end{equation}
where the two factors correspond to $u_\tau^\alpha(\bm{k})$ and $\psi_\tau^\alpha(\bm{k})$. The integration measure is 
\begin{equation}
\dd\omega_{(t_0,t]}=\prod_{\tau\in\Pi_{(t_0,t]}} \prod_{\bm{k}\in\mathbb{G}_+}\prod_{\alpha \in I}\frac{\dd\psi_\tau^\alpha(\bm{k})}{\pi L^3} \prod_{\beta \in I} \frac{\dd u_\tau^\beta(\bm{k})}{\pi L^3},
\label{def:path_measure}
\end{equation}
where
\begin{equation}
\int_{\mathbb{C}}f(z)\dd z:=\int_{\mathbb{R}^2}f(z)\,\dd\operatorname{Re}z\,\dd\operatorname{Im}z.
\label{def:int_dz}
\end{equation}
For compactness, we write $\dd\omega\equiv\dd\omega_{(t_0,t]}$. To state the discrete action, we introduce the backward difference
\begin{equation}
D_\tau u_\tau^\alpha(\bm{k}):=\frac{u_\tau^\alpha(\bm{k})-u_{\tau-\Delta t}^\alpha(\bm{k})}{\Delta t}.
\label{def:backward_difference}
\end{equation}
The discrete laminar action is
\begin{equation}
\begin{split}
S_{\mathrm{lam}}[\omega] =& \sum_{\tau\in\Pi_{(t_0,t]}}\sum_{\bm{k}\in\mathbb{G}_+}
\bigg(\Ov{\psi}_\tau^\alpha(\bm{k})F(\Vert\bm{k}\Vert)P_{\alpha\beta}(\bm{k})\psi_\tau^\beta(\bm{k})-\I\psi_\tau^\alpha(\bm{k})\left[D_\tau\Ov{u}_\tau^\alpha(\bm{k})+\nu_0k^2\Ov{u}_{\tau-\Delta t}^\alpha(\bm{k})\right]
\\
&\hspace{60mm}-\I\Ov{\psi}_\tau^\alpha(\bm{k})\left[D_\tau u_\tau^\alpha(\bm{k})+\nu_0k^2u_{\tau-\Delta t}^\alpha(\bm{k})\right]\bigg)\frac{\Delta t}{L^3},
\end{split}
\label{equ:Slam_discrete}
\end{equation}
while the discrete turbulent action is
\begin{equation}
\begin{split}
S_{\mathrm{tur}}[\omega]=\sum_{\tau\in\Pi_{(t_0,t]}}\sum_{\bm{k}\in\mathbb{G}_+}\I\bigg[\psi_\tau^\alpha(\bm{k})\Ov{Q_{\tau-\Delta t}^\alpha[u](\bm{k})}+\Ov{\psi}_\tau^\alpha(\bm{k})Q_{\tau-\Delta t}^\alpha[u](\bm{k})\bigg]\frac{\Delta t}{L^3}.
\end{split}
\label{equ:Stur_discrete}
\end{equation}
These expressions follow from the It\^{o} discretization of \eqref{equ:NavierStokesEq_stochastic}. The associated causal Jacobian is field-independent and is absorbed into the normalization of the path measure.

Using $u_\tau^\alpha(-\bm{k})=\Ov{u}_\tau^\alpha(\bm{k})$ and $\psi_\tau^\alpha(-\bm{k})=\Ov{\psi}_\tau^\alpha(\bm{k})$ with $\mathbb{G}=\mathbb{G}_+\cup\mathbb{G}_-$, the sums over $\mathbb{G}_+$ in \eqref{equ:Slam_discrete} and \eqref{equ:Stur_discrete} can be expressed as sums over the full lattice $\mathbb{G}$. The time sums then yield the compact continuum notation used in \eqref{equ:Slam} and \eqref{equ:Stur}.

\subsection{Discrete definition of the Green function}
\label{app:discrete_Green_function}

Perturbing Eq.~\eqref{equ:NavierStokesEq_stochastic} by an external force $P_{\alpha\beta}(\bm{k})h_t^\beta(\bm{k})$ adds the action 
\begin{equation}
S_{\mathrm{per}}[\omega,h]=\sum_{\tau\in\Pi_{(t_0,t]}}\sum_{\bm{k}\in\mathbb{G}_+}\I\left[\psi_\tau^\alpha(\bm{k})P_{\alpha\beta}(\bm{k})\Ov{h}_{\tau-\Delta t}^\beta(\bm{k})+\Ov{\psi}_\tau^\alpha(\bm{k})P_{\alpha\beta}(\bm{k})h_{\tau-\Delta t}^\beta(\bm{k})\right]\frac{\Delta t}{L^3}.
\label{equ:S_per}
\end{equation}
The perturbed expectation value is
\begin{equation}
\EVh{f}=\int_{\Omega_{(t_0,t]}}f(\omega)e^{-S[\omega]-S_{\mathrm{per}}[\omega,h]}\dd\omega,\qquad \EVh{1}=1.
\label{def:perturbed_expectation}
\end{equation}

Since $h_{s-\Delta t}$ acts during $[s-\Delta t,s]$, the It\^{o}-discretized Green function is
\begin{equation}
G_{t,s}^{\alpha\beta}(\bm{k},\bm{p}):=\lim_{\Delta t\to0}\left.\frac{L^3}{\Delta t}\PD{}{h_{s-\Delta t}^\beta(\bm{p})}\EVh{U_t^\alpha(\bm{k})}\right|_{h=0}=-\I\EV{U_t^\alpha(\bm{k})\Ov{\Psi}_s^\gamma(\bm{p})}P_{\gamma\beta}(\bm{p}).
\label{def:Green_function_discrete}
\end{equation}

The causal identities follow by temporarily coupling an auxiliary source $h^\alpha$ without the projector. This means that in $S_{\mathrm{per}}[\omega;\bm{h}]$, the substitution $P_{\alpha\beta}(\bm{k})h_\tau^\beta(\bm{k})\to h_\tau^\alpha(\bm{k})$ is made. For $s>t$, differentiation of the functional gives
\begin{align}
\EV{U_t^\alpha(\bm{k})\Ov{\Psi}_s^\beta(\bm{p})}&=\left.\frac{\I L^3}{\Delta t}\PD{}{h_{s-\Delta t}^\beta(\bm{p})}\int_{\Omega_{(t_0,s]}}u_t^\alpha(\bm{k})e^{-S[\omega]-S_{\mathrm{per}}[\omega,h]}\dd\omega\right|_{h=0}
\\
&=\left.\frac{\I L^3}{\Delta t}\PD{}{h_{s-\Delta t}^\beta(\bm{p})}\int_{\Omega_{(t_0,t]}}u_t^\alpha(\bm{k})e^{-S[\omega]-S_{\mathrm{per}}[\omega,h]}\dd\omega\right|_{h=0}=0.
\label{equ:response_causality_functional}
\end{align}
In the second line, the fields on $(t,s]$ have been integrated out using the normalization of the functional expectation value. The remaining functional on $\Omega_{(t_0,t]}$ is independent of the later source $h_{s-\Delta t}$, proving Eq.~\eqref{equ:response_causality}. An analogous calculation proves Eq.~\eqref{equ:EV_psi_psi}.

\subsection{Discrete derivation of the exact relations}
\label{app:exact_relations_discrete}

For a complex variable $z\in\mathbb{C}$, we use the Wirtinger derivatives
\begin{equation}
\PD{}{z}
=
\frac{1}{2}
\left(
\PD{}{\operatorname{Re}z}
-\I\PD{}{\operatorname{Im}z}
\right),
\qquad
\PD{}{\Ov{z}}
=
\frac{1}{2}
\left(
\PD{}{\operatorname{Re}z}
+\I\PD{}{\operatorname{Im}z}
\right).
\label{def:Wirtinger_derivatives}
\end{equation}
The variables $z$ and $\Ov{z}$ are treated as formally independent when these derivatives are applied.

Let $t,s\in\Pi_{(t_0,t_{\mathrm{f}}]}$, where $t_{\mathrm{f}}$ is the upper endpoint of the path interval, and let $\bm{k},\bm{p}\in\mathbb{G}_+$. Since the discretized path space is finite-dimensional, the required identities follow from ordinary integration. Integration with respect to $\Ov{\psi}_t^\alpha(\bm{k})$ gives
\begin{equation}
\int_{\Omega_{(t_0,t_{\mathrm f}]}}
\PD{}{\Ov{\psi}_t^\alpha(\bm{k})}
\left(
\Ov{u}_s^\beta(\bm{p})e^{-S[\omega]}
\right)
\dd\omega
=
0.
\label{equ:IBP_velocity}
\end{equation}
Applying the derivative to the discrete action in Eq.~\eqref{equ:IBP_velocity} yields
\begin{equation}
\begin{split}
\EV{
D_tU_t^\alpha(\bm{k})
\Ov{U}_s^\beta(\bm{p})
}
+
\nu_0k^2
\EV{
U_{t-\Delta t}^\alpha(\bm{k})
\Ov{U}_s^\beta(\bm{p})
}
&=
-\I F(\Vert\bm{k}\Vert)
P_{\alpha\gamma}(\bm{k})
\EV{
\Psi_t^\gamma(\bm{k})
\Ov{U}_s^\beta(\bm{p})
}
\\
&\qquad
+
\EV{
Q_{t-\Delta t}^\alpha[U](\bm{k})
\Ov{U}_s^\beta(\bm{p})
}.
\end{split}
\label{equ:Relation_1_discrete}
\end{equation}

Analogously, making the substitutions $\Ov{\psi}_t^\alpha(\bm{k})\to \psi_s^\beta(\bm{p})$  and  $\Ov{u}^\beta_s(\bm{p})\to u^\alpha_t(\bm{k})$ in Eq.~\eqref{equ:IBP_velocity}, and integrating with respect to $\psi_s^\beta(\bm{p})$ yields the relation for the time evolution of the second velocity field,
\begin{equation}
\begin{split}
\EV{
U_t^\alpha(\bm{k})
D_s\Ov{U}_s^\beta(\bm{p})
}
+
\nu_0p^2
\EV{
U_t^\alpha(\bm{k})
\Ov{U}_{s-\Delta t}^\beta(\bm{p})
}
&=
-\I F(\Vert\bm{p}\Vert)
\EV{
U_t^\alpha(\bm{k})
\Ov{\Psi}_s^\gamma(\bm{p})
}
P_{\gamma\beta}(\bm{p})
\\
&\qquad
+
\EV{
U_t^\alpha(\bm{k})
\Ov{Q}_{s-\Delta t}^\beta[U](\bm{p})
}.
\end{split}
\label{equ:Relation_2_discrete}
\end{equation}

To obtain the corresponding response relation, we insert an additional response field and use 
\begin{equation}
\int_{\Omega_{(t_0,t_{\mathrm f}]}}
\PD{}{\Ov{\psi}_t^\alpha(\bm{k})}
\left(
\Ov{\psi}_s^\beta(\bm{p})e^{-S[\omega]}
\right)
\dd\omega
=
0.
\label{equ:IBP_response}
\end{equation}
Expanding the derivative gives
\begin{equation}
\begin{split}
\EV{
D_tU_t^\alpha(\bm{k})
\Ov{\Psi}_s^\beta(\bm{p})
}
+
\nu_0k^2
\EV{
U_{t-\Delta t}^\alpha(\bm{k})
\Ov{\Psi}_s^\beta(\bm{p})
}
&=
\I L^3
\delta_{\alpha\beta}
\delta_{\bm{k},\bm{p}}
\frac{\delta_{t,s}}{\Delta t}
\\
&\qquad
+
\EV{
Q_{t-\Delta t}^\alpha[U](\bm{k})
\Ov{\Psi}_s^\beta(\bm{p})
}.
\end{split}
\label{equ:Relation_3_discrete}
\end{equation}
Here, $\delta_{t,s}$ denotes the Kronecker delta on the discrete time grid. The derivative acting on the inserted response field in Eq.~\eqref{equ:IBP_response} produces the contact term in Eq.~\eqref{equ:Relation_3_discrete}. The remaining contribution containing two response fields vanishes according to Eq.~\eqref{equ:EV_psi_psi}.

For $\bm{k}\in\mathbb{G}_-$, the same identities follow by applying the corresponding Wirtinger derivative to the independent variable at $-\bm{k}\in\mathbb{G}_+$ and using the reality conditions. Equations \eqref{equ:Relation_1_discrete}--\eqref{equ:Relation_3_discrete} therefore extend to the full lattice $\bm{k},\bm{p}\in\mathbb{G}$.

Setting $t=s$ in Eq.~\eqref{equ:Relation_3_discrete} determines the equal-time response. Causality implies
\begin{equation*}
\EV{
U_{t-\Delta t}^\alpha(\bm{k})
\Ov{\Psi}_t^\beta(\bm{p})
}
=
0.
\end{equation*}
For the same reason, the nonlinear term involving velocity fields at $t-\Delta t$ does not respond to an insertion at the later time $t$. Using the definition of the backward difference, Eq.~\eqref{equ:Relation_3_discrete} consequently reduces to
\begin{equation}
\frac{1}{\Delta t}
\EV{
U_t^\alpha(\bm{k})
\Ov{\Psi}_t^\beta(\bm{p})
}
=
\I L^3
\delta_{\alpha\beta}
\delta_{\bm{k},\bm{p}}
\frac{1}{\Delta t}.
\end{equation}
Hence,
\begin{equation}
\EV{
U_t^\alpha(\bm{k})
\Ov{\Psi}_t^\beta(\bm{p})
}
=
\I L^3
\delta_{\alpha\beta}
\delta_{\bm{k},\bm{p}},
\end{equation}
which gives Eq.~\eqref{equ:equal_time_response}.

For strictly separated times $t>s$, the continuum limits of Eqs.~\eqref{equ:Relation_1_discrete} and \eqref{equ:Relation_3_discrete} are regular. The advanced response in Eq.~\eqref{equ:Relation_1_discrete} vanishes because a velocity field at time $s$ cannot respond to a response-field insertion at the later time $t$. The resulting continuum identities are Eqs.~\eqref{equ:C2_C3_ts} and \eqref{equ:G2_G3_ts}.

The equal-time correlation requires additional care because its discrete time derivative obeys the product rule
\begin{equation}
\begin{split}
D_t
\EV{
U_t^\alpha(\bm{k})
\Ov{U}_t^\beta(\bm{p})
}
&=
\EV{
\left(
D_tU_t^\alpha(\bm{k})
\right)
\Ov{U}_t^\beta(\bm{p})
}
+
\EV{
U_{t-\Delta t}^\alpha(\bm{k})
D_t\Ov{U}_t^\beta(\bm{p})
}.
\end{split}
\label{equ:DerivativeTimeDiscreteUtUt}
\end{equation}
The time shift in the second term is essential. It implements the It\^{o} prescription and prevents the forcing contribution from being counted twice.

Combining Eqs.~\eqref{equ:Relation_1_discrete} and \eqref{equ:Relation_2_discrete} according to Eq.~\eqref{equ:DerivativeTimeDiscreteUtUt}, and subsequently taking the continuum limit, gives Eq.~\eqref{equ:C2_C3_tt}. The equal-time response contributes through the first term in Eq.~\eqref{equ:DerivativeTimeDiscreteUtUt}, whereas the response involving $U_{t-\Delta t}$ vanishes by causality. Consequently, the forcing contribution in Eq.~\eqref{equ:C2_C3_tt} appears only once.

\section{Derivation of the self-consistent Gaussian mean-field closure}
\label{app:mean_field_derivation}

Let $a=(\alpha,\bm{k},\tau)$ denote the compound index of a Cartesian component, a wavevector in $\mathbb{G}_+$, and a point on the discrete time grid.  We collect the corresponding path variables in the
vectors
\begin{equation}
\bm{u}=(u_a),
\qquad
\bm{\psi}=(\psi_a),
\qquad
\bm{\phi}
=
\begin{pmatrix}
\bm{u}\\
\bm{\psi}
\end{pmatrix}.
\end{equation}
The effective Gaussian action is defined by
\begin{equation}
S_{\mathrm{eff}}[\omega]
=
\Ov{\bm{\phi}}^{\intercal}
\bm{\Gamma}_{\mathrm{eff}}
\bm{\phi},
\label{def:S_eff}
\end{equation}
where $\bm{\Gamma}_{\mathrm{eff}}$ is an initially unknown quadratic vertex kernel. Its inverse,
\begin{equation}
\bm{\Gamma}_{\mathrm{eff}}^{-1}
=
\begin{pmatrix}
\EV{\bm{U}\otimes\Ov{\bm{U}}}_{\mathrm{eff}}
&
\EV{\bm{U}\otimes\Ov{\bm{\Psi}}}_{\mathrm{eff}}
\\
\EV{\bm{\Psi}\otimes\Ov{\bm{U}}}_{\mathrm{eff}}
&
0
\end{pmatrix},
\label{def:Gaussian_two_point_matrix}
\end{equation}
contains the effective Gaussian covariance and response kernels. These kernels are not identified with the full interacting two-point functions at the outset. Their self-consistent identification follows only after the moment expansion described below.

The associated Gaussian expectation value is
\begin{equation}
\EV{f}_{\mathrm{eff}}
=
\int_{\Omega_{(t_0,t]}}
f(\omega)e^{-S_{\mathrm{eff}}[\omega]}\dd\omega.
\label{def:ExpectedValue_eff}
\end{equation}
The effective response kernels retain the causal structure
\begin{equation}
\EV{
U_t^\alpha(\bm{k})
\Ov{\Psi}_s^\beta(\bm{p})
}_{\mathrm{eff}}
=
0,
\qquad
t<s,
\label{equ:effective_response_causality}
\end{equation}
and the equal-time condition
\begin{equation}
\EV{
U_t^\alpha(\bm{k})
\Ov{\Psi}_t^\beta(\bm{p})
}_{\mathrm{eff}}
=
\I L^3
\delta_{\alpha\beta}
\delta_{\bm{k},\bm{p}}.
\label{equ:effective_equal_time_response}
\end{equation}
No fluctuation--dissipation relation is imposed between the covariance and response blocks of $\bm{\Gamma}_{\mathrm{eff}}^{-1}$.

\subsection{Normalization of the effective Gaussian functional}
\label{app:Gaussian_normalization}

Let $d:=3N_t N_k$ denote the dimension of the complex vectors $\bm{u}$ and $\bm{\psi}$. The combined vector $\bm{\phi}$ has the dimension $2d$. The Gaussian integral is first defined for vertex matrices whose Hermitian part is positive definite and is subsequently analytically continued to the complex MSRDJ vertex kernel $\bm{\Gamma}_{\mathrm{eff}}$. This gives
\begin{equation}
\EV{1}_{\mathrm{eff}}
=
\det\left(
\pi\bm{\Gamma}_{\mathrm{eff}}^{-1}
\right)
\left(
\frac{1}{\pi^2L^6}
\right)^d.
\label{equ:Gaussian_normalization}
\end{equation}
The mixed blocks of $\bm{\Gamma}_{\mathrm{eff}}^{-1}$ are triangular as a consequence of causality, and their diagonal entries are fixed by Eq.~\eqref{equ:effective_equal_time_response}. The block-determinant identity therefore yields
\begin{equation}
\begin{split}
\det\left(
\pi\bm{\Gamma}_{\mathrm{eff}}^{-1}
\right)
&=
(-1)^d
\det\left(
\pi\EV{\bm{U}\otimes\Ov{\bm{\Psi}}}_{\mathrm{eff}}
\right)
\det\left(
\pi\EV{\bm{\Psi}\otimes\Ov{\bm{U}}}_{\mathrm{eff}}
\right)
\\
&=
(-1)^d
\left(
\pi\I L^3
\right)^{2d}
=
\pi^{2d}L^{6d}.
\end{split}
\label{equ:det_Gaussian_two_point_matrix}
\end{equation}
Consequently, $\EV{1}_{\mathrm{eff}}=1$. This normalization is understood through analytic continuation. The effective MSRDJ functional remains a complex Gaussian functional and is not a positive probability measure.

\subsection{Moment-wise expansion around the effective action}

We define
\begin{equation}
\Delta S[\omega]
:=
S[\omega]-S_{\mathrm{eff}}[\omega]
\label{def:DeltaS}
\end{equation}
and formally expand
\begin{equation}
\EV{f}
=
\int_{\Omega_{(t_0,t]}}
f(\omega)e^{-\Delta S[\omega]-S_{\mathrm{eff}}[\omega]}\dd\omega
=
\EV{f}_{\mathrm{eff}}
-
\EV{f\Delta S}_{\mathrm{eff}}
+
O\!\left(\Delta S^2\right).
\label{equ:DeltaS_expansion}
\end{equation}
For each moment, the expansion is truncated after its first non-vanishing contribution. Different moments are therefore retained at different formal orders in $\Delta S$. This prescription is not a controlled perturbative expansion. No small parameter ensures that the neglected terms are uniformly smaller than the retained contributions. Instead, it defines a self-consistent mean-field closure of the moment hierarchy.

The Gaussian contributions to the two-point functions are already nonzero. Hence,
\begin{align}
\EV{
U_t^\alpha(\bm{k})
\Ov{U}_s^\beta(\bm{p})
}
&\approx
\EV{
U_t^\alpha(\bm{k})
\Ov{U}_s^\beta(\bm{p})
}_{\mathrm{eff}},
\label{equ:matching_C}
\\
\EV{
U_t^\alpha(\bm{k})
\Ov{\Psi}_s^\beta(\bm{p})
}
&\approx
\EV{
U_t^\alpha(\bm{k})
\Ov{\Psi}_s^\beta(\bm{p})
}_{\mathrm{eff}},
\label{equ:matching_R}
\\
\EV{
\Psi_t^\alpha(\bm{k})
\Ov{U}_s^\beta(\bm{p})
}
&\approx
\EV{
\Psi_t^\alpha(\bm{k})
\Ov{U}_s^\beta(\bm{p})
}_{\mathrm{eff}}.
\label{equ:matching_A}
\end{align}
Only at this stage are the corresponding blocks of $\bm{\Gamma}_{\mathrm{eff}}^{-1}$ identified self-consistently with approximations to the full interacting two-point functions.

Because $S_{\mathrm{eff}}$ is centred and quadratic, all odd Gaussian moments vanish. Moreover, the insertion of the quadratic difference $S_{\mathrm{lam}}-S_{\mathrm{eff}}$ into a cubic observable produces an odd Gaussian moment and therefore vanishes. The first nonzero contributions to the required three-point functions are consequently
\begin{align}
\EV{
U_t^\gamma(\bm{k}-\bm{q})
U_t^\rho(\bm{q})
\Ov{U}_s^\beta(\bm{p})
}
&\approx
-\EV{
U_t^\gamma(\bm{k}-\bm{q})
U_t^\rho(\bm{q})
\Ov{U}_s^\beta(\bm{p})
S_{\mathrm{tur}}
}_{\mathrm{eff}},
\label{equ:three_point_mean_field}
\\
\EV{
U_t^\gamma(\bm{k}-\bm{q})
U_t^\rho(\bm{q})
\Ov{\Psi}_s^\beta(\bm{p})
}
&\approx
-\EV{
U_t^\gamma(\bm{k}-\bm{q})
U_t^\rho(\bm{q})
\Ov{\Psi}_s^\beta(\bm{p})
S_{\mathrm{tur}}
}_{\mathrm{eff}}.
\label{equ:response_three_point_mean_field}
\end{align}
With the discrete turbulent action $S_{\mathrm{tur}}$ and the abbreviation
$\tau^-=\tau-\Delta t$, this gives for the third velocity moment
\begin{equation}
\begin{split}
\EV{
U_t^\gamma(\bm{k}-\bm{q})
U_t^\rho(\bm{q})
\Ov{U}_s^\beta(\bm{p})
}\approx
&
-\frac{\I\Delta t}{L^6}
\sum_{\tau\in\Pi_{(t_0,t]}}
\sum_{\bm{q}_0,\bm{q}_1\in\mathbb{G}}
M_{\mu\nu\sigma}(\bm{q}_0)
\\
&\qquad\quad\times
\EV{
U_t^\gamma(\bm{k}-\bm{q})
U_t^\rho(\bm{q})
\Ov{U}_s^\beta(\bm{p})
\Ov{\Psi}_\tau^\mu(\bm{q}_0)
U_{\tau^-}^\nu(\bm{q}_0-\bm{q}_1)
U_{\tau^-}^\sigma(\bm{q}_1)
}_{\mathrm{eff}}.
\end{split}
\label{equ:velocity_three_point_discrete}
\end{equation}
and for the third response moment
\begin{equation}
\begin{split}
\EV{
U_t^\gamma(\bm{k}-\bm{q})
U_t^\rho(\bm{q})
\Ov{\Psi}_s^\beta(\bm{p})
}\approx
&
-\frac{\I\Delta t}{L^6}
\sum_{\tau\in\Pi_{(t_0,t]}}
\sum_{\bm{q}_0,\bm{q}_1\in\mathbb{G}}
M_{\mu\nu\sigma}(\bm{q}_0)
\\
&\qquad\quad\times
\EV{
U_t^\gamma(\bm{k}-\bm{q})
U_t^\rho(\bm{q})
\Ov{\Psi}_s^\beta(\bm{p})
\Ov{\Psi}_\tau^\mu(\bm{q}_0)
U_{\tau^-}^\nu(\bm{q}_0-\bm{q}_1)
U_{\tau^-}^\sigma(\bm{q}_1)
}_{\mathrm{eff}}.
\end{split}
\label{equ:response_three_point_discrete}
\end{equation}
Since the effective measure is Gaussian, the six-point function can
be decomposed into products of two-point functions using Wick's
theorem. Any contraction of $\Ov{\Psi}_\tau(\bm{q}_0)$ with one of the velocity fields at $\tau^-$  vanishes by causality. In contrast, equal-time velocity
contractions at $\tau^-$ remain finite at this stage because
homogeneity has not yet been imposed.

Inserting the approximations \eqref{equ:velocity_three_point_discrete} and \eqref{equ:response_three_point_discrete} into the exact relations \eqref{equ:C2_C3_ts}, \eqref{equ:G2_G3_ts} and \eqref{equ:C2_C3_tt} closes the hierarchy and yields the nonlinear DIA kernels in the continuum limit
\begin{equation}
\begin{split}
H_{t,s,\tau}^{\alpha\beta}[C,G]
(\bm{k},\bm{p})&=
\sum_{\bm{q},\bm{q}_0,\bm{q}_1\in\mathbb{G}}
M_{\alpha\gamma\rho}(\bm{k})
M_{\mu\nu\sigma}(\bm{q}_0)
\\
&\qquad\qquad\times
\bigg[
G_{s,\tau}^{\beta\mu}(-\bm{p},\bm{q}_0) C_{tt}^{\rho\gamma}(\bm{q},\bm{q}-\bm{k}) C^{\sigma\nu}_{\tau,\tau}(\bm{q}_1,\bm{q}_1-\bm{q}_0)
\\
&\qquad\qquad\qquad
+2 G_{t,\tau}^{\rho\mu}(\bm{q},\bm{q}_0)C^{\gamma\beta}_{t,s}(\bm{k}-\bm{q},\bm{p}) C_{\tau,\tau}^{\sigma\nu}(\bm{q}_1,\bm{q}_1-\bm{q}_0)
\\
&\qquad\qquad\qquad
+ 4 G_{t,\tau}^{\gamma\mu}(\bm{k}-\bm{q},\bm{q}_0)C_{\tau,s}^{\sigma\beta}(\bm{q}_1,\bm{p})C_{t,\tau}^{\rho\nu}(\bm{q},\bm{q}_1-\bm{q}_0)
\\
&\qquad\qquad\qquad
+2G_{s,\tau}^{\beta\mu}(-\bm{p},\bm{q}_0) C_{t,\tau}^{\gamma\nu}(\bm{k}-\bm{q},\bm{q}_1-\bm{q}_0)C_{t,\tau}^{\rho\sigma}(\bm{q},-\bm{q}_1)
\bigg] \frac{1}{L^9}
\end{split}
\label{def:H_continuum}
\end{equation}
and
\begin{equation}
\begin{split}
L_{t,s,\tau}^{\alpha\beta}[C,G]
(\bm{k},\bm{p}) &=
\sum_{\bm{q},\bm{q}_0,\bm{q}_1\in\mathbb{G}}
M_{\alpha\gamma\rho}(\bm{k})
M_{\mu\nu\sigma}(\bm{q}_0)
\\
&\qquad\qquad\quad\times
\bigg[
2 G_{t,\tau}^{\rho\mu}(\bm{q},\bm{q}_0)G^{\gamma\beta}_{t,s}(\bm{k}-\bm{q},\bm{p}) C_{\tau,\tau}^{\sigma\nu}(\bm{q}_1,\bm{q}_1-\bm{q}_0)
\\
&\qquad\qquad\qquad
+ 4 G_{t,\tau}^{\gamma\mu}(\bm{k}-\bm{q},\bm{q}_0)G_{\tau,s}^{\sigma\beta}(\bm{q}_1,\bm{p})C_{t,\tau}^{\rho\nu}(\bm{q},\bm{q}_1-\bm{q}_0)
\bigg] \frac{1}{L^9}.
\end{split}
\label{def:L_continuum}
\end{equation}
For the second kernel, we use the identity $P_{\mu\kappa}(\bm{q}_0)M_{\kappa\nu\sigma}(\bm{q}_0) = M_{\mu\nu\sigma}(\bm{q}_0)$, to rewrite the kernel as a functional of $C$ and $G$.

For homogeneous turbulence, the correlation and Green functions are
diagonal in Fourier space. The terms containing $C_{\tau,\tau}^{\nu\sigma}
(\bm{q}_0-\bm{q}_1,-\bm{q}_1)$ then vanish because momentum conservation would require $\bm{q}_0=\bm{0}$, which is excluded from $\mathbb{G}$.

\section{Approximations for the memory kernels}

\subsection{Wavenumber scaling of the initial response kernel}
\label{app:approximation_Mg0}

We first establish the asymptotic scaling $M(k,0)\sim v_{\mathrm{rms}}^2k^2$. Expanding the vertex $V(k,p,\mu)$ entering $M(k,0)$ in the limits $k\ll\kappa_c$ and $k\gg\kappa_c$ gives the corresponding prefactors.

For $k\ll\kappa_c$, one obtains
\begin{equation}
M(k,0)\approx\frac{1}{2}\int_0^\infty E(p)\int_{-1}^{1}\left[kp\mu(1-\mu^2)+2k^2(1-\mu^2)\mu^2\right]\dd\mu\,\dd p=\frac{2}{5}k^2v_{\mathrm{rms}}^2.
\label{equ:approxMg0_k_small}
\end{equation}
This approximation is justified because $E(p)\sim p^2$ at small $p$, which suppresses contributions from $p\ll\kappa_c$ at least quadratically. For $k\ll\kappa_c$, the integral is therefore dominated by $p\gg k$.

Conversely, for $k\gg\kappa_c$,
\begin{equation}
M(k,0)\approx\frac{1}{2}\int_0^\infty E(p)\int_{-1}^{1}k^2(1-\mu^2)\dd\mu\,\dd p=k^2v_{\mathrm{rms}}^2.
\label{equ:approxMg0_k_large}
\end{equation}
Here, the decay $E(p)\sim p^{-\beta}$ with $\beta>1$ suppresses contributions from $p\gg\kappa_c$. The dominant contribution consequently arises from $p\ll k$. Numerical evaluation confirms that Eqs.~\eqref{equ:approxMg0_k_small} and \eqref{equ:approxMg0_k_large} accurately describe the two asymptotic regimes.

\subsection{Time dependence at large wavenumbers}
\label{app:approximation_Mg_Mc}

We next derive the approximation $M(k,s)\simeq M(k,0)g(k,s)$ for $k\gg\kappa_c$. As in Eq.~\eqref{equ:approxMg0_k_large}, the integral is dominated by $p\ll k$. Consequently, $q\simeq k$ and $g(q,s)\simeq g(k,s)$. Moreover, on the timescale over which the large-wavenumber response $g(k,s)$ decays, the correlation function at the energy-containing wavenumbers varies only weakly. Hence, $c(p,s)\simeq c(p,0)=1$ in the dominant integration region. It follows that
\begin{equation}
M(k,s)\simeq g(k,s)\frac{1}{2}\int_0^\infty E(p)\int_{-1}^{1} V(k,p,\mu)\dd\mu\,\dd p=M(k,0)g(k,s).
\end{equation}

The same separation between slowly varying small-wavenumber modes and rapidly varying large-wavenumber modes gives $N(k,s)\simeq N(k,0)c(k,s)$ for $k\gg \kappa_c$. Numerical evaluation confirms that Eq.~\eqref{equ:approx_Mct_Mgt_k_large} accurately reproduces the full memory kernels in the large-wavenumber regime.

\section{Energy conservation}\label{app:EnergyConservation}

The following proof of energy conservation in the DIA is based on the derivation in Ref.~\cite{mccomb2014}. To demonstrate energy conservation, it is sufficient to show that
\begin{equation*}
\int_0^\infty 2\Phi(k,0)\,\mathrm{d}k = 0 .
\end{equation*}
Using the definition of $\Phi(k,t)$ in \eqref{equ:DIA_Phikt} together with the memory kernels $M(k,\tau)$ and $N(k,\tau)$ in \eqref{equ:DIA_Mg} and \eqref{equ:DIA_Mc}, we write
\begin{equation*}
\begin{split}
\Phi(k,0)
&= k^2 \int_0^\infty N(k,s)\, g(k,s)\,\mathrm{d}s
   - E(k) \int_0^\infty M(k,s)\, c(k,s)\,\mathrm{d}s \\
&= \varphi_1(k) - \varphi_2(k),
\end{split}
\end{equation*}
with
\begin{align*}
\varphi_1(k)
&=\frac{1}{2}\int_0^\infty\!\int_0^\infty
  k^2 g(k,s) E(p)c(p,s)
  \int_{-1}^1 V(k,p,\mu)c(q,s)\frac{E(q)}{q^2}\,
  \mathrm{d}\mu\,\mathrm{d}p\,\mathrm{d}s,\\[0.2em]
\varphi_2(k)
&=\frac{1}{2}\int_0^\infty\!\int_0^\infty
  E(k)c(k,s) E(p)c(p,s)
  \int_{-1}^1 V(k,p,\mu)g(q,s)\,
  \mathrm{d}\mu\,\mathrm{d}p\,\mathrm{d}s .
\end{align*}
Energy conservation follows once we have shown that
\begin{equation*}
\int_0^\infty \varphi_1(k)\,\mathrm{d}k
 = \int_0^\infty \varphi_2(k)\,\mathrm{d}k .
\end{equation*}
The vertex function is defined by
\begin{equation*}
V(\Vert \bm{k}\Vert,\Vert \bm{p}\Vert,\mu)= b(\bm{k},\bm{p})=-2M^{\alpha\gamma\rho}(\bm{k})M^{\gamma\alpha\sigma}(\bm{k}-\bm{p})P^{\rho\sigma}(\bm{p}). 
\end{equation*}

Rewriting the angular integrals in Cartesian coordinates gives
\begin{align*}
\begin{split}
\varphi_1(k)&=\int_0^\infty  \int_{\mathbb{R}^3} \frac{\Vert \bm{k}\Vert^2}{4\pi \Vert \bm{p}\Vert^2}b(\bm{k},\bm{p}) g(\Vert \bm{k}\Vert,s) c(\Vert \bm{p}\Vert,s)c(\Vert \bm{k}-\bm{p}\Vert,s) \\
& \qquad\qquad\qquad\qquad\qquad\qquad\qquad\qquad \times \frac{E(\Vert \bm{k}-\bm{p}\Vert)E(\Vert \bm{p}\Vert)}{\Vert \bm{k}-\bm{p}\Vert^2} \dd \bm{p} \dd s\\
\varphi_2(k)&=\int_0^\infty \int_{\mathbb{R}^3} \frac{1}{4\pi \Vert \bm{p}\Vert^2}b(\bm{k},\bm{p}) c(\Vert\bm{k}\Vert,s) c(\Vert \bm{p}\Vert,s)g(\Vert \bm{k}-\bm{p}\Vert,s)\\
& \qquad\qquad\qquad\qquad\qquad\qquad\qquad\qquad \times E(\Vert \bm{k}\Vert)E(\Vert \bm{p}\Vert) \dd \bm{p} \dd s
\end{split}
\end{align*}
Applying the substitution $\bm{p} \rightarrow \bm{k}-\bm{p}$ yields 
\begin{align*}
\varphi_1(k)&=\int_0^\infty  \int_{\mathbb{R}^3} \frac{\Vert \bm{k}\Vert^2}{4\pi}b(\bm{k},\bm{k}-\bm{p}) g(\Vert \bm{k}\Vert,s)c(\Vert \bm{k}- \bm{p}\Vert,s)c(\Vert \bm{p}\Vert,s) \\
& \qquad\qquad\qquad\qquad\qquad\qquad\qquad\qquad \times \frac{E(\Vert \bm{k}-\bm{p}\Vert)E(\Vert\bm{p}\Vert)}{\Vert \bm{k}-\bm{p}\Vert^2 \Vert \bm{p}\Vert^2}   \dd \bm{p} \dd s\\
\varphi_2(k)&=\int_0^\infty \int_{\mathbb{R}^3} \frac{1}{4\pi} b(\bm{k},\bm{k}-\bm{p}) c(\Vert\bm{k}\Vert,s)c(\Vert\bm{k}- \bm{p}\Vert,s)g(\Vert \bm{p}\Vert,s) \\
& \qquad\qquad\qquad\qquad\qquad\qquad\qquad\qquad \times \frac{E(\Vert \bm{k}-\bm{p}\Vert)E(\Vert \bm{k}\Vert)}{\Vert \bm{k}-\bm{p}\Vert^2} \dd \bm{p} \dd s.
\end{align*}

The function $b(\cdot,\cdot)$ satisfies the relation
\begin{equation*}
b(\bm{k},\bm{k}-\bm{p})=-2M^{\alpha\gamma\rho}(\bm{k})M^{\gamma\alpha\sigma}(\bm{p})P^{\rho\sigma}(\bm{k}-\bm{p})=b(\bm{p},\bm{p}-\bm{k}).
\end{equation*}
Using this symmetry and interchanging the variables $\bm{k}$ and $\bm{p}$ inside the triple integral gives
\begin{equation*}
\begin{split}
&\int_0^\infty \varphi_2(k)\dd k =\int_{\mathbb{R}^3} \frac{\varphi_2(\Vert \bm{k}\Vert)}{4\pi \Vert \bm{k}\Vert^2}\dd \bm{k}\\
&\quad=\int_0^\infty \int_{\mathbb{R}^3} \int_{\mathbb{R}^3} \frac{1}{16\pi^2} b(\bm{k},\bm{k}-\bm{p}) c(\Vert\bm{k}\Vert,s)c(\Vert\bm{k}- \bm{p}\Vert,s)g(\Vert \bm{p}\Vert,s) \\
& \qquad\qquad\qquad\qquad\qquad\qquad\qquad\qquad \times \frac{E(\Vert \bm{k}-\bm{p}\Vert)E(\Vert \bm{k}\Vert)}{\Vert \bm{k}-\bm{p}\Vert^2 \Vert \bm{k}\Vert^2} \dd \bm{p} \dd \bm{k} \dd s\\
&\quad = \int_{\mathbb{R}^3} \frac{\varphi_1(\Vert \bm{k}\Vert)}{4\pi \Vert \bm{k}\Vert^2}\dd \bm{k}=\int_0^\infty \varphi_1(k)\dd k
\end{split}
\end{equation*}
which completes the proof. In the third equality, we have interchanged the integration variables $\bm{k}$ and $\bm{p}$ and used the symmetry of $b(\bm{k},\bm{k}-\bm{p})$.

\section{Dissipation-range exponent analysis of the energy spectrum}
\label{app:dissipation_range_analysis}

Following the commonly used stretched-exponential representation of the dissipation-range spectrum \cite{Khurshid2018}, we consider
\begin{equation}
E(k)=\gamma_0\varepsilon^{2/3}\kappa_d^{-5/3}x^{\gamma_1}\exp\left(-\gamma_2x^{\gamma_3}\right),\qquad x=\frac{k}{\kappa_d}.
\label{equ:DR_Modelfunction_Ek}
\end{equation}
For a constant stretched-exponential exponent, this form implies
\begin{equation}
\frac{\dd^2\log E}{\dd(\log x)^2}=-\gamma_2\gamma_3^2x^{\gamma_3}.
\label{equ:Second_Derivative_logE}
\end{equation}
We therefore define the local exponent by
\begin{equation}
\gamma_3^{\mathrm{loc}}(x)=\frac{\dd}{\dd\log x}\log\left[-\frac{\dd^2\log E}{\dd(\log x)^2}\right].
\label{def:gamma3_local}
\end{equation}

In the near-dissipation range $0.2\lesssim x\lesssim4$, $\gamma_3^{\mathrm{loc}}(x)$ decreases continuously from approximately $2.5$ to $1$. Thus, a single stretched-exponential exponent does not describe the entire crossover. For $x\gtrsim4$, however, $\gamma_3^{\mathrm{loc}}(x)$ is indistinguishable from unity within numerical accuracy. We therefore set $\gamma_3=1$ in the far-dissipation range, yielding a power law multiplied by a simple exponential as anticipated by Kraichnan \cite{Kraichnan1959}.

\begin{figure}
\centerline{\includegraphics{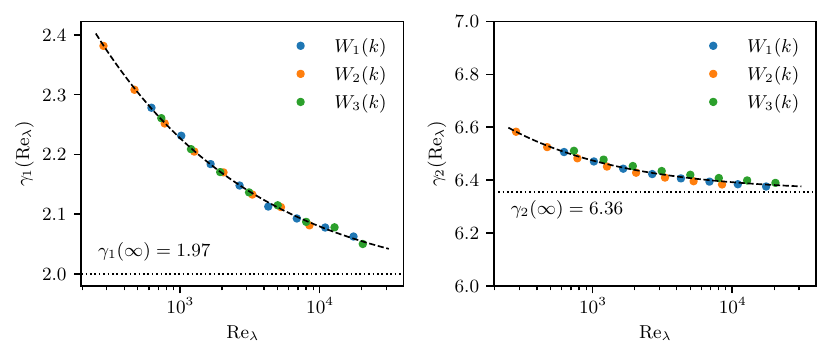}}
\caption{Reynolds-number dependence of the far-dissipation-range parameters for the three work spectra $W_i$. The left panel shows $\gamma_1(\Rey_\lambda)$ and the right panel shows $\gamma_2(\Rey_\lambda)$. Dashed lines represent the empirical fits \eqref{equ:gamma1_Re} and \eqref{equ:gamma2_Re}. In determining both parameters, the stretched-exponential exponent is fixed to $\gamma_3=1$.}
\label{fig:gamma1_gamma2}
\end{figure}

The Reynolds-number dependence of the remaining fitted exponents is described empirically by
\begin{align}
\gamma_1(\Rey_\lambda)&\simeq1.97+4.08\,\Rey_\lambda^{-0.37},
\label{equ:gamma1_Re}\\
\gamma_2(\Rey_\lambda)&\simeq6.36+4.42\,\Rey_\lambda^{-0.52}.
\label{equ:gamma2_Re}
\end{align}
Figure~\ref{fig:gamma1_gamma2} shows the fitted parameters for all three work spectra. The different forcing shapes follow the same Reynolds-number dependence within numerical accuracy and approach
\begin{equation}
\gamma_1\to1.97\simeq2,\qquad \gamma_2\to6.36
\end{equation}
as $\Rey_\lambda\to\infty$. The limiting values $\gamma_1\simeq2$ and $\gamma_3=1$ are consistent with Kraichnan's asymptotic prediction. The parameter $\gamma_2(\Rey_\lambda)$ is already approximately constant for $\Rey_\lambda\gtrsim10^3$, where the numerical values scatter around the empirical fit with an uncertainty of approximately $0.24\%$.

\section{Discretisation and numerical solution of the DIA equations}\label{app:NumericalDiscretisation}

To discretise Eqs. \eqref{equ:DIA_ckt}--\eqref{equ:DIA_Ek}, suitable partitions must be chosen for the wavenumber $k\in [k_{\mathrm{min}},k_{\mathrm{max}}]$, the angle integration variable $\mu \in [-1,1]$ and the time $t\in [0,t_{\mathrm{max}}]$. On the interval $[k_{\mathrm{min}},k_{\mathrm{max}}]$ we select the partition $\Pi_k$ with the grid points
\begin{equation}
k_m=k_{\mathrm{min}} \frac{b^{m+1}-1}{b-1} \qquad  m\in \{0,1,2,...,\mathsf{N}_k\}
\end{equation}
where $\mathsf{N}_k=\max\{m:4 k_d >k_m\}$ applies. We set $k_{\mathrm{min}}=0.1 \kappa_c$ for the minimum wavenumber and $b=1.02571$, which gives $k_{400}\simeq10^5 \kappa_c$. This resolution of $\Pi_k$ is sufficient.

For the integration variable $\mu \in [-1,1]$, we select the partition $\Pi_\mu$ with the grid points
\begin{equation}
\mu_n=\cos \vartheta_n \qquad \text{where} \qquad \vartheta_n=\pi\left(1-\frac{n}{\mathsf{N}_\mu} \right)  \qquad n\in \{0,1,...,\mathsf{N}_\mu\}.
\end{equation}
The angle $\vartheta$ is therefore sampled on an equidistant grid with spacing $\Delta\vartheta=\pi/\mathsf{N}_\mu$.

\begin{figure}
\centerline{\includegraphics{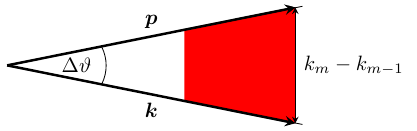}}
\caption{Schematic illustration to derive a lower limit for the angular step size $\Delta \vartheta$ as a function of the wavenumber step size $k_m-k_{m-1}$.}
\label{fig:PartitionMu}
\end{figure}

To obtain comparable angular and radial resolutions in wavevector space, we choose $\mathsf{N}_\mu$ as a function of $b$. This dependence is discussed in the following. Figure \ref{fig:PartitionMu} shows the wavevector $\bm{k}$, which is rotated by the angle $\Delta \vartheta$, resulting in the wavevector $\bm{p}$. Both wavevectors have the norm $\Vert \bm{k}\Vert=\Vert \bm{p}\Vert =k_m$ with the choice $k_m\in \Pi_k$. For the distance between $\bm{k}$ and $\bm{p}$ we now require $\Vert \bm{k}-\bm{p}\Vert=k_m-k_{m-1}$, which then results in
\begin{equation}
\Vert \bm{k}-\bm{p}\Vert=\sqrt{k_m^2-2k_m^2 \cos(\Delta \vartheta)+k_m^2}=2k_m \sin\left( \frac{\Delta \vartheta}{2}\right)=k_m-k_{m-1}=k_{\mathrm{min}}b^m.
\end{equation}
Substituting the angular increment $\Delta\vartheta=\pi/\mathsf{N}_\mu$ then gives
\begin{equation}
\frac{\pi}{2\mathsf{N}_\mu}=\arcsin\left(\frac{\Delta k}{2k_m} \right)=\arcsin\left( \frac{1}{2}\frac{b-1}{b-b^{-m}}\right)\geq \frac{1}{2}\frac{b-1}{b-b^{-m}} \geq \frac{1}{2}\frac{b-1}{b}.
\end{equation}
This means that the partition $\Pi_\mu$ with the choice
\begin{equation}
\mathsf{N}_\mu=\bigg\lceil \frac{\pi b}{b-1} \bigg\rceil \label{equ:estimate_Nmu}
\end{equation}
is sufficiently fine, where $\lceil \cdot \rceil$ is the ceiling function. The values of all numerical constants are listed in table \ref{tab:NumericalParameters}.

\begin{table}
  \centering
  \caption{Numerical parameters for the partitions $\Pi_k$, $\Pi_\mu$ and $\Pi_\tau$.}
  \label{tab:NumericalParameters}
  \begin{tabular}{ccccc}
    \hline\hline
      $k_{\mathrm{min}}$  & $b$   &   $\mathsf{N}_\mu$ & $\mathsf{N}_t$ & $\Delta \tau$ \\ \hline
       $0.1 \kappa_c$   &  $1.02571$ & $126$ & $1600$ & $0.02$\\
    \hline\hline
  \end{tabular}
\end{table}

According to the equations \eqref{equ:DIA_ckt}--\eqref{equ:DIA_Ek}, the temporal decay of the correlation functions $c(k,t)$ and $g(k,t)$ depends strongly on the wavenumber $k$. Therefore, we rescale the time $t=\xi(k)\tau$ with the characteristic time
\begin{equation}
\xi(k)= M(k,0)^{-1/2}. \label{equ:Def_tc}
\end{equation}
As shown in Sec.~\ref{subsec:discussion_timescales}, $M(k,0)\simeq v_{\mathrm{rms}}^2k^2$ for $k\gg\kappa_c$. Hence, $\xi(k)\simeq \tau_{\mathrm{con}}(k)=1/(v_{\mathrm{rms}}k)$.

The rescaled correlation functions are then represented by 
\begin{equation}
\tilde{c}(k,\tau):=c(k,\xi(k)\tau) \qquad \text{and} \qquad \tilde{g}(k,\tau):=g(k,\xi(k)\tau).
\end{equation}

For the rescaled time $\tau$, we select the partition $\Pi_\tau$ with the equidistant grid 
\begin{equation}
\tau_j = j \Delta \tau  \qquad  j\in \{0,1,...,\mathsf{N}_t\}.
\end{equation}

Section~\ref{sec:Numerical_errors} shows that the partitions $\Pi_\tau$ and $\Pi_k$ are sufficiently fine when the parameters in Table~\ref{tab:NumericalParameters} are used. The number $\mathsf{N}_t$ is chosen large enough so that $\tilde{c}(k,\tau)$ and $\tilde{g}(k,\tau)$ decay sufficiently to zero. The step size $\Delta \tau$ and the number $\mathsf{N}_t$ are listed in table \ref{tab:NumericalParameters}. As usual, we set $\tilde{c}_{i,j}:=\tilde{c}(k_i,\tau_j)$ and $\tilde{g}_{i,j}: =\tilde{g}(k_i,\tau_j)$ and collect the elements $\tilde{c}_{i,j}$ and $\tilde{g}_{i,j}$ into the matrices $\tilde{\bm{c}}$ and $\tilde{\bm{g}}$. 

For $\tau\notin\Pi_\tau$ or $k\notin\Pi_k$, we use linear interpolation between neighbouring grid points. Linear interpolation is defined below for the one-dimensional function $\varphi(x)$ and the two-dimensional function $\psi(x,y(x))$, using the partitions $\Pi_x=(x_0,x_1,\ldots,x_n)$ for $x$ and $\Pi_y=(y_0,y_1,\ldots,y_m)$ for $y$.

The one-dimensional interpolation function $\mathcal{I}[\varphi](x)$ for $x_0<x<x_n$ is given by
\begin{equation}
\mathcal{I}[\varphi](x):= \varphi(x_\eta)(1-\lambda_x)+\varphi(x_{\eta+1})\lambda_x
\end{equation}
with the constants
\begin{align}
\eta&:=\max \{i\in \mathbb{N}:x-x_i \geq 0\}  \\
\lambda_x&:=\frac{x-x_\eta}{x_{\eta+1}-x_{\eta}}
\end{align}
The two-dimensional interpolation function $\mathcal{I}[\psi](x,y(x))$ for $x_0<x<x_n$ and $y_0<y<y_m$ is defined by
\begin{equation}
\begin{split}
\mathcal{I}[\psi](x,y(x))&=[\psi(x_\eta,y_{\varrho_\eta})(1-\lambda_{y_\eta})+\psi(x_\eta,y_{\varrho_\eta+1})\lambda_{y_\eta}](1-\lambda_x)\\
&\quad + [\psi(x_{\eta+1},y_{\varrho_{\eta+1}})(1-\lambda_{y_{\eta+1}})+\psi(x_{\eta+1},y_{\varrho_{\eta+1}+1})\lambda_{y_{\eta+1}}]\lambda_x
\end{split}
\end{equation}
with the constants
\begin{align}
\eta&:=\max \{i\in \mathbb{N}:x-x_i \geq 0\}  \\
\lambda_x&:=\frac{x-x_\eta}{x_{\eta+1}-x_{\eta}} \\
\varrho_r&:=\max\{i\in \mathbb{N}:y(x_r)-y_i\geq 0\}, \qquad r\in\{0,1,...,n\}\\
\lambda_{y_r}&:=\frac{y(x_r)-y_{\varrho_r}}{y_{\varrho_r+1}-y_{\varrho_r}}
\end{align}
The case $x=x_i\in \Pi_x$ yields $\eta=i$ and $\lambda_x=0$. If $x=k\in [k_{\mathrm{min}},k_{\mathrm{max}})$ and $y(x)=\tau(k)\in [0,\Delta \tau \mathsf{N}_t)$, $\eta$ and $\varrho_\eta$ can be explicitly expressed by
\begin{align}
\eta&:= \bigg\lfloor\,  \frac{\log(k(b-1)/k_{\mathrm{min}}+1)}{\log b}-1  \bigg\rfloor  \label{equ:Def_eta} \\
\varrho_r&:= \bigg\lfloor \frac{\tau(k_r)}{\Delta \tau} \bigg\rfloor \qquad r\in \{0,1,...,\mathsf{N}_k\},
\end{align}
where $\lfloor \cdot \rfloor$ is the floor function. The case $k\notin [k_{\mathrm{min}},k_{\mathrm{max}})$ or $\tau(k)\notin [0,\Delta \tau \mathsf{N}_t)$ is specified below.

The purely temporal structure of the equations \eqref{equ:DIA_ckt} and \eqref{equ:DIA_gkt} corresponds to the form
\begin{equation}
\varphi^{\prime}(t)+a\varphi(t)=\psi(t). \label{equ:time_structure_DIA}
\end{equation}
With the choice $\varphi(t)=f(t)e^{-at}$, the differential equation  
\begin{equation}
e^{-at}f^{\prime}(t)=\psi(t)
\end{equation}
follows. We discretise the time derivative $f^{\prime}(t)$ using the backward differentiation formula
\begin{align}
\text{first order} \qquad f^{\prime}(t) &\longleftarrow \frac{f(t_j)-f(t_{j-1})}{\Delta t}  \\
\text{second order} \qquad f^{\prime}(t) &\longleftarrow \frac{3}{2\Delta t}\left( f(t_j)-\frac{4}{3}f(t_{j-1})+\frac{1}{3}f(t_{j-2})\right) 
\end{align}
with the equidistant time grid $t_j=j\Delta t+t_0$. If these discretisations are used, the discrete formulations
\begin{align}
\text{first order} \qquad &\frac{\varphi(t_j)-e^{-a\Delta t}\varphi(t_{j-1})}{\Delta t}= \psi(t_j) \label{equ:time_structure_DIA_BDF1} \\
\text{second order} \qquad & \frac{3}{2\Delta t}\left( \varphi(t_j)-\frac{4}{3}e^{-a\Delta t}\varphi(t_{j-1})+\frac{1}{3}e^{-2a\Delta t}\varphi(t_{j-2})\right) =\psi(t_j) \label{equ:time_structure_DIA_BDF2}
\end{align}
result for Eq.~\eqref{equ:time_structure_DIA}. The great advantage of this discretisation is that the linear term $a\varphi(t)$ is integrated exactly. The first time step, $j=1$, is computed using the first-order scheme \eqref{equ:time_structure_DIA_BDF1}. The second-order scheme \eqref{equ:time_structure_DIA_BDF2} is used for all subsequent time steps, $j\geq2$.

We introduce the notation $\Delta t(k_i)=\xi(k_i)\Delta \tau $. For the sum of the time derivative and the dissipative term at time point $t=\xi(k_i)\tau_j$ the discretisations
\begin{align}
    \mathcal{D}\tilde{c}_{i,j} &\longleftarrow \pd{t} c(k_i,t)+\nu_0k^2_i c(k_i,t)\\
    \mathcal{D}\tilde{g}_{i,j} &\longleftarrow \pd{t} g(k_i,t)+\nu_0k^2_i g(k_i,t)
\end{align}
result, where $\mathcal{D}$ is a discrete differential operator defined by
\begin{equation}
    \mathcal{D} A_{i,j}=
    \begin{cases}
    \frac{1}{\Delta t(k_i)}(A_{i,j}-e^{-\nu_0k_i^2 \Delta t(k_i)}A_{i,j-1})&  \text{if} \quad j=1\\ 
    \frac{3}{2\Delta t(k_i)}\left( A_{i,j}-\frac{4}{3}e^{-\nu_0k_i^2 \Delta t(k_i)}A_{i,j-1}+\frac{1}{3}e^{-2 \nu_0k_i^2 \Delta t(k_i)}A_{i,j-2}\right) & \text{if} \quad j \geq 2
    \end{cases}
\end{equation}
for an arbitrary matrix $A_{i,j}$.

All integrals in Eqs. \eqref{equ:DIA_ckt}--\eqref{equ:DIA_Mc} are discretised using the trapezoidal rule. The notation of it is defined below. Let $\varphi(x)$ be a one-dimensional function on the interval $[x_{\mathrm{min}},x_{\mathrm{max}}]$, with partition $\Pi_x=(x_m,x_{m+1},x_{m+2},\ldots,x_n)$ and boundary points $x_m=x_{\mathrm{min}}$ and $x_n=x_{\mathrm{max}}$. The trapezoidal rule replaces the integral by
\begin{equation}
\int_{x_{\mathrm{min}}}^{x_{\mathrm{max}}} \varphi(x)\dd x \longleftarrow \sum_{i=m}^n \varphi(x_i)w^x_i(m,n),
\end{equation}
where the weight is given by
\begin{equation}
w^x_i(m,n)=\frac{1}{2}\begin{cases}
x_{m+1}-x_{m} & \text{if} \quad i=m\\
x_{i+1}-x_{i-1} & \text{if} \quad m < i < n\\
x_{n}-x_{n-1} & \text{if} \quad i=n
\end{cases}
\end{equation}
for $m<n$ and $w_m^x(m,m)=0$ for $m=n$.

The energy spectrum can be written in the general form 
\begin{equation}
    E(k_i)=E(\zeta(k_i))=\varepsilon^{2/3}\kappa_c^{-5/3} e^{-\zeta(k_i)}.
\end{equation}

Again we summarise the elements $\zeta(k_i)$ into the vector $\bm{\zeta}$. For the memory kernels $M(k,t)$ and $N(k,t)$ in the equations \eqref{equ:DIA_Mg} and \eqref{equ:DIA_Mc}, the discrete memory kernels follow using the trapezoidal rule
\begin{equation}
\begin{split}
\tilde{N}_{i,j}&=\frac{1}{2} \sum_{m=0}^{\mathsf{N}_k} E(\zeta(k_m)) \mathcal{I}[\tilde{c}(k_m,\cdot)](\xi(k_i)\tau_j/\xi(k_m)) w_m^k(0,\mathsf{N}_k)  \\
&\qquad \times \sum_{n=1}^{\mathsf{N}_\mu-1} V(k_i,k_m,\mu_n) \mathcal{I}[\tilde{c}(\cdot,\cdot)](q,\xi(k_i)\tau_j/\xi(q))\frac{E(\mathcal{I}[\zeta](q))}{q^2}  w^\mu_n(0,\mathsf{N}_\mu)  \label{equ:DIA_discret_Mc}   
\end{split}
\end{equation}
and
\begin{equation}
\begin{split}
\tilde{M}_{i,j}&=\frac{1}{2} \sum_{m=0}^{\mathsf{N}_k} E(\zeta(k_m)) \mathcal{I}[\tilde{c}(k_m,\cdot)](\xi(k_i)\tau_j/\xi(k_m)) w_m^k(0,\mathsf{N}_k)  \\
&\qquad\quad  \times \sum_{n=1}^{\mathsf{N}_\mu-1} V(k_i,k_m,\mu_n) \mathcal{I}[\tilde{g}(\cdot,\cdot)](q,\xi(k_i)\tau_j/\xi(q)) w^\mu_n(0,\mathsf{N}_\mu)\label{equ:DIA_discret_Mg} 
\end{split}
\end{equation}
with the wavenumber $q=\sqrt{k_i^2-2\mu_nk_i k_m+k_m^2}$ and the logarithmic interpolation
\begin{equation}
E(\mathcal{I}[\zeta](q))= \varepsilon^{2/3}\kappa_c^{-5/3} e^{-\mathcal{I}[\zeta](q)} 
\end{equation}
of $E(q)$. In the case of $q\notin [k_{\mathrm{min}},k_{\mathrm{max}})$ or $\tau \notin [0,\Delta \tau \mathsf{N}_t)$ we set
\begin{equation}
\mathcal{I}[\tilde{c}(\cdot,\cdot)](q,\tau)=\mathcal{I}[\tilde{g}(\cdot,\cdot)](q,\tau)=0, \qquad \frac{E(\mathcal{I}[\zeta](q))}{q^2}=0.
\end{equation}
The boundary points $n=0$ and $n=\mathsf{N}_\mu$ are omitted because the vertex $V(k_i,k_m,\mu_n)$ vanishes there. If this step is not carried out, indeterminate expressions of the form $0/0$ result for $n=\mathsf{N}_\mu$ and $i=m$, which leads to numerical problems. We have parallelised the calculation of the matrix elements $\tilde{M}_{i,j}$ and $\tilde{N}_{i,j}$ in order to save computing time.

The discrete form $\tilde{\Phi}_{i,j}$ of the auxiliary function $\Phi(k_i,\xi(k_i)\tau_j)$ is given by
\begin{equation}
\tilde{\Phi}_{i,j}=k_i^2 \sum_{l=0}^{\mathsf{N}_t-j} \tilde{N}_{i,j+l}\tilde{g}_{i,l} \xi(k_i) w^\tau_l(0,\mathsf{N}_t-j)-E(\zeta(k_i))\sum_{l=0}^{\mathsf{N}_t-j} \tilde{M}_{i,j+l} \tilde{c}_{i,l} \xi(k_i) w^\tau_l(0,\mathsf{N}_t-j). 
\end{equation}

The DIA equations for homogeneous, isotropic, and stationary turbulence can now be written in discrete form
\begin{align}
\mathcal{D}\tilde{c}_{i,j}&=-\sum_{l=0}^j \tilde{M}_{i,j-l}\tilde{c}_{i,l}\xi(k_i) w^\tau_l(0,j)+\varepsilon^{-2/3}\kappa_c^{5/3} e^{\zeta(k_i)}\tilde{\Phi}_{i,j}, \label{equ:DIA_discret_ckt} \\
\mathcal{D}\tilde{g}_{i,j}&=-\sum_{l=0}^j \tilde{M}_{i,j-l}\tilde{g}_{i,l}\xi(k_i) w^\tau_l(0,j), \label{equ:DIA_discret_gkt} \\
\varepsilon^{2/3}\kappa_c^{-5/3} e^{-\zeta(k_i)}&=\frac{W(k_i)+2k_i^2 \sum_{l=0}^{\mathsf{N}_t} \tilde{N}_{i,l}\tilde{g}_{i,l}\xi (k_i) w^\tau_l(0,\mathsf{N}_t)}{2\nu_0k_i^2 + 2 \sum_{l=0}^{\mathsf{N}_t} \tilde{M}_{i,l} \tilde{c}_{i,l}\xi(k_i) w^\tau_l(0,\mathsf{N}_t)}. \label{equ:DIA_discret_chi} 
\end{align}
The discretised DIA equations \eqref{equ:DIA_discret_ckt} and \eqref{equ:DIA_discret_gkt} are linear equations in $\tilde{c}_{i,j}$ and  $\tilde{g}_{i,j}$ when $\tilde{N}_{i,j}$, $\tilde{M}_{i,j}$, and $\tilde{\Phi}_{i,j}$ are held fixed. We therefore choose the initial functions 
\begin{equation}
\tilde{c}_{i,j}^0=\tilde{g}_{i,j}^0=e^{-j/200}, \qquad \zeta^0(k_i)=\log(1.5\,k_i/\kappa_c)
\end{equation}
for all $i\in\{0,1,2,...,\mathsf{N}_k\}$ and $j\in \{0,1,2,...,\mathsf{N}_t\}$ and use them to calculate the matrices $\tilde{N}_{i,j}^{0}$, $\tilde{M}_{i,j}^{0}$ and $\tilde{\Phi}_{i,j}^0$. Using them, we now calculate the new functions $\tilde{c}_{i,j}^1$ and $\tilde{g}_{i,j}^1$ from equations \eqref{equ:DIA_discret_ckt} and \eqref{equ:DIA_discret_gkt}. This results in the following fixed-point iterations 
\begin{align}
\mathcal{D}\tilde{c}_{i,j}^{\rho+1}&=-\sum_{l=0}^j \tilde{M}^{\rho}_{i,j-l}\tilde{c}_{i,l}^{\rho+1}\xi^{\rho}(k_i) w^\tau_l(0,j)+\varepsilon^{-2/3}\kappa_c^{5/3} e^{\zeta^{\rho}(k_i)}\tilde{\Phi}_{i,j}^{\rho},\\
\mathcal{D}\tilde{g}_{i,j}^{\rho+1}&=-\sum_{l=0}^j \tilde{M}^{\rho}_{i,j-l}\tilde{g}_{i,l}^{\rho+1}\xi^{\rho}(k_i) w^\tau_l(0,j),\\
\varepsilon^{2/3}\kappa_c^{-5/3}  e^{-\zeta^{\rho+1}(k_i)}&=\frac{W(k_i)+2k_i^2 \sum_{l=0}^{\mathsf{N}_t} \tilde{N}^{\rho}_{i,l}\tilde{g}_{i,l}^{\rho+1}\xi^{\rho}(k_i) w^\tau_l(0,\mathsf{N}_t)}{2\nu_0k_i^2 + 2 \sum_{l=0}^{\mathsf{N}_t} \tilde{M}^{\rho}_{i,l} \tilde{c}_{i,l}^{\rho+1}\xi^{\rho}(k_i) w^\tau_l(0,\mathsf{N}_t)}.
\end{align}
which generate the sequences $(\tilde{c}_{i,j}^\rho)_{\rho \in \mathbb{N}}$, $(\tilde{g}_{i,j}^\rho)_{\rho\in \mathbb{N}}$ and $(\zeta^\rho(k_i))_{\rho \in \mathbb{N}}$. We use $\Delta t(k_i)=\xi^\rho(k_i)\Delta \tau$ in the discrete differential operator $\mathcal{D}$. The fixed point iteration is stopped if
\begin{equation}
\Vert \tilde{\bm{c}}^{\rho+1}-\tilde{\bm{c}}^\rho\Vert_{\max}<\epsilon, \qquad \Vert \tilde{\bm{g}}^{\rho+1}-\tilde{\bm{g}}^\rho\Vert_{\max}<\epsilon, \qquad \Vert \bm{\zeta}^{\rho+1}-\bm{\zeta}^\rho \Vert_{\max} < \epsilon  
\end{equation}
where $\Vert \cdot \Vert_{\max}$ denotes the matrix maximum norm. The choice $\epsilon=10^{-8}$ is sufficient. For the initial functions tested, the iteration converges to the same numerical solution.

\section{Analysis of numerical errors}\label{sec:Numerical_errors}

The resolution of the partitions $\Pi_k$ and $\Pi_\tau$ is uniquely determined by the numerical parameters $k_{\mathrm{min}}$, $b$ and $\Delta \tau$ according to section \ref{app:NumericalDiscretisation}. The fineness of $\Pi_\mu$ is also determined by $b$ according to equation \eqref{equ:estimate_Nmu}. In this section, we show that the choice of the numerical parameters $b$ and $\Delta \tau$ in table \ref{tab:NumericalParameters} is sufficient. For the work spectra $W_i(k)$ used, $k_{\mathrm{min}}=0.1\kappa_c$ is small enough. 

We solve the DIA numerically using four levels of refinement of the partitions $\Pi_k^r$ and $\Pi_\tau^r$ with the index $r\in \{1,2,3,4\}$ for the dissipation wavenumber $k_d=6400\kappa_c$ and the work spectrum $W_1(k)$. The numerical parameters $\Delta \tau^r$ and $b^r$ of the partitions $\Pi_k^r$ and $\Pi_\tau^r$ are listed in table \ref{tab:Partition_tau_k}. These parameters are chosen so that the grid resolution doubles at each refinement step $r\to r+1$.

\begin{table}
  \centering
  \caption{Numerical parameters $\Delta \tau^r$ and $b^r$ for the partitions $\Pi_k^r$ and $\Pi_\tau^r$, with the grid resolution doubling at each refinement step $r\to r+1$.}
  \label{tab:Partition_tau_k}
  \begin{tabular}{ccc}
    \hline\hline
      r  & $\Delta \tau^r$   &   $b^r$   \\ \hline
       $1$   &  $0.16$ & $1.2787$ \\
       $2$   &  $0.08$ & $1.12304$ \\
       $3$   &  $0.04$ & $1.05589$ \\
       $4$   &  $0.02$ & $1.02571$ \\
    \hline\hline
  \end{tabular}
\end{table}

We start with the numerical errors of the energy spectrum for the partitions $\Pi_k^r$ and $\Pi_\tau^r$. In the interval $k\in [0,4k_d]$, the energy spectrum covers in total $47$ decades. A conventional representation in a log–log plot cannot therefore reveal the exact numerical errors. For the four ranges  large-scale range $\mathrm{(lr)}$, energy-containing range $\mathrm{(er)}$,  inertial range $\mathrm{(ir)}$ and  dissipation range $\mathrm{(dr)}$, we therefore define the four weight functions
\begin{align}
\rho_E^{\mathrm{lr}}(k)&=(k/\kappa_c)^{-2} \label{equ:rho_E_lr}\\
\rho_E^{\mathrm{er}}(k)&=1 \\
\rho_E^{\mathrm{ir}}(k)&=(k/\kappa_c)^{3/2} \\
\rho_E^{\mathrm{dr}}(k)&=(k/\kappa_d)^{-2.152} e^{6.411 k/\kappa_d}, \label{equ:rho_E_dr}
\end{align}
which highlight the respective spectral ranges. We use $\kappa_d=1588\kappa_c$, the value obtained for $r=4$.

\begin{figure}
\centerline{\includegraphics{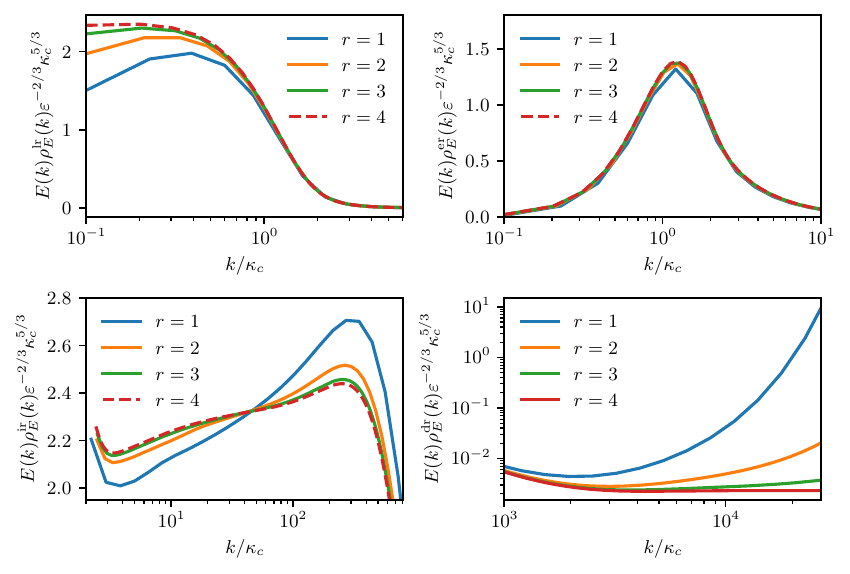}}
\caption{Numerical energy spectra for the partitions $\Pi_k^r$ and $\Pi_\tau^r$, computed for $k_d=6400\kappa_c$ and the work spectrum $W_1(k)$. In the upper left panel, the energy spectrum is weighted with $\rho_E^{\mathrm{lr}}(k)$, in the upper right panel with $\rho_E^{\mathrm{er}}(k)$, in the lower left panel with $\rho_E^{\mathrm{ir}}(k)$, and in the lower right panel with $\rho_E^{\mathrm{dr}}(k)$. The weighting functions are listed in equations \eqref{equ:rho_E_lr}---\eqref{equ:rho_E_dr}.}
\label{fig:Err_Ek}
\end{figure}

In the upper left panel of Fig.~\ref{fig:Err_Ek}, the energy spectra are weighted with $\rho_E^{\mathrm{lr}}(k)$ in order to resolve the numerical errors in the large-scale range. The curves converge rapidly as $r$ increases. The upper right panel shows the energy-containing range of the spectra. For $r\geq 2$, the curves already lie very close to each other. The lower left panel shows the compensated spectra in the inertial range. Here, too, the curves converge rapidly with increasing $r$. The maximum deviation between the curves for $r=3$ and $r=4$ is less than $0.021$. The lower right panel shows the spectra weighted with $\rho_E^{\mathrm{dr}}(k)$ in order to resolve the numerical errors in the dissipation range. The curves for $r=1$ and $r=2$ differ by up to approximately three orders of magnitude, indicating a substantial multiplicative discrepancy. However, since the spectra in the dissipation range $0.2 \kappa_d \lesssim k \lesssim 4k_d$ decrease by roughly $41$ decades, the corresponding relative error remains small. Moreover, the curves again converge rapidly as $r$ increases.

\begin{figure}
\centerline{\includegraphics{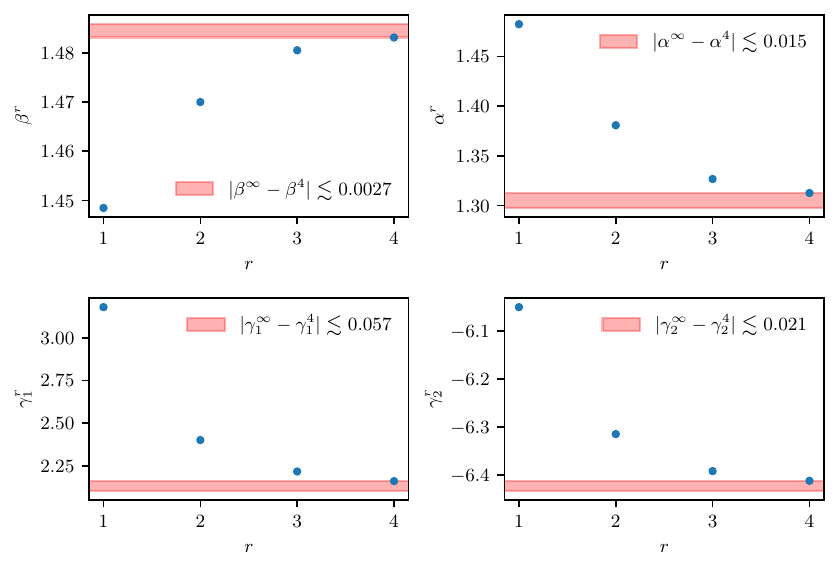}}
\caption{Numerical constants $\alpha^r$, $\beta^r$, $\gamma_1^r$ and $\gamma_2^r$, defined in equations \eqref{equ:Ek_powerlaw_ir} and \eqref{equ:DR_Modelfunction_Ek}, as functions of the refinement index $r$, which determines the resolution of the partitions. The coarse error estimate in equation \eqref{equ:Err_varphi} is shown as red bars in the figures.}
\label{fig:Err_Constants}
\end{figure}

For a more detailed assessment of the numerical errors in the inertial range, it is useful to examine the dependence of the prefactor $\alpha$ and the exponent $\beta$ in equation \eqref{equ:Ek_powerlaw_ir} on the partitions $\Pi_k^r$ and $\Pi_\tau^r$. The upper left and right panels of Fig.~\ref{fig:Err_Constants} show the corresponding values $\alpha^r$ and $\beta^r$ as blue dots. The estimated parameters approach limiting values as $r$ increases. An analogous analysis can be performed in the dissipation range by considering the dependence of the constants $\gamma_1$ and $\gamma_2$ in equation \eqref{equ:DR_Modelfunction_Ek} on the partitions $\Pi_k^r$ and $\Pi_\tau^r$. The results are shown in the lower left and right panels of Fig.~\ref{fig:Err_Constants}. Here as well, a clear saturation with increasing $r$ is observed.

To estimate the remaining error at $r=4$, we compare the computed constants with their limiting values as $r\to\infty$. We use the following contraction argument. For a scalar sequence $(\varphi^r)_{r\in \mathbb{N}}$, the property 
\begin{equation}
\frac{|\varphi^{r+2}-\varphi^{r+1}|}{|\varphi^{r+1}-\varphi^{r}|}=\eta^r_\varphi \label{def:eta_varphi_r}
\end{equation}
is assumed to hold, with $\eta^{\mathrm{max}}_\varphi:=\sup_{r\geq1}\eta_\varphi^r<1$. Then, for the sequence  $(\varphi^r)_{r\in \mathbb{N}}$, the error estimate 
\begin{equation}
|\varphi^{\infty}-\varphi^r| \leq \frac{\eta^{\mathrm{max}}_\varphi}{1-\eta^{\mathrm{max}}_\varphi}|\varphi^{r}-\varphi^{r-1}| \label{equ:Err_varphi}
\end{equation}
can be obtained using the geometric series.

For each choice $\varphi \in \{\alpha,\beta,\gamma_1,\gamma_2\}$, the contraction factors $\eta^r_\varphi$ are shown in table \ref{tab:Err_Constants}. All computed contraction factors are approximately $0.5$ or smaller. If we further assume that $\eta^r_\varphi \lesssim 0.5$ holds for all $r\geq 1$, then the distance between $\varphi^\infty$ and $\varphi^4$ can be roughly estimated by $\vert \varphi^\infty- \varphi^4\vert \lesssim  \vert \varphi^4- \varphi^3\vert$. The rough estimates for each choice $\varphi \in \{\alpha,\beta,\gamma_1,\gamma_2\}$ are shown in Fig.~\ref{fig:Err_Constants} as red horizontal bars. The actual numerical error is probably smaller than the rough estimate indicates.

\begin{table}
  \centering
  \caption{Values of the contraction factors $\eta_\varphi^r$, defined in equation \eqref{def:eta_varphi_r}, for the choice $\varphi \in \{\alpha,\beta,\gamma_1,\gamma_2\}$.}
  \label{tab:Err_Constants}
  \begin{tabular}{ccccc}
    \hline\hline
      r  & $\eta_\beta^r$   &   $\eta_\alpha^r$ &  $\eta_{\gamma_1}^r$ &  $\eta_{\gamma_2}^r$ \\ \hline
       $1$   &  $0.49$ & $0.54$ & $0.24$  & $0.30$  \\
       $2$   &  $0.25$ & $0.27$ & $0.31$  & $0.27$  \\
    \hline\hline
  \end{tabular}
\end{table}

\begin{figure}
\centerline{\includegraphics{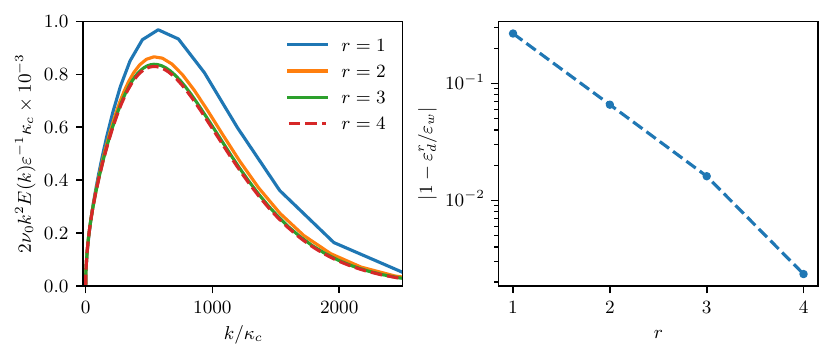}}
\caption{The left panel shows dissipation spectra $2\nu_0 k^2E(k)$ for the partitions $\Pi_k^r$ and $\Pi_\tau^r$. The right panel shows the numerical error $\vert 1-\varepsilon_d^r/\varepsilon_w\vert$ measuring the deviation from the stationary energy balance $\varepsilon_d=\varepsilon_w$.}
\label{fig:Err_epsilon}
\end{figure}

A very sensitive observable for numerical errors is the dissipation rate $\varepsilon_d$, which, according to equation \eqref{def:epsilon_d}, is determined by the integral of the dissipation spectrum $2\nu_0 k^2E(k)$ over all wavenumbers $k\in [0,\infty)$. The left panel of Fig.~\ref{fig:Err_epsilon} shows the dissipation spectra $2\nu_0 k^2E(k)$ for all $r$, where the curves for increasing $r$ also converge quite quickly. In the following, we denote by $\varepsilon_d^r$ the dissipation rate that was determined numerically with the partitions $\Pi_k^r$ and $\Pi_\tau^r$. Due to energy conservation, $\varepsilon_d=\varepsilon_w$ must hold in the stationary case, where $\varepsilon_w$ is the mean work rate defined in equation \eqref{def:epsilon_w}. A good measure of the numerical error of the dissipation spectrum is therefore $\vert 1-\varepsilon_d^r/\varepsilon_w\vert$, which is shown in the right panel of Fig.~\ref{fig:Err_epsilon}. The error $\vert 1-\varepsilon_d^r/\varepsilon_w\vert$ decreases at least exponentially, and at the finest resolution, $r=4$, it is $\vert 1-\varepsilon_d^4/\varepsilon_w\vert=0.00234$.

The characteristic decay times $\xi_c(k)$ and $\xi_g(k)$ provide sensitive measures of the temporal decay of $c(k,t)$ and $g(k,t)$. In order to determine the error of the time-dependent correlation functions for the partitions $\Pi_k^r$ and $\Pi_\tau^r$, we therefore discuss the characteristic decay times $\xi_c(k)$ and $\xi_g(k)$. Since the characteristic decay times have a strong wavenumber dependence, we weight $\xi_c(k)$ with the function 
\begin{equation}
\rho_c(k)=
\begin{cases}
k^2/(2\kappa_c)^2 & k\in [0,2\kappa_c) \\
k/(2\kappa_c) & k\in [2\kappa_c,\kappa_d/5) \\
\frac{\sqrt{k\kappa_d/5}}{2\kappa_c} & k\in [\kappa_d/5,\infty)
\end{cases} \label{def:rho_c}
\end{equation}
and $\xi_g(k)$ with the function
\begin{equation}
\rho_g(k)=
\begin{cases}
k^2/(2\kappa_c)^2 & k\in [0,2\kappa_c) \\
k/(2\kappa_c) & k\in [2\kappa_c,\infty) 
\end{cases} \label{def:rho_g}
\end{equation}
where, similar to the energy spectrum, $\kappa_d=1588\kappa_c$ is used for $r=4$.

\begin{figure}
\centerline{\includegraphics{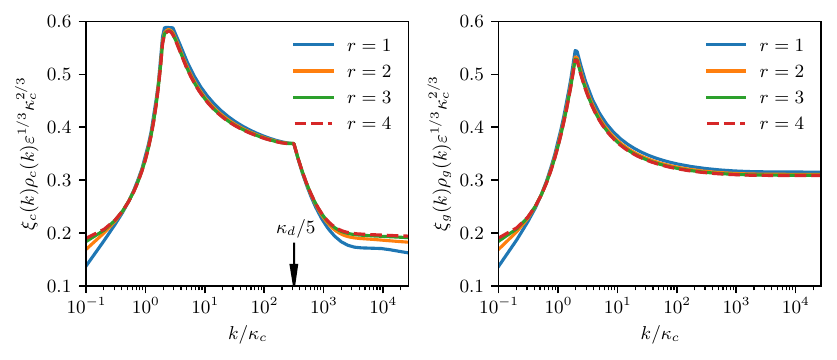}}
\caption{The left panel shows the characteristic decay time $\xi_c(k)$ for the partitions $\Pi_k^r$ and $\Pi_\tau^r$ weighted with the function $\rho_c(k)$, defined in equation \eqref{def:rho_c}. The right panel shows the characteristic decay time $\xi_g(k)$ weighted with the function $\rho_g(k)$, given in equation \eqref{def:rho_g}.} 
\label{fig:Err_xic_xig}
\end{figure}

In the left panel of Fig.~\ref{fig:Err_xic_xig}, the characteristic decay time  $\xi_c(k)$ is shown weighted by $\rho_c(k)$ for the partitions $\Pi_k^r$ and $\Pi_\tau^r$. For $r\geq2$, the curves nearly coincide over most of the wavenumber range. Only for $k\lesssim 0.4\kappa_c$ and $k\gtrsim \kappa_d$ are there still moderate deviations, but these decrease rapidly as $r$ increases. The characteristic decay time $\xi_g(k)$ shows similar convergence. For $r\geq2$, moderate deviations remain only at $k\lesssim 0.4\kappa_c$.

As an additional check, we refined the temporal partition $\Pi_\tau$ while keeping the wavenumber partition $\Pi_k$ fixed, and vice versa. Both tests showed consistent convergence, supporting the results obtained by refining the two partitions simultaneously. These supplementary results are not shown here.

\section{Local and nonlocal contributions to the FDR violation}
\label{app:InteractionsFDRviolation}

We rewrite Eq.~\eqref{equ:explicit_integrated_fdr_violation} for the FDR violation $\varphi(k)$ by transforming from $(p,\mu)$ to the bipolar coordinates $(p,q)$. This yields
\begin{equation}
\varphi(k)
=\int_0^\infty\int_{\lvert k-p\rvert}^{k+p}
R(k,p,q)\dd q\dd p,
\end{equation}
where the contribution from an individual triad is given by
\begin{equation}
\begin{split}
R(k,p,q)
={}&\frac{1}{2}V_q(k,p,q)\frac{q}{kp}E(p)
\int_0^\infty\int_0^\infty c(p,t+s)
\\
&\times\left[
\frac{k^2E(q)}{q^2E(k)}c(q,t+s)g(k,s)
-g(q,t+s)c(k,s)
\right]\dd s\dd t.
\end{split}
\label{equ:FDR_violation_Rkpq}
\end{equation}
Here, $V_q(k,p,q):=V(k,p,\mu(k,p,q))$, with $\mu(k,p,q)=(k^2+p^2-q^2)/(2kp)$.

\begin{figure}
\centerline{\includegraphics{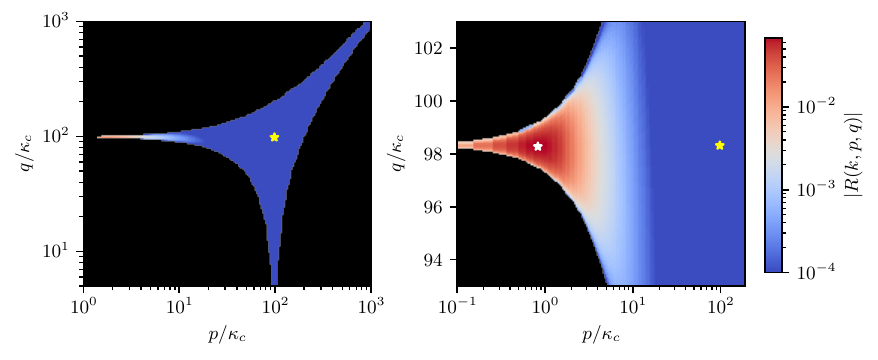}}
\caption{Contribution of individual triads to the integrated FDR violation for the work spectrum $W_1(k)$, with $k_d=6400\kappa_c$, evaluated at the fixed inertial-range wavenumber $k=98.3\kappa_c$. The left panel shows $\lvert R(k,p,q)\rvert$ over the admissible triad domain $\lvert k-p\rvert\leq q\leq k+p$ and a black background outside this region. The yellow star indicates the local triad $p=q=k$. The right panel provides a magnified view of the dominant nonlocal region, $p\sim\kappa_c$ and $q\sim k$. The white star marks the location of the maximum, where $R(k,p,q)\simeq0.069/\kappa_c^2$.}
\label{fig:Triads}
\end{figure}

The function $R(k,p,q)$ allows us to identify the triads that contribute most strongly to $\varphi(k)$. The left panel of Fig.~\ref{fig:Triads} shows $\lvert R(k,p,q)\rvert$ for the work spectrum $W_1(k)$, with dissipation wavenumber $k_d=6400\kappa_c$, at the fixed wavenumber $k= 98.3\kappa_c$ in the inertial range. The yellow star indicates the local triad $k=p=q$. Appreciable contributions are confined almost entirely to the nonlocal region $p\sim\kappa_c$ and $q\sim k$; throughout the remainder of the admissible $(p,q)$ domain, the contributions are negligibly small. We observe the same behaviour for all wavenumbers examined in the inertial and dissipation ranges. The right panel provides a magnified view of the dominant region $p\sim\kappa_c$ and $q\sim k$. The white star indicates $\max_{p,q}R(k,p,q)\simeq 0.069/\kappa_c^2$. 

\section{Comparison of the stationary DIA and DNS}
\label{app:forcing_DNS}

\subsection{Gaussian mean-field approximation for the negative-damping forcing}

For the DIA calculation, we make the replacement $L_c\to L$ in the definition of $\mathcal{E}_B(t)$ in Eq.~\eqref{def:Etot_in_band_b}, since we later consider the continuum limit $L\to\infty$. Introducing the fluctuation $\Delta\mathcal{E}_B(t)=\mathcal{E}_B(t)-\EV{\mathcal{E}_B(t)}$, we expand the field-dependent coefficient $\gamma[U](t)$ about the mean energy in the forced modes
\begin{equation}
\begin{split}
\gamma[U](t)=\frac{\varepsilon}{2\mathcal{E}_B(t)}&=\frac{\varepsilon}{2}\frac{1}{\EV{\mathcal{E}_B(t)}+\Delta\mathcal{E}_B(t)}\\
&\approx\frac{\varepsilon}{2\EV{\mathcal{E}_B(t)}}-\frac{\varepsilon}{2\EV{\mathcal{E}_B(t)}^2}\Delta\mathcal{E}_B(t)+O\left(\Delta\mathcal{E}_B(t)^2\right).
\end{split}
\end{equation}

We show below that, within the simplest nontrivial Gaussian approximation introduced in Sec.~\ref{sec:mean_field}, the zeroth-order term provides the leading contribution, whereas the contribution generated by the linear fluctuation term vanishes relative to it as $L\to\infty$. Adding the deterministic forcing term $\gamma[U](t)\chi_B(k)U^\alpha_t(\bm{k})$ to the stochastic Navier--Stokes equation \eqref{equ:NavierStokesEq_stochastic} modifies the quadratic part $S_{\mathrm{lam}}$ of the action in Eq.~\eqref{equ:Slam} through the formal replacement $\nu_0k^2\to\nu_0k^2-\gamma[U](\tau)\chi_B(k)$. The exact moment equations \eqref{equ:C2_C3_ts}, \eqref{equ:G2_G3_ts}, and \eqref{equ:C2_C3_tt} consequently become
\begin{equation}
\begin{split}
\left(\pd{t}+\nu_0 k^2 \right) \EV{U_t^\alpha(\bm{k}) \Ov{U}_s^\beta(\bm{p})} = \sum_{\bm{q}\in\mathbb{G}} M_{\alpha\gamma\rho}(\bm{k}) \EV{U_t^\gamma(\bm{k}-\bm{q}) U_t^\rho(\bm{q}) \Ov{U}_s^\beta(\bm{p})}\frac{1}{L^3}+\chi_B(k)\EV{\gamma[U](t)U_t^\alpha(\bm{k}) \Ov{U}_s^\beta(\bm{p})}.
\end{split}
\label{equ:C2_C3_ts_DNS}
\end{equation}
Similarly,
\begin{equation}
\begin{split}
\left(\pd{t}+\nu_0 k^2\right)\EV{U_t^\alpha(\bm{k})\Ov{\Psi}_s^\beta(\bm{p})}=\sum_{\bm{q}\in\mathbb{G}} M_{\alpha\gamma\rho}(\bm{k})\EV{U_t^\gamma(\bm{k}-\bm{q})U_t^\rho(\bm{q})\Ov{\Psi}_s^\beta(\bm{p})}\frac{1}{L^3}+\chi_B(k)\EV{\gamma[U](t)U_t^\alpha(\bm{k}) \Ov{\Psi}_s^\beta(\bm{p})},
\end{split}
\label{equ:G2_G3_ts_DNS}
\end{equation}
and
\begin{equation}
\begin{split}
\left(\pd{t}+\nu_0k^2+\nu_0p^2\right)\EV{U_t^\alpha(\bm{k})\Ov{U}_t^\beta(\bm{p})}
&=F(\Vert\bm{k}\Vert)L^3P_{\alpha\beta}(\bm{k})\delta_{\bm{k},\bm{p}}+(\chi_B(k)+\chi_B(p))\EV{\gamma[U](t)U_t^\alpha(\bm{k})\Ov{U}_t^\beta(\bm{p})}
\\
&\qquad+\sum_{\bm{q}\in\mathbb{G}}M_{\alpha\gamma\rho}(\bm{k})\EV{U_t^\gamma(\bm{k}-\bm{q})U_t^\rho(\bm{q})\Ov{U}_t^\beta(\bm{p})}\frac{1}{L^3}
\\
&\qquad+\sum_{\bm{q}\in\mathbb{G}}\Ov{M}_{\beta\gamma\rho}(\bm{p})\EV{U_t^\alpha(\bm{k})\Ov{U}_t^\gamma(\bm{p}-\bm{q})\Ov{U}_t^\rho(\bm{q})}\frac{1}{L^3}.
\end{split}
\label{equ:C2_C3_tt_DNS}
\end{equation}

Using the expansion of $\gamma[U](t)$, the additional negative-damping contribution in Eq.~\eqref{equ:C2_C3_ts_DNS} is approximated by
\begin{equation}
\begin{split}
\EV{\gamma[U](t)U_t^\alpha(\bm{k})\Ov{U}_s^\beta(\bm{p})}\approx\frac{\varepsilon}{2\EV{\mathcal{E}_B(t)}}\EV{U_t^\alpha(\bm{k})\Ov{U}_s^\beta(\bm{p})}-\frac{\varepsilon}{2\EV{\mathcal{E}_B(t)}^2}\EV{U_t^\alpha(\bm{k})\Ov{U}_s^\beta(\bm{p})\Delta\mathcal{E}_B(t)}.
\end{split}
\label{equ:approx_forcing_DNS}
\end{equation}
Applying the Gaussian mean-field approximation from Sec.~\ref{sec:mean_field} to the second term gives
\begin{equation}
\frac{\varepsilon}{2\EV{\mathcal{E}_B(t)}^2}\EV{U_t^\alpha(\bm{k})\Ov{U}_s^\beta(\bm{p})\Delta\mathcal{E}_B(t)}\approx \frac{\varepsilon}{2\EV{\mathcal{E}_B(t)}^2}\sum_{\bm{q}\in\mathbb{G}}\chi_B(q)\EV{U_t^\mu(\bm{q})\Ov{U}_s^\beta(\bm{p})}\EV{U_t^\alpha(\bm{k})\Ov{U}_t^\mu(\bm{q})}\frac{1}{L^6}.
\end{equation}
Indeed, Wick's theorem yields
\begin{equation}
\begin{split}
\EV{U_t^\alpha(\bm{k})\Ov{U}_s^\beta(\bm{p})\mathcal{E}_B(t)}_{\mathrm{eff}}=\EV{U_t^\alpha(\bm{k})\Ov{U}_s^\beta(\bm{p})}_{\mathrm{eff}}\EV{\mathcal{E}_B(t)}_{\mathrm{eff}}+\sum_{\bm{q}\in\mathbb{G}}\chi_B(q)\EV{U_t^\mu(\bm{q})\Ov{U}_s^\beta(\bm{p})}_{\mathrm{eff}}\EV{U_t^\alpha(\bm{k})\Ov{U}_t^\mu(\bm{q})}_{\mathrm{eff}}\frac{1}{L^6}.
\end{split}
\end{equation}

According to Eq.~\eqref{equ:isotropic_C}, a homogeneous two-point function at $\bm{k}=\bm{p}$ scales as $\EV{U\Ov{U}}\sim O(L^3)$. The leading term in Eq.~\eqref{equ:approx_forcing_DNS} therefore scales as $O(L^3)$. In the connected correction, homogeneity constrains the wavevectors in both two-point functions and collapses the sum over $\bm{q}$. The two factors of $L^3$ are then cancelled by the explicit factor $L^{-6}$, leaving an $O(L^0)$ contribution. The correction is consequently suppressed by $O(L^{-3})$ relative to the leading term.

The same counting applies to Eqs.~\eqref{equ:G2_G3_ts_DNS} and \eqref{equ:C2_C3_tt_DNS}, with the formal replacements $\Ov{U}_s^\beta(\bm{p})\to\Ov{\Psi}_s^\beta(\bm{p})$ and $s\to t$, respectively. Higher-order fluctuation terms are likewise suppressed by self-averaging in the Gaussian approximation. Hence, in the limit $L\to\infty$, the negative-damping coefficient reduces at leading mean-field order to $\gamma_{\mathrm{DIA}}(t)=\varepsilon/(2\EV{\mathcal{E}_B(t)}_{\mathrm{eff}})$, since $\EV{\mathcal{E}_B(t)}\approx \EV{\mathcal{E}_B(t)}_{\mathrm{eff}}$. For a stationary solution, $\EV{\mathcal{E}_B(t)}_{\mathrm{eff}}$ is independent of time, and $\gamma_{\mathrm{DIA}}(t)$ reduces to the constant coefficient $\gamma_{\mathrm{DIA}}$ defined in Eq.~\eqref{equ:gamma_dia}.

In the DNS considered here, randomly initialised velocity fields evolve deterministically according to the forced Navier--Stokes equations, and observables are averaged over the resulting ensemble. In our MSRDJ formulation, the velocity field instead starts from the fixed initial condition $U_{t_0}=0$ and evolves under stochastic forcing with spectrum $F(k)$, as specified in Eq.~\eqref{equ:NavierStokesEq_stochastic}. The expectation $\EV{\cdot}$ therefore averages over realisations of the stochastic dynamics. To compare these expectations with the DNS averages $\langle\cdot\rangle_{\mathrm{DNS}}$, we use the following procedure:
\begin{itemize}
\item[(i)] At the initial time $t_0$, we impose the initial condition $U_{t_0}=0$, as in the stochastic Navier--Stokes equation \eqref{equ:NavierStokesEq_stochastic}, and apply an additive stochastic forcing with spectrum $F(k)$ supported in the neighbourhood of $\kappa_c$, as in Sec.~\ref{sec:NumericalSolution}.
\item[(ii)] Once the stochastic forcing has generated a non-vanishing velocity field, we adiabatically switch on the deterministic negative-damping force $\chi_B(k)\gamma_{\mathrm{DIA}}(t)U_t^\alpha(\bm{k})$ at a time $t_1>t_0$, with $\gamma_{\mathrm{DIA}}(t)=\varepsilon/(2\EV{\mathcal{E}_B(t)}_{\mathrm{eff}})$. Introducing $t_1$ avoids the singularity of the negative-damping coefficient at the initial condition $U_{t_0}=0$.
\item[(iii)] At a time $t_0^{*}>t_1$ sufficiently far from the initial time, we adiabatically switch off the additive stochastic forcing by taking $F(k)\to0$. The stationary correlation and Green functions are subsequently evaluated at times $t,s\gg t_0^{*}$.
\end{itemize}

\subsection{Numerical parameter choice}

Since $\chi_B(k)$ is discontinuous at $k=\kappa_c$ and $k=2.5\kappa_c$,  the wavenumber grid $\Pi_k$ must be sufficiently fine near the forced band. For the work spectrum \eqref{equ:dia_negative_damping_Wk}, we therefore use the geometrically spaced grid points
\begin{equation}
k_m=k_{\mathrm{min}}b^m,\qquad m\in\{0,1,2,\ldots,\mathsf{N}_k\}.
\label{equ:Pik_DNS_DIA}
\end{equation}
For this work spectrum, convergence tests show that the cutoffs $k_{\mathrm{min}}=0.5\kappa_c$ and $\mathsf{N}_k=\max\{m:k_d >k_m\}$ are sufficient. The partition $\Pi_\mu$ is chosen as in Sec.~\ref{app:NumericalDiscretisation}. The estimate of $\mathsf{N}_\mu$ in Eq.~\eqref{equ:estimate_Nmu} also applies to the geometrically spaced wavenumber grid \eqref{equ:Pik_DNS_DIA}. The DIA equations incorporating the replacement $\nu_0k^2\to\nu_0k^2-\gamma_{\mathrm{DIA}}\chi_B(k)$ are then discretised and solved as described in Sec.~\ref{app:NumericalDiscretisation}. For the wavenumber partition \eqref{equ:Pik_DNS_DIA}, we performed a numerical error analysis analogous to that presented in Sec.~\ref{sec:Numerical_errors}. The numerical parameters are listed in Table~\ref{tab:NumericalParameters_DNS_DIA}.

\begin{table}
  \centering
\caption{Numerical parameters for the partitions $\Pi_k$, $\Pi_\mu$, and $\Pi_\tau$ used in the comparison between DNS and DIA. The number of wavenumber points $\mathsf{N}_k$ depends on $k_d$ through $k_{\mathrm{max}}$.}
\label{tab:NumericalParameters_DNS_DIA}
  \begin{tabular}{cccccc}
    \hline\hline
$k_{\mathrm{min}}$ & $k_{\mathrm{max}}$ & $b$ & $\mathsf{N}_\mu$ & $\mathsf{N}_t$ & $\Delta\tau$ \\ \hline
$0.5\kappa_c$ & $k_d$ & $1.01454$ & $220$ & $600$ & $0.02$ \\
    \hline\hline
  \end{tabular}
\end{table}

\subsection{Shell-averaged energy spectrum and time-dependent functions}
\label{app:dns_shell-averaging}

For $n\in\mathbb{N}$, the stationary and isotropic DNS energy spectrum, correlation function, and Green function are defined on the discrete wavenumber shells
\begin{equation}
\mathbb I_n=\left\{\bm{k}\in\mathbb G_c:k_n\leq\lVert\bm{k}\rVert<k_{n+1}\right\},
\qquad
\mathbb G_c=\kappa_c\mathbb Z^3\setminus\{\bm{0}\},
\qquad
k_n=n\kappa_c.
\end{equation}
In our Fourier-space convention, the stationary isotropic energy spectrum is given by
\begin{equation}
E^{\mathrm{DNS}}_n=\frac{1}{\kappa_c}\sum_{\bm{k}\in\mathbb I_n}\frac{1}{2}
\left\langle u^\alpha_t(\bm{k})\Ov{u}^\alpha_t(\bm{k})\right\rangle_{\mathrm{DNS}}\frac{1}{L_c^6}.
\label{eq:dns_shell_spectrum}
\end{equation}
To compare the continuous DIA solution with the shell-averaged DNS data, we associate each discrete shell $\mathbb I_n$ with the continuous shell
\begin{equation}
\mathcal I_n=\left\{\bm{k}\in\mathbb R^3:k_n\leq\lVert\bm{k}\rVert<k_{n+1}\right\}.
\end{equation}
Stationarity and isotropy then reduce the corresponding shell averages of the DIA energy spectrum, correlation function, and response function to 
\begin{align}
E^{\mathrm{DIA}}_n&= \frac{1}{\kappa_c} \int_{\mathcal I_n} \frac{1}{2}C_{\mathrm{DIA}}^{\alpha\alpha} (\bm{k},t,t)\,\frac{\dd \bm{k}}{(2\pi)^3}  = \frac{1}{\kappa_c} \int_{k_n}^{k_{n+1}} E_{\mathrm{DIA}}(k)\,\dd k, \label{eq:dia_shell_spectrum}\\
c^{\mathrm{DIA}}_n(t) &= \frac{ \displaystyle \int_{\mathcal I_n} C_{\mathrm{DIA}}^{\alpha\alpha} (\bm{k},t+s,s)\,\dd \bm{k}}{\displaystyle \int_{\mathcal I_n} C_{\mathrm{DIA}}^{\alpha\alpha} (\bm{k},s,s)\,\dd \bm{k}} = \frac{ \displaystyle \int_{k_n}^{k_{n+1}} E_{\mathrm{DIA}}(k)c_{\mathrm{DIA}}(k,t)\,\dd k}{\displaystyle \int_{k_n}^{k_{n+1}} E_{\mathrm{DIA}}(k)\,\dd k },\label{eq:dia_shell_correlation}
\\
g^{\mathrm{DIA}}_n(t) &= \frac{ \displaystyle \int_{\mathcal I_n} G_{\mathrm{DIA}}^{\alpha\alpha} (\bm{k},t+s,s)\,\dd \bm{k} }{\displaystyle \int_{\mathcal I_n} G_{\mathrm{DIA}}^{\alpha\alpha}(\bm{k},s,s)\,\dd \bm{k}}= \frac{\displaystyle \int_{k_n}^{k_{n+1}} k^2g_{\mathrm{DIA}}(k,t)\,\dd k}{\displaystyle \int_{k_n}^{k_{n+1}} k^2\,\dd k}, \label{eq:dia_shell_response}
\end{align}
where $C^{\alpha\beta}(\mathbf k,t,s):=P_{\alpha\beta}(\mathbf k)C(k,t,s)$ and $G^{\alpha\beta}(\mathbf k,t,s):=P_{\alpha\beta}(\mathbf k)G(k,t,s).$



\bibliography{apssamp}

@PREAMBLE{
 "\providecommand{\noopsort}[1]{}" 
 # "\providecommand{\singleletter}[1]{#1}%" 
}

@article{Reynolds1895,
  author  = {Reynolds, O.},
  title   = {On the dynamical theory of incompressible viscous fluids and the determination of the criterion},
  journal = {Philos. Trans. R. Soc. Lond. A},
  volume  = {186},
  pages   = {123--164},
  year    = {1895}
}

@article{Guioth2022,
  title={Path large deviations for the kinetic theory of weak turbulence},
  author={Guioth, J. and Bouchet, F. and Eyink, G. L.},
  journal={J. Stat. Phys.},
  volume={189},
  number={2},
  pages={20},
  year={2022},
  publisher={Springer}
}

@article{Onuki2024,
  title={Dynamical Large Deviations for an Inhomogeneous Wave Kinetic Theory: {Linear} Wave Scattering by a Random Medium},
  author={Onuki, Y. and Guioth, J. and Bouchet, F.},
  journal={Ann. Henri Poincaré},
  volume={25},
  number={1},
  pages={1215--1259},
  year={2024}
}

@article{Tomassini1997,
title = {An exact renormalization group analysis of 3D well developed turbulence},
journal = {Phys. Lett. B},
volume = {411},
number = {1},
pages = {117-126},
year = {1997},
author = {Tomassini, P.},
}

@article{Pomeau1986,
  title={Front motion, metastability and subcritical bifurcations in hydrodynamics},
  author={Pomeau, Y.},
  journal={Physica D},
  volume={23},
  number={1-3},
  pages={3--11},
  year={1986},
  publisher={Elsevier}
}

@article{Hof2023,
  title={Directed percolation and the transition to turbulence},
  author={Hof, B.},
  journal={Nat. Rev. Phys.},
  volume={5},
  number={1},
  pages={62--72},
  year={2023},
  publisher={Nature Publishing Group UK London}
}

@article{Cugliandolo1996,
  title = {Large Time Out-of-Equilibrium Dynamics of a Manifold in a Random Potential},
  author = {Cugliandolo, L. F. and Kurchan, J. and Le Doussal, P.},
  journal = {Phys. Rev. Lett.},
  volume = {76},
  issue = {13},
  pages = {2390--2393},
  numpages = {0},
  year = {1996},
  publisher = {American Physical Society}
}

@article{Kim2001,
  title={The dynamics of the spherical p-spin model: {From} microscopic to asymptotic},
  author={Kim, B. and Latz, A.},
  journal={EPL},
  volume={53},
  number={5},
  pages={660--666},
  year={2001}
}

@article{hinrichsen2000,
  title={Non-equilibrium critical phenomena and phase transitions into absorbing states},
  author={Hinrichsen, H.},
  journal={Adv. Phys.},
  volume={49},
  number={7},
  pages={815--958},
  year={2000},
  publisher={Taylor \& Francis}
}

@article{Koniges1987,
  title={Statistical closure approximations and the fluctuation-dissipation theorem for drift-wave interaction},
  author={Koniges, A. E. and Leith, C. E.},
  journal={Phys. Fluids},
  volume={30},
  number={10},
  pages={3065--3074},
  year={1987},
  publisher={American Institute of Physics}
}

@article{Edwards1964,
  title={The statistical dynamics of homogeneous turbulence},
  volume={18}, 
  number={2}, 
  journal={J. Fluid Mech.}, 
  author={Edwards, S. F.},
  year={1964}, 
  pages={239–273}
}

@article{Bandak2022,
  title = {Dissipation-range fluid turbulence and thermal noise},
  author = {Bandak, D. and Goldenfeld, N. and Mailybaev, A. A. and Eyink, G.},
  journal = {Phys. Rev. E},
  volume = {105},
  issue = {6},
  pages = {065113},
  numpages = {33},
  year = {2022},
  month = {Jun},
  publisher = {American Physical Society},
  doi = {10.1103/PhysRevE.105.065113}
}

@article{Martin1973,
  title = {Statistical Dynamics of Classical Systems},
  author = {Martin, P. C. and Siggia, E. D. and Rose, H. A.},
  journal = {Phys. Rev. A},
  volume = {8},
  issue = {1},
  pages = {423--437},
  numpages = {0},
  year = {1973},
  month = {Jul},
  publisher = {American Physical Society},
  doi = {10.1103/PhysRevA.8.423},
  url = {https://link.aps.org/doi/10.1103/PhysRevA.8.423}
}

@article{Janssen1976,
  title={On a {Lagrangean} for classical field dynamics and renormalization group calculations of dynamical critical properties},
  author={Janssen, Hans-Karl},
  journal={Z. Phys. B},
  volume={23},
  number={4},
  pages={377--380},
  year={1976},
  publisher={Springer}
}

@article{Dominicis1976,
  title={Techniques de renormalisation de la th{\'e}orie des champs et dynamique des ph{\'e}nomenes critiques},
  author={Dominicis, C De},
  journal={J. Phys. (Paris) Colloq.},
  volume={37},
  number={C1},
  pages={247--253},
  year={1976}
}

@book{Helias2020,
  title={{Statistical Field Theory for Neural Networks}},
  author={Helias, Moritz and Dahmen, David},
  volume={970},
  year={2020},
  publisher={Springer}
}

@book{Kamenev2011,
  place={Cambridge}, 
  title={{Field Theory of Non-Equilibrium Systems}},
  publisher={Cambridge University Press},
  author={Kamenev, Alex},
  year={2011}
}

@article{kraichnan1959,
  title={The structure of isotropic turbulence at very high {Reynolds} numbers},
  author={Kraichnan, Robert H},
  journal={J. Fluid Mech.},
  volume={5},
  number={4},
  pages={497--543},
  year={1959},
  publisher={Cambridge University Press}
}

@article{Karman1938,
    author = {de Karman, Theodore and Howarth, Leslie},
    title = {On the Statistical Theory of Isotropic Turbulence},
    journal = {Philos. Trans. R. Soc. Lond. A.},
    volume = {164},
    number = {917},
    pages = {192-215},
    year = {1938},
    month = {01}
}

@book{mccomb2014,
  title={{Homogeneous, Isotropic Turbulence: Phenomenology, Renormalization and Statistical Closures}},
  author={McComb, W David},
  volume={162},
  year={2014},
  publisher={Oxford University Press}
}

@book{mccomb1990,
    author = {McComb, W. D.},
    title = {{The Physics of Fluid Turbulence}},
    publisher = {Oxford University Press},
    year = {1990}
}

@article{kaneda2007,
  title={Lagrangian renormalized approximation of turbulence},
  author={Kaneda, Yukio},
  journal={Fluid Dyn. Res.},
  volume={39},
  number={7},
  pages={526},
  year={2007},
  publisher={IOP Publishing}
}

@article{Khurshid2018,
  title = {Energy spectrum in the dissipation range},
  author = {Khurshid, Sualeh and Donzis, Diego A. and Sreenivasan, K. R.},
  journal = {Phys. Rev. Fluids},
  volume = {3},
  issue = {8},
  pages = {082601},
  numpages = {10},
  year = {2018},
  month = {Aug},
  publisher = {American Physical Society},
}

@article{Cugliandolo1997,
  title = {Energy flow, partial equilibration, and effective temperatures in systems with slow dynamics},
  author = {Cugliandolo, Leticia F. and Kurchan, Jorge and Peliti, Luca},
  journal = {Phys. Rev. E},
  volume = {55},
  issue = {4},
  pages = {3898--3914},
  numpages = {0},
  year = {1997},
  month = {Apr},
  publisher = {American Physical Society},
}

@incollection{Cugliandolo2002,
  author    = {Cugliandolo, L. F.},
  title     = {Dynamics of Glassy Systems},
  booktitle = {Slow relaxations and nonequilibrium dynamics in condensed matter},
  editor    = {Barrat, J. L. and Feigelman, M. and Kurchan, J. and Dalibard, J.},
  series    = {Les Houches Session},
  number    = {LXXVII},
  publisher = {Springer},
  year      = {2003},
  pages     = {367--521}
}

@article{Sollich2002,
title = {Fluctuation-dissipation relations and effective
temperatures in simple non-mean field systems},
author = {Peter Sollich and Suzanne Fielding and Peter Mayer},
journal = {J. Phys. Condens. Matter},
volume = {14},
number = {7},
pages = {1683},
year = {2002},
month = {Feb},
publisher = {},
}

@article{Buhot2002,
  title = {Fluctuation-Dissipation Relations in the Activated Regime of Simple Strong-Glass Models},
  author = {Buhot, Arnaud and Garrahan, Juan P.},
  journal = {Phys. Rev. Lett.},
  volume = {88},
  issue = {22},
  pages = {225702},
  numpages = {4},
  year = {2002},
  month = {May},
  publisher = {American Physical Society},
}

@article{Umberto2008,
title = {Fluctuation–dissipation: {Response} theory in statistical physics},
author = {Umberto Marini Bettolo Marconi and Andrea Puglisi and Lamberto Rondoni and Angelo Vulpiani},
journal = {Phys. Rep.},
volume = {461},
number = {4},
pages = {111-195},
year = {2008},
}

@article{Kolmogorov1941,
  title={The local structure of turbulence in incompressible viscous fluid for very large {Reynolds} numbers},
  author={Kolmogorov, Andrey Nikolaevich},
  journal={Dokl. Akad. Nauk SSSR},
  volume={30},
  pages={301},
  year={1941}
}

@article{Kraichnan1965,
    author = {Kraichnan, Robert H.},
    title = {Lagrangian‐History Closure Approximation for Turbulence},
    journal = {Phys. Fluids},
    volume = {8},
    number = {4},
    pages = {575-598},
    year = {1965},
    month = {04},
}

@article{Kraichnan1966,
    author = {Kraichnan, Robert H.},
    title = {Isotropic Turbulence and Inertial‐Range Structure},
    journal = {Phys. Fluids},
    volume = {9},
    number = {9},
    pages = {1728-1752},
    year = {1966},
    month = {09},
}

@article{Kraichnan1971a,
  title={An {almost-Markovian} {Galilean-invariant} turbulence model},
  author={Kraichnan, Robert H.},
  journal={J. Fluid Mech.},
  volume={47},
  number={3},
  pages={513--524},
  year={1971},
  publisher={Cambridge University Press}
}

@article{Kraichnan1971b,
  title={Inertial-range transfer in two-and three-dimensional turbulence},
  author={Kraichnan, Robert H},
  journal={J. Fluid Mech.},
  volume={47},
  number={3},
  pages={525--535},
  year={1971},
  publisher={Cambridge University Press}
}

@article{Orszag1970,
  title={Analytical theories of turbulence},
  author={Orszag, Steven A},
  journal={J. Fluid Mech.},
  volume={41},
  number={2},
  pages={363--386},
  year={1970},
  publisher={Cambridge University Press}
}

@article{Canet2022,
  title={Functional renormalisation group for turbulence},
  author={Canet, L{\'e}onie},
  journal={J. Fluid Mech.},
  volume={950},
  pages={P1},
  year={2022},
  publisher={Cambridge University Press}
}

@article{Matsumoto2021, 
title={Correlation function and linear response function of homogeneous isotropic turbulence in the {Eulerian} and {Lagrangian} coordinates}, 
author={Matsumoto, Takeshi and Otsuki, Michio and Ooshida, Takeshi and Goto, Susumu},
journal={J. Fluid Mech.},
volume={919}, 
year={2021},
pages={A9}}

@article{Kraichnan1964,
  title={Kolmogorov's hypotheses and {Eulerian} turbulence theory},
  author={Kraichnan, Robert H},
  journal = {Phys. Fluids},
  volume = {7},
  number = {11},
  pages = {1723–1734},
  year = {1964},
  month = {11}
}

@article{Kraichnan1964steady,
  title={Approximations for Steady-State Isotropic Turbulence},
  author={Kraichnan, Robert H},
  journal={Phys. Fluids},
  volume={7},
  number={8},
  pages={1163--1168},
  year={1964},
  publisher={AIP Publishing}
}

@book{Leslie1973,
  title={{Developments in the Theory of Turbulence}},
  author={Leslie, David Clement},
  year={1973},
  place={Oxford}, 
  publisher={Clarendon Press},
}

@article{Kuchler2023,
  title={Universal velocity statistics in decaying turbulence},
  author={K{\"u}chler, Christian and Bewley, Gregory P and Bodenschatz, Eberhard},
  journal={Phys. Rev. Lett.},
  volume={131},
  number={2},
  pages={024001},
  year={2023},
  publisher={APS}
}

@article{MydlarskiWarhaft1996, 
author={Mydlarski, L. and Warhaft, Z.}, 
title={On the onset of {high-Reynolds-number} grid-generated wind tunnel turbulence}, 
volume={320}, 
journal={J. Fluid Mech.}, 
year={1996}, 
pages={331–368}}

@article{Ishihara2016,
  title={Energy spectrum in high-resolution direct numerical simulations of turbulence},
  author={Ishihara, Takashi and Morishita, Koji and Yokokawa, Mitsuo and Uno, Atsuya and Kaneda, Yukio},
  journal={Phys. Rev. Fluids},
  volume={1},
  number={8},
  pages={082403},
  year={2016},
  publisher={APS}
}

@article{ishihara2020,
  title={Second-order velocity structure functions in direct numerical simulations of turbulence with $\mathrm{{R}}_\lambda$ up to 2250},
  author={Ishihara, Takashi and Kaneda, Yukio and Morishita, Koji and Yokokawa, Mitsuo and Uno, Atsuya},
  journal={Phys. Rev. Fluids},
  volume={5},
  number={10},
  pages={104608},
  year={2020},
  publisher={APS}
}

@article{Kraichnan1964decay,
  title={Decay of isotropic turbulence in the direct-interaction approximation},
  author={Kraichnan, Robert H},
  journal={Phys. Fluids},
  volume={7},
  number={7},
  pages={1030--1048},
  year={1964},
  publisher={American Institute of Physics}
}

@article{Mccomb2017,
  title={A formal derivation of the local energy transfer {(LET)} theory of homogeneous turbulence},
  author={McComb, W. D. and Yoffe, S. R.},
  journal={J. Phys. A},
  volume={50},
  number={37},
  pages={375501},
  year={2017},
  publisher={IOP Publishing}
  
}

@article{Mccomb2003,
  title={Two-point, two-time closures applied to forced isotropic turbulence},
  author={McComb, W. D. and Quinn, A. P.},
  journal={Physica A},
  volume={317},
  number={3-4},
  pages={487--508},
  year={2003},
  publisher={Elsevier}
}

@article{Kida1997,
  title={A {Lagrangian} direct-interaction approximation for homogeneous isotropic turbulence},
  volume={345},
  journal={J. Fluid Mech.}, 
  author={Kida, Shigeo and Goto, Susumu}, 
  year={1997},
  pages={307–345}
}

@article{Okamura2018, 
  title={Closure model for homogeneous isotropic turbulence in the {Lagrangian} specification of the flow field},
  volume={841},
  journal={J. Fluid Mech.},
  author={Okamura, Makoto},
  year={2018}, 
  pages={521–551}
}

@article{Inagaki2021,
  title={Scale-similar structures of homogeneous isotropic non-mirror-symmetric turbulence based on the {Lagrangian} closure theory},
  author={Inagaki, Kazuhiro},
  journal={J. Fluid Mech.},
  volume={926},
  pages={A14},
  year={2021},
  publisher={Cambridge University Press}
}

@article{Kaneda1981,
  title={Renormalized expansions in the theory of turbulence with the use of the {Lagrangian} position function},
  author={Kaneda, Yukio},
  journal={J. Fluid Mech.},
  volume={107},
  pages={131--145},
  year={1981},
  publisher={Cambridge University Press}
}

@article{Gorbunova2020,
  title = {Analysis of the dissipative range of the energy spectrum in grid turbulence and in direct numerical simulations},
  author = {Gorbunova, Anastasiia and Balarac, Guillaume and Bourgoin, Micka\"el and Canet, L\'eonie and Mordant, Nicolas and Rossetto, Vincent},
  journal = {Phys. Rev. Fluids},
  volume = {5},
  issue = {4},
  pages = {044604},
  numpages = {19},
  year = {2020},
  month = {04},
  publisher = {American Physical Society}
}

@article{Carini2010,
  title = {Direct-numerical-simulation-based measurement of the mean impulse response of homogeneous isotropic turbulence},
  author = {Carini, Marco and Quadrio, Maurizio},
  journal = {Phys. Rev. E},
  volume = {82},
  issue = {6},
  pages = {066301},
  numpages = {10},
  year = {2010},
  month = {Dec},
  publisher = {American Physical Society},
}

@article{Mccomb1989, 
  title={Velocity-derivative skewness and two-time velocity correlations of isotropic turbulence as predicted by the {LET} theory}, 
  author={Mccomb, W. D. and Shanmugasundaram, V. and Hutchinson, P.},
  volume={208},
  journal={J. Fluid Mech.},
  year={1989}, 
  pages={91–114}
}

@article{Zhou2021,
  title = {Turbulence theories and statistical closure approaches},
  author = {Ye Zhou},
  journal = {Phys. Rep.},
  volume = {935},
  pages = {1-117},
  year = {2021},
  issn = {0370-1573}
}

@article{Grant1962,
  title={Turbulence spectra from a tidal channel}, 
  volume={12}, 
  number={2},
  journal={J. Fluid Mech.}, 
  author={Grant, H. L. and Stewart, R. W. and Moilliet, A.},
  year={1962},
  pages={241–268}
}

@article{Alexakis2023,
  title={Fluctuation relations at large scales in three-dimensional hydrodynamic turbulence},
  author={Alexakis, A. and Chibbaro, S. and Michel, G.},
  journal={EPL},
  volume={144},
  number={4},
  pages={43001},
  year={2023},
}

@article{Carnevale1983, 
title={Viscosity renormalization based on direct-interaction closure}, 
volume={131}, 
journal={J. Fluid Mech.}, 
author={Carnevale, George F. and Frederiksen, Jorgen S.},
year={1983},
pages={289–303}
}

@book{Pope2000,
  title={{Turbulent Flows}},
  author={Pope, Stephen B},
  year={2000},
  publisher={Cambridge University Press}
}

@book{Davidson2004,
  title={{Turbulence: An Introduction for Scientists and Engineers}},
  author={Davidson, Peter Alan},
  year={2004},
  publisher={Oxford University Press}
}

@article{Orszag1972,
  title={Numerical simulation of three-dimensional homogeneous isotropic turbulence},
  author={Orszag, Steven A and Patterson Jr, GS},
  journal={Phys. Rev. Lett.},
  volume={28},
  number={2},
  pages={76},
  year={1972},
  publisher={APS}
}

@article{Leith1975,
  title={Climate response and fluctuation dissipation},
  author={Leith, Cecil E},
  journal={J. Atmos. Sci.},
  volume={32},
  number={10},
  pages={2022--2026},
  year={1975},
  publisher={American Meteorological Society}
}

@article{Leith1978,
  title={Predictability of climate},
  author={Leith, C. E.},
  journal={Nature},
  volume={276},
  number={5686},
  pages={352--355},
  year={1978},
  publisher={Nature Publishing Group UK London}
}

@article{Bell1980,
  title={Climate sensitivity from fluctuation dissipation: {Some} simple model tests},
  author={Bell, Thomas L},
  journal={J. Atmos. Sci.},
  volume={37},
  number={8},
  pages={1700--1707},
  year={1980}
}

@article{Kraichnan1959FDR,
  title={Classical fluctuation-relaxation theorem},
  author={Kraichnan, Robert H},
  journal={Phys. Rev.},
  volume={113},
  number={5},
  pages={1181},
  year={1959}
}

@article{Kraichnan2000FDR,
  title={Deviations from fluctuation--relaxation relations},
  author={Kraichnan, Robert H},
  journal={Physica A},
  volume={279},
  number={1-4},
  pages={30--36},
  year={2000},
  publisher={Elsevier}
}

@article{McComb2005,
  title = {Eulerian spectral closures for isotropic turbulence using a time-ordered fluctuation-dissipation relation},
  author = {McComb, W. D. and Kiyani, K.},
  journal = {Phys. Rev. E},
  volume = {72},
  issue = {1},
  pages = {016309},
  numpages = {12},
  year = {2005},
  month = {Jul}
}

@article{North1993,
  title={Fluctuation dissipation in a general circulation model},
  author={North, Gerald R and Bell, Robert E and Hardin, James W},
  journal={Clim. Dyn.},
  volume={8},
  number={6},
  pages={259--264},
  year={1993},
  publisher={Springer}
}

@article{Lacorata2007,
  title={Fluctuation-Response Relation and modeling in systems with fast and slow dynamics},
  author={Lacorata, Guglielmo and Vulpiani, Angelo},
  journal={Nonlinear Process. Geophys.},
  volume={14},
  number={5},
  pages={681--694},
  year={2007},
  publisher={Copernicus Publications G{\"o}ttingen, Germany}
}

@article{Woodruff1992,
  title={Dyson equation analysis of inertial-range turbulence},
  author={Woodruff, S. L.},
  journal={Phys. Fluids A},
  volume={4},
  number={5},
  pages={1077--1079},
  year={1992},
  publisher={American Institute of Physics}
}

@book{Zarate2006,
  title={{Hydrodynamic Fluctuations in Fluids and Fluid Mixtures}},
  author={José M. Ortiz de Zárate and Jan V. Sengers},
  year={2006},
  publisher={Elsevier}
}

@article{Wu2020,
  title={Nonequilibrium thermodynamics of turbulence and stochastic fluid systems},
  author={Wu, Wei and Wang, Jin},
  journal={New J. Phys.},
  volume={22},
  number={11},
  pages={113017},
  year={2020},
  publisher={IOP Publishing}
}

@article{Berera2013,
  title = {Eulerian field-theoretic closure formalisms for fluid turbulence},
  author = {Berera, Arjun and Salewski, Matthew and McComb, W. D.},
  journal = {Phys. Rev. E},
  volume = {87},
  issue = {1},
  pages = {013007},
  numpages = {25},
  year = {2013},
  month = {Jan},
  publisher = {American Physical Society}
}

@article{Alexakis2019,
  title={On the thermal equilibrium state of large-scale flows},
  author={Alexakis, Alexandros and Brachet, Marc-Etienne},
  journal={J. Fluid Mech.},
  volume={872},
  pages={594--625},
  year={2019},
  publisher={Cambridge University Press}
}

@article{Kraichnan1973,
  title={Helical turbulence and absolute equilibrium},
  author={Kraichnan, Robert H},
  journal={J. Fluid Mech.},
  volume={59},
  number={4},
  pages={745--752},
  year={1973},
  publisher={Cambridge University Press}
}

@article{Dallas2015,
  title = {Statistical Equilibria of Large Scales in Dissipative Hydrodynamic Turbulence},
  author = {Dallas, V. and Fauve, S. and Alexakis, A.},
  journal = {Phys. Rev. Lett.},
  volume = {115},
  issue = {20},
  pages = {204501},
  numpages = {5},
  year = {2015},
  month = {Nov},
  publisher = {American Physical Society},
  doi = {10.1103/PhysRevLett.115.204501},
  url = {https://link.aps.org/doi/10.1103/PhysRevLett.115.204501}
}

@article{Matsumoto2014,
  title = {Response function of turbulence computed via fluctuation-response relation of a {Langevin} system with vanishing noise},
  author = {Matsumoto, Takeshi and Otsuki, Michio and Takeshi, Ooshida and Goto, Susumu and Nakahara, Akio},
  journal = {Phys. Rev. E},
  volume = {89},
  issue = {6},
  pages = {061002(R)},
  numpages = {5},
  year = {2014},
  month = {Jun},
  publisher = {American Physical Society},
  doi = {10.1103/PhysRevE.89.061002},
  url = {https://link.aps.org/doi/10.1103/PhysRevE.89.061002}
}

@article{Biferale2001,
  title = {Fluctuation-response relation in turbulent systems},
  author = {Biferale, L. and Daumont, I. and Lacorata, G. and Vulpiani, A.},
  journal = {Phys. Rev. E},
  volume = {65},
  issue = {1},
  pages = {016302},
  numpages = {7},
  year = {2001},
  month = {Dec},
  publisher = {American Physical Society},
  doi = {10.1103/PhysRevE.65.016302},
  url = {https://link.aps.org/doi/10.1103/PhysRevE.65.016302}
}

@article{Tarpin2018,
  title={Breaking of scale invariance in the time dependence of correlation functions in isotropic and homogeneous turbulence},
  author={Tarpin, Malo and Canet, L{\'e}onie and Wschebor, Nicol{\'a}s},
  journal={Phys. Fluids},
  volume={30},
  number={5},
  year={2018},
  publisher={AIP Publishing}
}

@article{Canet2017,
  title = {Spatiotemporal velocity-velocity correlation function in fully developed turbulence},
  author = {Canet, L\'eonie and Rossetto, Vincent and Wschebor, Nicol\'as and Balarac, Guillaume},
  journal = {Phys. Rev. E},
  volume = {95},
  issue = {2},
  pages = {023107},
  numpages = {8},
  year = {2017},
  month = {Feb},
  publisher = {American Physical Society},
  doi = {10.1103/PhysRevE.95.023107}
}

@article{Gorbunova2021,
  title={Spatio-temporal correlations in three-dimensional homogeneous and isotropic turbulence},
  author={Gorbunova, Anastasiia and Balarac, Guillaume and Canet, L{\'e}onie and Eyink, Gregory and Rossetto, Vincent},
  journal={Phys. Fluids},
  volume={33},
  number={4},
  year={2021},
  publisher={AIP Publishing}
}

@article{Gorce2022,
  title = {Statistical Equilibrium of Large Scales in Three-Dimensional Hydrodynamic Turbulence},
  author = {Gorce, Jean-Baptiste and Falcon, Eric},
  journal = {Phys. Rev. Lett.},
  volume = {129},
  issue = {5},
  pages = {054501},
  numpages = {6},
  year = {2022},
  month = {Jul},
  publisher = {American Physical Society}
}

@article{Tennekes1975,
  title={Eulerian and {Lagrangian} time microscales in isotropic turbulence},
  author={Tennekes, Henk},
  journal={J. Fluid Mech.},
  volume={67},
  number={3},
  pages={561--567},
  year={1975},
  publisher={Cambridge University Press}
}

@article{Chen1989,
  title={Sweeping decorrelation in isotropic turbulence},
  author={Chen, Shiyi and Kraichnan, Robert H},
  journal={Phys. Fluids A},
  volume={1},
  number={12},
  pages={2019--2024},
  year={1989},
  publisher={American Institute of Physics}
}

@article{He2017,
  title={Space-time correlations and dynamic coupling in turbulent flows},
  author={He, Guowei and Jin, Guodong and Yang, Yue},
  journal={Annu. Rev. Fluid Mech.},
  volume={49},
  pages={51--70},
  year={2017},
  publisher={Annual Reviews}
}

\end{document}